\documentclass[11pt]{article}
\pdfoutput=1

\usepackage{jheppub}
\usepackage{url,comment}
\usepackage{times}
\usepackage{latexsym}
\usepackage{graphicx, graphics, hyperref, amsmath, amssymb, slashed, color, bbm,amsthm, array}
\usepackage[usenames,dvipsnames]{xcolor}
 \usepackage{subfigure}
\usepackage{mdwlist}%for compressed lists
\usepackage{multirow}
\usepackage[normalem]{ulem}
\usepackage[e]{esvect}%longer vector arrow with \vv{#}  command
\usepackage[linewidth=1.5pt, roundcorner=2pt, innertopmargin=10pt, leftmargin=0pt, rightmargin=0pt, innerleftmargin=1pt]{mdframed}
\usepackage{cancel}
\usepackage{pstricks}
\usepackage[toc,page]{appendix}

\newcommand{\cref}[1]{Chapter~\ref{c.#1}}

\newcommand{\bea}{\begin{eqnarray}}
\newcommand{\eea}{\end{eqnarray}}

\def\simgt{\mathrel{\lower2.5pt\vbox{\lineskip=0pt\baselineskip=0pt
           \hbox{$>$}\hbox{$\sim$}}}}
\def\simlt{\mathrel{\lower2.5pt\vbox{\lineskip=0pt\baselineskip=0pt
           \hbox{$<$}\hbox{$\sim$}}}}
           
\newcommand{\FDM}{F_{\rm DM}}
\newcommand{\muchiN}{\mu_{\scriptscriptstyle \chi N}}

\graphicspath{{figs/}}

\title{Direct Detection of sub-GeV Dark Matter with Semiconductor Targets}

\author[a]{Rouven Essig, }
\emailAdd{rouven.essig@stonybrook.edu}
\affiliation[a]{C.N. Yang Institute for Theoretical Physics, Stony Brook University, Stony Brook, NY 11794-3800}

\author[b,c]{Marivi Fern\'andez-Serra, }
\emailAdd{maria.fernandez-serra@stonybrook.edu}
\affiliation[b]{Department of Physics and Astronomy, Stony Brook University, Stony Brook, NY 11794-3800}
\affiliation[c]{Institute for Advanced Computational Sciences, Stony Brook University, Stony Brook, NY 11794-3800}

\author[d]{Jeremy Mardon, }
\emailAdd{jmardon@stanford.edu}
\affiliation[d]{Stanford Institute for Theoretical Physics, Department of Physics, Stanford University, Stanford, CA 94305}

\author[b,c]{Adri\'an Soto, }
\emailAdd{adrian.soto-cambres@stonybrook.edu}

\author[e]{Tomer Volansky, }
\emailAdd{tomerv@post.tau.ac.il}
\affiliation[e]{Raymond and Beverly Sackler School of Physics and Astronomy, Tel-Aviv University, Tel-Aviv 69978, Israel}

\author[a]{Tien-Tien Yu}
\emailAdd{chiu-tien.yu@stonybrook.edu}

\date{\today}

\preprint{YITP-SB-15-30}

\abstract{
Dark matter in the sub-GeV mass range is a theoretically motivated but largely unexplored paradigm. 
Such light masses are out of reach for conventional nuclear recoil direct detection experiments, but may be detected through the small ionization signals caused by dark matter-electron scattering. 
Semiconductors are well-studied and are particularly promising target materials because their ${\cal O}(1~\rm{eV})$ band gaps allow for ionization signals from dark matter particles as light as a few hundred keV.
Current direct detection technologies are being adapted for dark matter-electron scattering. 
In this paper, we provide the theoretical calculations for dark matter-electron scattering rate in semiconductors, overcoming several complications that stem from the many-body nature of the problem. 
We use density functional theory to numerically calculate the rates for dark matter-electron scattering in silicon and germanium, and estimate the sensitivity for upcoming experiments such as DAMIC and SuperCDMS. We find that the reach for these upcoming experiments has the potential to be orders of magnitude beyond current direct detection constraints and that sub-GeV dark matter has a sizable modulation signal.
We also give the first direct detection limits on sub-GeV dark matter from its scattering off electrons in a semiconductor target (silicon) based on published results from DAMIC.  
We make available publicly our code, \href{http://ddldm.physics.sunysb.edu}{\tt QEdark}, with which we calculate our results.
Our results can be used by experimental collaborations to calculate their own sensitivities based on their specific setup.  
The searches we propose will probe vast new regions of unexplored dark matter model and parameter space.  
}
\begin{document}
\maketitle

%%%%%%%%%%%%%%%%%%%%%%%%%%%%%%%%%%%%%%%%%%%%%%%%%
\section{Introduction}
%%%%%%%%%%%%%%%%%%%%%%%%%%%%%%%%%%%%%%%%%%%%%%%%%

%%%%%%%%%%%%%%%%%%%%%%%%%%%%%%%%%%
\subsection{The search for dark matter}

There has been tremendous progress in the last three decades in the direct detection search for weak-scale dark matter (DM) using underground detectors.
The original aim was to probe the scattering through $Z$-exchange of  DM candidates with roughly weak-scale mass against nuclei~\cite{Goodman:1984dc}.
Now, experiments searching for these DM-induced nuclear recoils~\cite{Aprile:2012nq,Akerib:2013tjd,Agnese:2014aze} are sensitive to scattering cross sections many orders of magnitude below the $Z$-exchange cross section, for candidates in the $\mathcal O$(10\,GeV--10\,TeV) mass range.
The motivation behind this incredible experimental achievement has been 
the theoretically appealing, and dominant, Weakly Interacting Massive Particle (WIMP) paradigm: 
DM as a weak-scale thermal relic associated with new physics that solves the hierarchy problem.
However, the era of this paradigm's preeminence appears to be ending due to both the lack of a DM discovery, which excludes 
significant regions of WIMP parameter space, and the absence of non-Standard Model (SM) physics at colliders, 
which has undermined the theoretical motivation behind it. 
More importantly, several other theoretically motivated candidates exist for resolving this great mystery of particle physics.  

Motivated particle-DM candidates have been proposed over a vast range of masses, from ultra-light bosonic fields such as a 
QCD axion~\cite{Preskill:1982cy, Dine:1982ah, Abbott:1982af}, to non-thermal GUT-scale relics~\cite{Kolb:1998ki}.
While these have inspired a diverse array of experimental searches, 
techniques for probing them are far less developed than the WIMP search program.
One well-motivated candidate that has received increased attention recently and is the focus of this paper is light dark matter (LDM), with 
DM masses in the MeV to GeV range.  
LDM is often motivated by production mechanisms that go beyond the standard freeze-out and may be found in several frameworks in which the sub-GeV mass scale arises naturally.  
In addition, the origin for the DM relic density can be naturally addressed by several mechanisms that suggest that LDM interacts with SM particles via, for example, an exchange of a light ``dark photon'', an axion, or through an electromagnetic dipole moment.
There is a large range of parameter space of such models that evades both laboratory and astrophysical bounds~\cite{Essig:2011nj,Boehm:2003hm,Strassler:2006im,Hooper:2008im,Cholis:2008vb,ArkaniHamed:2008qn,Pospelov:2008jd,Essig:2010ye,Morrissey:2009ur, Feng:2008ya, Cohen:2010kn,Lin:2011gj, Loeb:2010gj,Tulin:2013teo, MarchRussell:2012hi, Chu:2011be, Graham:2012su, Kaplinghat:2013yxa, Boddy:2014yra, Boddy:2014qxa, Hochberg:2014dra, Hochberg:2014kqa}. 

Investigating LDM is an important and natural direction to pursue in the DM search effort.
An essential part of this pursuit is extending direct detection searches to this low mass range.
Several possible ways to do this were described in~\cite{Essig:2011nj}.  
Fortunately, much of the impressive technology being developed for the Weak-scale direct detection program can be readily adapted to search for LDM.  
An example of this was described in~\cite{Essig:2012yx}, obtaining the first direct detection limits on DM with masses as low as a few MeV using published XENON10 data.
In this work, we study in detail the even more promising possibility of semiconductor-based LDM searches, significantly expanding the preliminary work done in~\cite{Essig:2011nj}. 
Other, complementary techniques to search for LDM have been discussed in ~\cite{Bird:2004ts, McElrath:2005bp, Fayet:2006sp, Bird:2006jd,Rubin:2006gc,Tajima:2006nc, Kahn:2007ru, Fayet:2007ua,Fayet:2009tv,Yeghiyan:2009xc, delAmoSanchez:2010ac, Badin:2010uh,Echenard:2012iq,Borodatchenkova:2005ct, Essig:2009nc,Reece:2009un,Dreiner:2009ic,Aubert:2008as,Essig:2013lka,Essig:2013vha,Izaguirre:2013uxa,Boehm:2013jpa,Nollett:2013pwa,Battaglieri:2014qoa, Izaguirre:2014bca,Batell:2014mga,Va'vra:2014tia,Izaguirre:2015yja,Kahn:2014sra,Hochberg:2015pha}.

%%%%%%%%%%%%%%%%%%%%%%%%%%%%%%%%%%
\subsection{Direct detection of sub-GeV dark matter}

Current direct detection experiments are limited to probing DM masses above a few GeV due to the high energy thresholds required
for detecting nuclear recoils. 
The challenge in probing lower DM masses is twofold: for lower masses, not only is the total kinetic energy of the DM particle decreased, but so is the fraction of energy that is transferred to the nucleus. As a result, the energy of the nuclear recoils is much lower and one must drastically reduce the threshold energies to detect it. This is an experimentally challenging task, although it may be possible to probe masses down to a few hundred MeV, see~\cite{Formaggio:2011jt,Cushman:2013zza}. 
Instead, as discussed in~\cite{Essig:2011nj}, scattering channels other than elastic nuclear recoil are likely to be far more fruitful. 

A very promising avenue is to search for the small ionization signals caused directly by DM-electron scattering. The lightness of the electron and the inelastic nature of the DM-electron scattering process allow DM particles to transfer a large fraction of their kinetic energy to the electron when they scatter, enabling DM as light as $\sim$1\,MeV to cause an ionization signal. Furthermore, detecting small ionization signals is already a well-developed part of direct detection technology. 
In fact, the XENON10 experiment was already sensitive to the ionization of a single electron~\cite{Aprile:2010bt}, and results of a short single-electron-sensitive run~\cite{Angle:2011th} were used in~\cite{Essig:2012yx} to place direct detection bounds on DM with masses as low as a few MeV.  
This serves as a proof-of-principle, motivating dedicated LDM searches in other dual-phase noble liquid experiments such as XENON100 and LUX.  
However, semiconductor targets have the potential to probe even smaller cross sections. 
 In  semiconductors such as silicon or germanium, the band gap (the threshold to ``ionize'' an electron by exciting it from a valence band to a conduction band) is $\sim$1\,eV 
-- a factor of 10 to 20 times lower than the ionization threshold in liquid xenon.  
The consequences of this lower energy threshold are significant.
Not only could this allow sensitivity to DM  down to masses below an MeV, but it would also mean a substantial increase in event rate for all DM masses~\cite{Essig:2011nj, Graham:2012su}.  
The reason for this is that, given the characteristic velocities of DM particles and electrons, 
$\sim$1\,eV recoil energies are typical, while recoil energies of $\sim$10\,eV require velocities that are only found on the tails of the DM and electron velocity distributions.  
Moreover, although the background that causes an ionization signal at such low energies is still poorly understood, it is reasonable to expect that background event rates in semiconductors may be significantly lower than in xenon-based detectors~\cite{PyleFigueroa} (especially since they may be operated cryogenically)\footnote{Unlike for traditional WIMP searches, nuclear recoils are not an important background for our electron recoil signal as their rates are expected to be much lower than background-induced electron recoils.}.
There is currently an active program in, for example, both the SuperCDMS and DAMIC collaborations to develop germanium- and silicon-based detectors that are sensitive to single electron-hole pairs~\cite{PyleFigueroa, EstradaTiffenberg}, enabling a leap forward in 
LDM detection.

This developing experimental program presents a new theoretical challenge: the calculation of the expected signal rate. 
Unlike for elastic nuclear recoils, this calculation is highly non-trivial. 
In this paper, we tackle this calculation head on and present detailed new results for germanium and silicon targets.

%%%%%%%%%%%%%%%%%%%%%%%%%%%%%%%%%%
\subsection{The challenge of calculating event rates}

Several factors complicate the calculation of DM-electron scattering rates. 
Bound electrons in dense media have 
\textit{a}) typical speeds of order $\alpha\approx 1/137$ or greater, much faster than DM particles (with $v \!\sim\! 10^{-3}$), 
\textit{b}) indefinite momentum, with even very large momenta having non-zero probability, 
and \textit{c}) a complicated structure of energy-levels. 
This greatly modifies the scattering kinematics and breaks the simple link between momentum transfer and energy deposition. 
As we discuss in more detail below, event rates can be highly sensitive to the energy-level structure and the tails of the electrons' momentum distributions. 
In addition, the quantum nature of both the initial and final electron states is important,
and they cannot be correctly treated classically.
As a result, approximate calculations which do not fully account for these details may not give accurate results. 
This becomes even more important for the large energy depositions, well above $\mathcal O$(eV), since these rely on the tails of the  electron's momentum distribution. 
Once correctly calculated, the effect of all these complications can be completely encoded in 
an atomic form factor~\cite{Essig:2011nj}. 
This function is different for each specific target material, but is independent of the DM model. Once it is known, event rates can be calculated relatively simply.

When the target is an isolated noble gas atom, the combination of spherical symmetry and previously-compiled bound-state electron wavefunctions makes calculation of the ionization form factor relatively straightforward. 
Refs~\cite{Essig:2011nj, Essig:2012yx} used this as an approximation for the form factor of a liquid xenon target. 
However, calculating the form factor for a crystal target (such as a semiconductor) is far more challenging. 
A periodic crystal lattice is a complex multi-body system, with outer-shell (valence) electrons delocalized and occupying a complicated energy band-structure. 
Accurate wavefunctions of the valence electrons cannot be found analytically, but must be computed numerically with an expansion in a discrete set of plane waves\footnote{Note that inner-shell (core) electrons, which are important in some cases, are more localized so that their wavefunctions are closer to those computed assuming an isolated atom.}. 
Taking this approach, a first calculation was done in~\cite{Essig:2011nj}, assuming a single-electron threshold in a germanium target. 
A second approach was taken in~\cite{Graham:2012su}, which succeeded in simplifying the calculation until it was analytically tractable.
However, the approximations required for this were so extensive that the result might be considered only as an order-of-magnitude estimate.
A third, semi-analytic approach was taken in~\cite{Lee:2015qva} (see ``Note Added''), where numerical bound-state wavefunctions for free germanium 
and silicon atoms are used and the outgoing electrons are described by plane waves.  The latter approach gives answers much 
closer to our full numerical calculation, but important differences remain.  

%%%%%%%%%%%%%%%%%%%%%%%%%%%%%%
\subsection{Overview of the paper}

\begin{figure}[!t]
%\begin{mdframed}
%\begin{center} {\bf Prospects for Upcoming DM--Electron Scattering Searches} \end{center}
\vspace{4mm}
\includegraphics[width=0.50\textwidth]{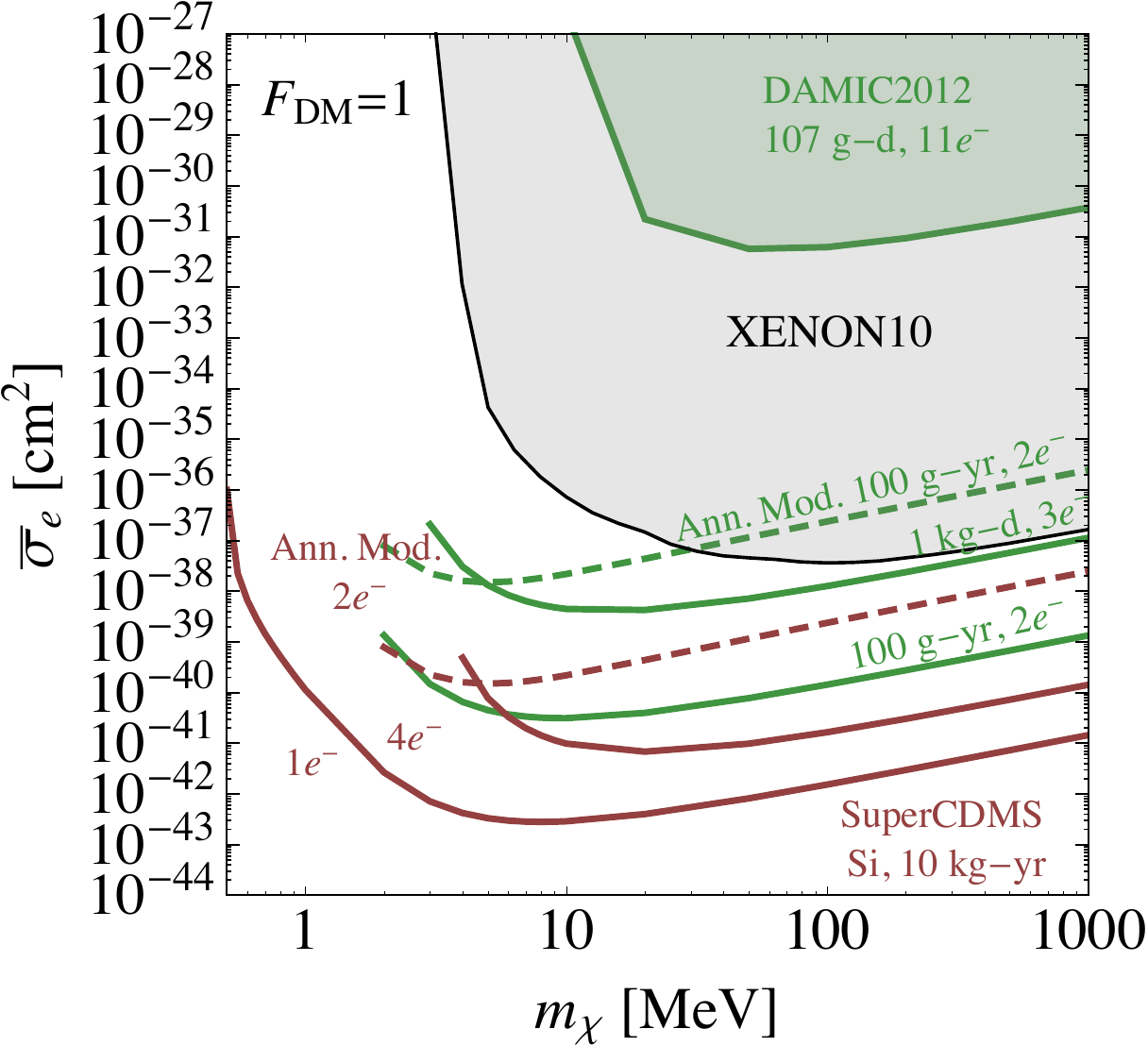}
~\includegraphics[width=0.50\textwidth]{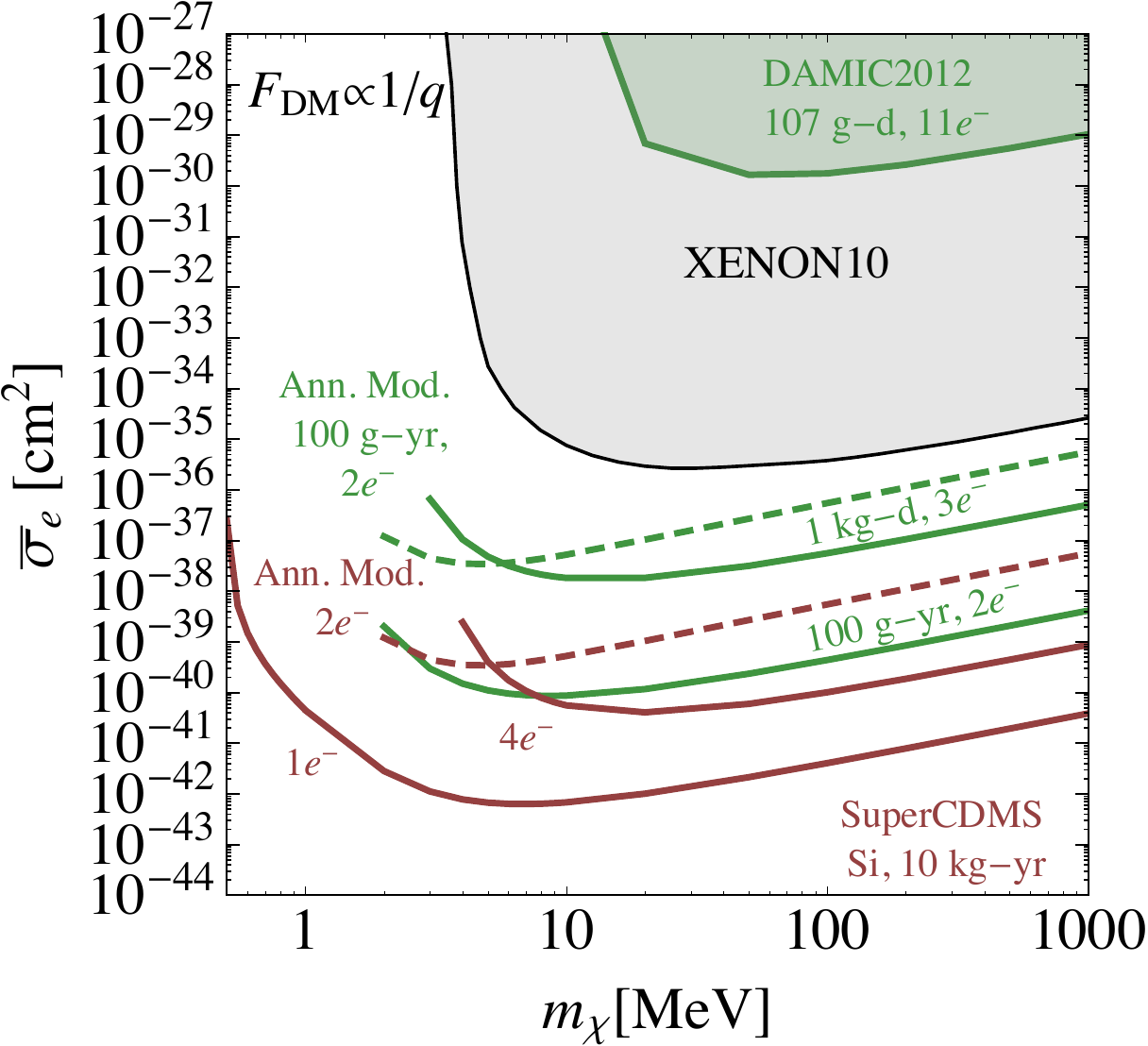}
\vspace{2mm}
\\
\begin{minipage}[t]{0.5\textwidth}
\mbox{}\\[-\baselineskip]
\includegraphics[width=\textwidth]{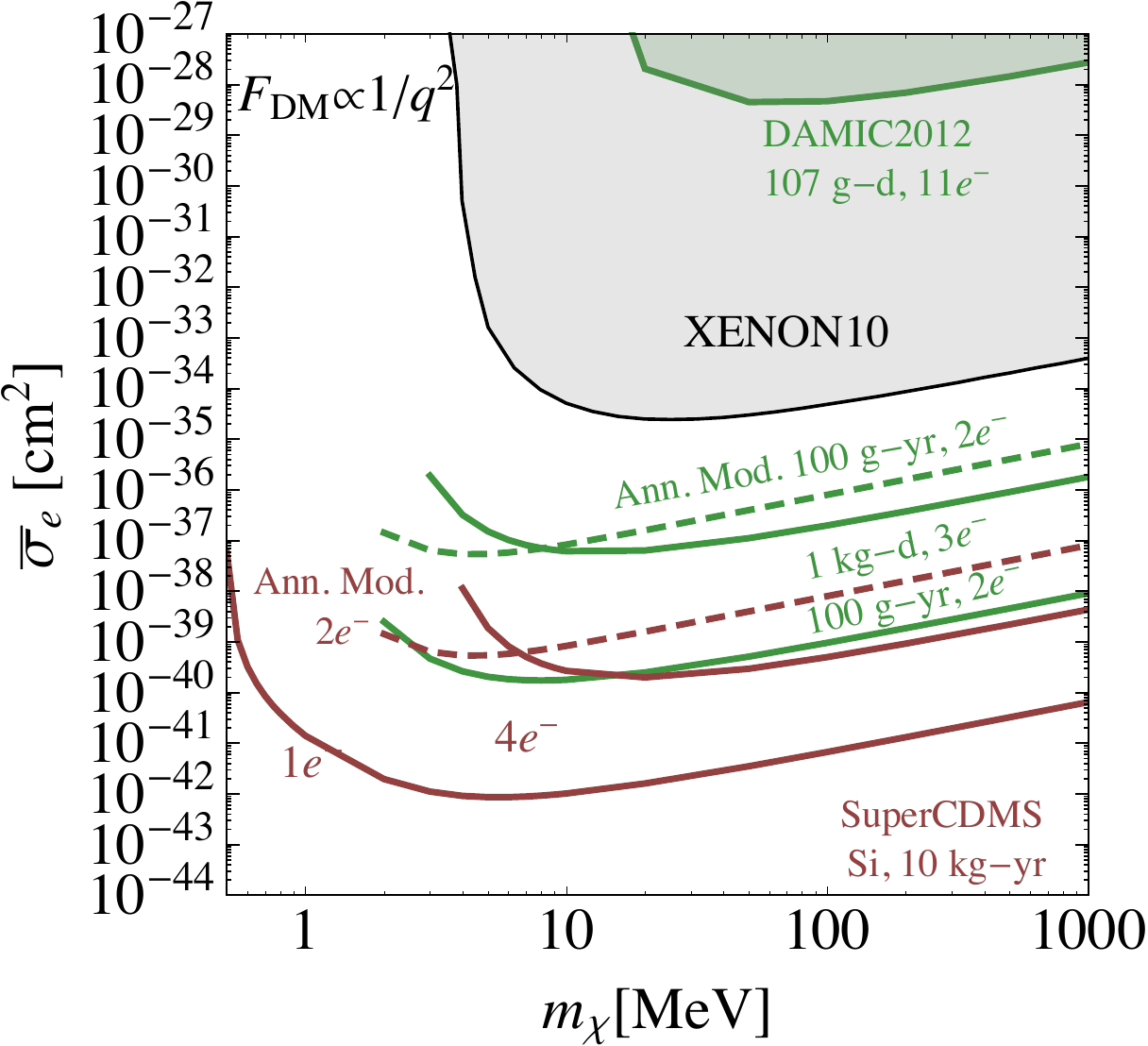}
\end{minipage}
\hfill
\begin{minipage}[t]{0.48\textwidth}
\mbox{}\\[-\baselineskip]
\vspace{-23.5pt}
\caption{\footnotesize{\bf Prospects for Upcoming DM--Electron Scattering Searches:}
Selected near-term projections of DM-electron scattering cross-section $\overline\sigma_e$ as a function of DM mass $m_\chi$ for the DAMIC (green curves) and SuperCDMS-silicon (dark red curves) experiments, 
for 
different 
ionization thresholds and (background-free) exposures, as indicated. 
Solid curves show the 
95\%~C.L.~exclusion 
reach from simple counting searches, while dashed curves show the 
$5\sigma$-discovery 
reach from annual modulation searches. 
The gray shaded region shows the current XENON10 bound~\cite{Essig:2012yx}, while the shaded green region shows the 
estimated bound from 2012 DAMIC data with a $\sim$11-electron-hole pair threshold.
The projections for SuperCDMS-germanium (not shown) are comparable to silicon. 
See \S\ref{subsec:upcoming-prospects} for more details. 
The three plots show results for the different indicated DM form factors, corresponding to different DM models. 
}
\label{fig:rates_DAMIC}
\end{minipage}
%\end{mdframed}
\end{figure}

In this paper, we present the results of a detailed numerical computation of DM-electron scattering rates in germanium and silicon targets as a function of the electron recoil energy. This significantly expands on the previous calculation in~\cite{Essig:2011nj}. 
Higher recoil energies for the scattered electron allow a larger number of additional electron-hole pairs to be promoted via secondary scattering. 
Using a semi-empirical understanding of these secondary scattering processes, 
we convert our calculated differential event rate to an estimated event rate as a function of the number of observed electron-hole pairs. 
These results will allow several experimental collaborations, such as  DAMIC and SuperCDMS,
to calculate their projected sensitivity to the DM-electron scattering cross-section, given their specific experimental setups and thresholds.
It will also allow them to derive limits on this cross section in the absence of a signal, or the preferred cross section value should there be a signal, in forthcoming data.  
Achieving low ionization thresholds could allow these experiments to probe large regions of LDM parameter space in the near future, as illustrated in Fig.~\ref{fig:rates_DAMIC}.

In \S\ref{sec:models}, we briefly discuss the direct detection prospects for a few popular LDM models.  
We will see that the upcoming generation of experiments with semiconductor targets play an essential role in testing 
these models.  
In \S\ref{sec:calculating-rates}, we outline how to calculate the rate for DM to scatter off bound electrons.  
We provide an intuitive understanding of the scattering kinematics.   
Our discussion is general and applicable to both electrons bound to (free) atoms as well as electrons in semiconductor targets.  
The details of this calculation as well as comprehensive formulas are contained in Appendix~\ref{sec:full-derivation}, significantly 
expanding on the information contained in~\cite{Essig:2011nj,Essig:2012yx}.  
We then focus on semiconductor targets, and describe the numerical computation of the scattering rates in \S\ref{sec:numerical}. 
We describe our code {\tt QEdark}, which is an additional module to the publicly available code {\tt Quantum ESPRESSO}~\cite{QE-2009}.  The latter calculates the band structure and all electron wavefunctions using density functional theory (DFT) and pseudopotentials, two established condensed matter computational tools, to calculate the Bloch wavefunction coefficients for the initial and final state electrons. 
In {\tt QEdark}, we use this information to calculate the crystal form factor for DM-electron scattering as well as the scattering rates.  
{\tt QEdark} and the crystal form factors will be publicly available at \href{http://ddldm.physics.sunysb.edu}{this link}. 
 In \S\ref{sec:conversion-to-ionization-size}, we discuss the conversion from the energy of the primary scattered electron to the size of the final ionization signal. We present a conversion formula and discuss the uncertainty associated with it.
In \S\ref{sec:results}, we present the results of our computation,  showing the cross-section sensitivity as a function of detector threshold, as well as the potential discovery reach using annual modulation.  
We also provide detailed sensitivity estimates for two representative, near-term experiments that may soon reach the required 
sensitivity to detect LDM, namely DAMIC and SuperCDMS.  
We conclude in \S\ref{sec:conclusions}.  
 The appendices contain additional technical details: 
Appendix~\ref{sec:full-derivation} provides a detailed derivation of the formulae for the scattering rate and crystal form-factor, 
Appendix~\ref{sec:etavmin} describes our choice of local DM velocity distribution, 
Appendix~\ref{sec:convergence} discusses the convergence of our numerical results, Appendix~\ref{sec:3d-shell} studies the effects of inner-shell electrons on the overall scattering rate, Appendix~\ref{sec:MC-model} presents details of the systematic study of secondary interactions, and Appendix~\ref{sec:DFT-review} gives a brief review of DFT and pseudopotentials. 

We note that our main results are contained in Figs.~\ref{fig:rates_DAMIC}, \ref{fig:A'-models}, \ref{fig:rates_ERcuts}, and \ref{fig:ann-mod-discovery-reach} and described in \S\ref{sec:models} and \S\ref{sec:results}.

%%%%%%%%%%%%%%%%%%%%%%%%%%%%%%%%%%%%
%%%%%%%%%%%%%%%%%%%%%%%%%%%%%%%%%%%%
%%%%%%%%%%%%%%%%%%%%%%%%%%%%%%%%%%%%
%%%%%%%%%%%%%%%%%%%%%%%%%%%%%%%%%%%%
\section{Models of Light Dark Matter}
\label{sec:models}
%%%%%%%%%%%%%%%%%%%%%%%%%%%%%%%%%%%%%%%%%%%%%%%

Theories of LDM have been receiving increased attention in recent years. 
Here we illustrate with just a few benchmark LDM models how the upcoming generation of experiments with 
semiconductor targets, including SuperCDMS and DAMIC, play an essential role in the search for LDM.  
Classes of models that are probed by LDM direct detection include DM that scatters through a dark-photon mediator   
or through a dipole moment interaction.  
We focus on DM coupled to a dark photon, leaving a discussion of dipole moment interactions~\cite{Sigurdson:2004zp,Graham:2012su}, the SIMP~\cite{Hochberg:2014dra,Hochberg:2014kqa}, and other models that can be constrained by electron recoils to an upcoming 
publication~\cite{futureModels}. 

For illustration, we consider models of LDM based on the vector-portal, 
in which the dark sector (and the DM particle, $\chi$) communicates with  the SM through a $U(1)_D$ gauge boson $A'$. 
The $A'$ is kinetically mixed with the SM hypercharge $U(1)_Y$ via the interaction 
\begin{eqnarray}
\mathcal{L} \supset \frac{\epsilon}{2 \cos\theta_W} F^{\mu\nu}_Y F'_{\mu\nu} \, ,
\end{eqnarray}
causing it to couple dominantly to electrically charged particles at low energies. 
Here $\epsilon$ is the kinetic mixing parameter, $\theta_W$ is the Weinberg mixing angle, and $F_Y^{\mu\nu}$ ($F'_{\mu\nu}$) is the $U(1)_Y$ ($U(1)_D$) field strength.

DM particles can scatter off electrons in direct-detection experiments through $A'$ exchange. 
In the notation of \S\ref{sec:rate calculation} below, the DM-electron reference cross section is given by
\bea \label{eq:sigma-e-A'}
\overline\sigma_e = \frac{16\pi\mu^2_{\chi e} \alpha \epsilon^2\alpha_D}{(m_{A'}^2+\alpha^2 m_e^2)^2}
\simeq
\begin{cases}
\frac{16 \pi \mu_{\chi e}^2 \alpha \epsilon^2 \alpha_D}{m_{A'}^4}\,, & m_{A'} \gg \alpha m_e \\
\frac{16 \pi \mu_{\chi e}^2 \alpha \epsilon^2 \alpha_D}{(\alpha \, m_e)^4}\,, & m_{A'} \ll \alpha m_e
\end{cases}\,,
\eea
where $\mu_{\chi e}$ is the DM-electron reduced mass and $\alpha_D\equiv g_D^2/4\pi$ 
(with $g_D$ the $U(1)_D$ gauge coupling). 
We note that this expression is the same for DM that is a complex scalar or a fermion.
The corresponding DM form factor is 
\begin{equation}
\label{eq:vectorFDM}
 F_{DM}(q) = \frac{m_{A'}^2+\alpha^2m_e^2}{m_{A'}^2+q^2} \simeq
\begin{cases}
1\,, & m_{A'} \gg \alpha m_e \\
\frac{\alpha^2 m_e^2}{q^2}\,, & m_{A'} \ll \alpha m_e
\end{cases}
\end{equation}
where $q$ is the momentum transfer between the DM and electron.

\begin{figure}
%\begin{mdframed}
%\begin{center} {\bf Prospects for Benchmark Models}\end{center}
\vspace{4mm}
\includegraphics[width=0.5\textwidth]{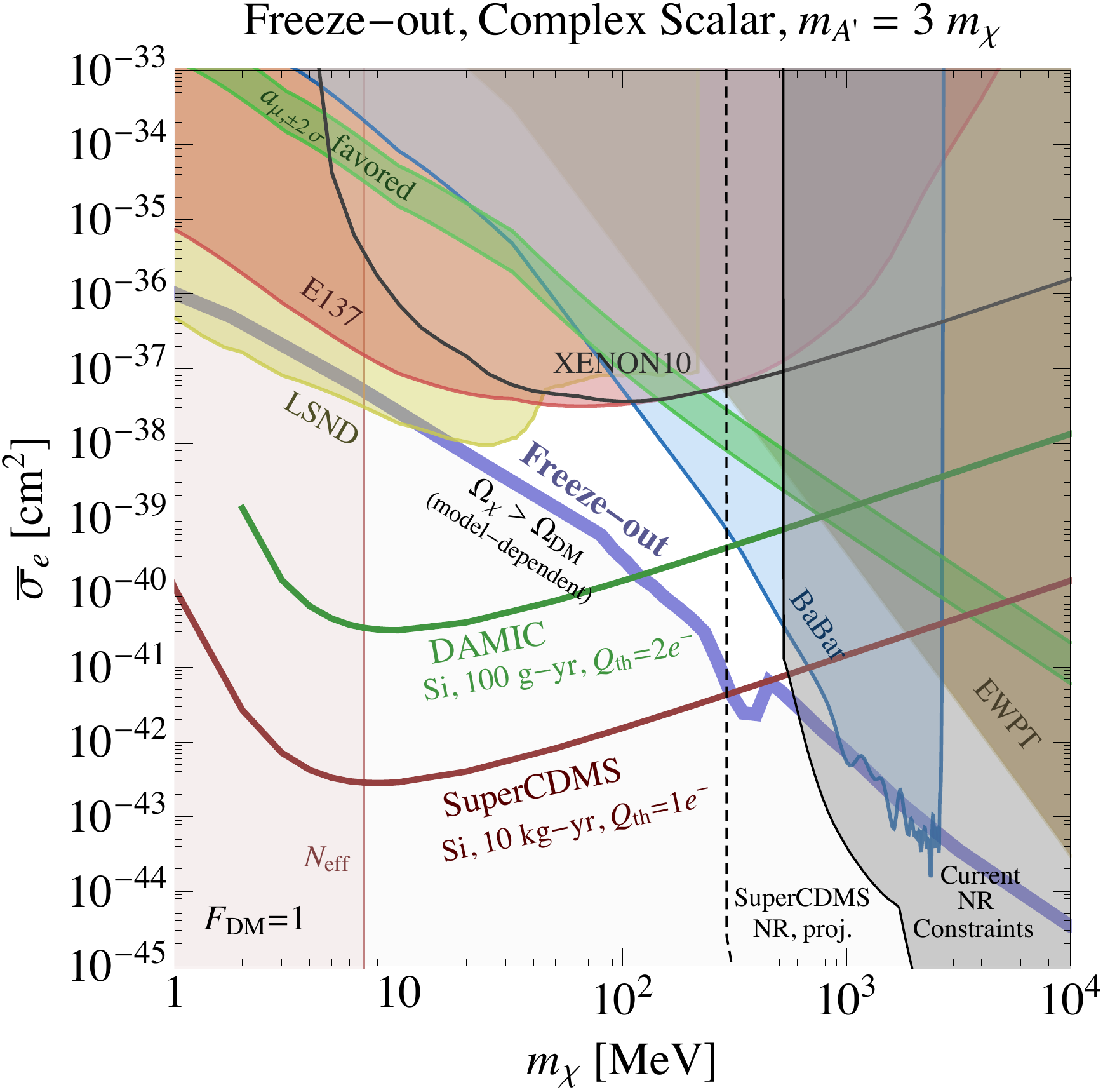}
~\includegraphics[width=0.5\textwidth]{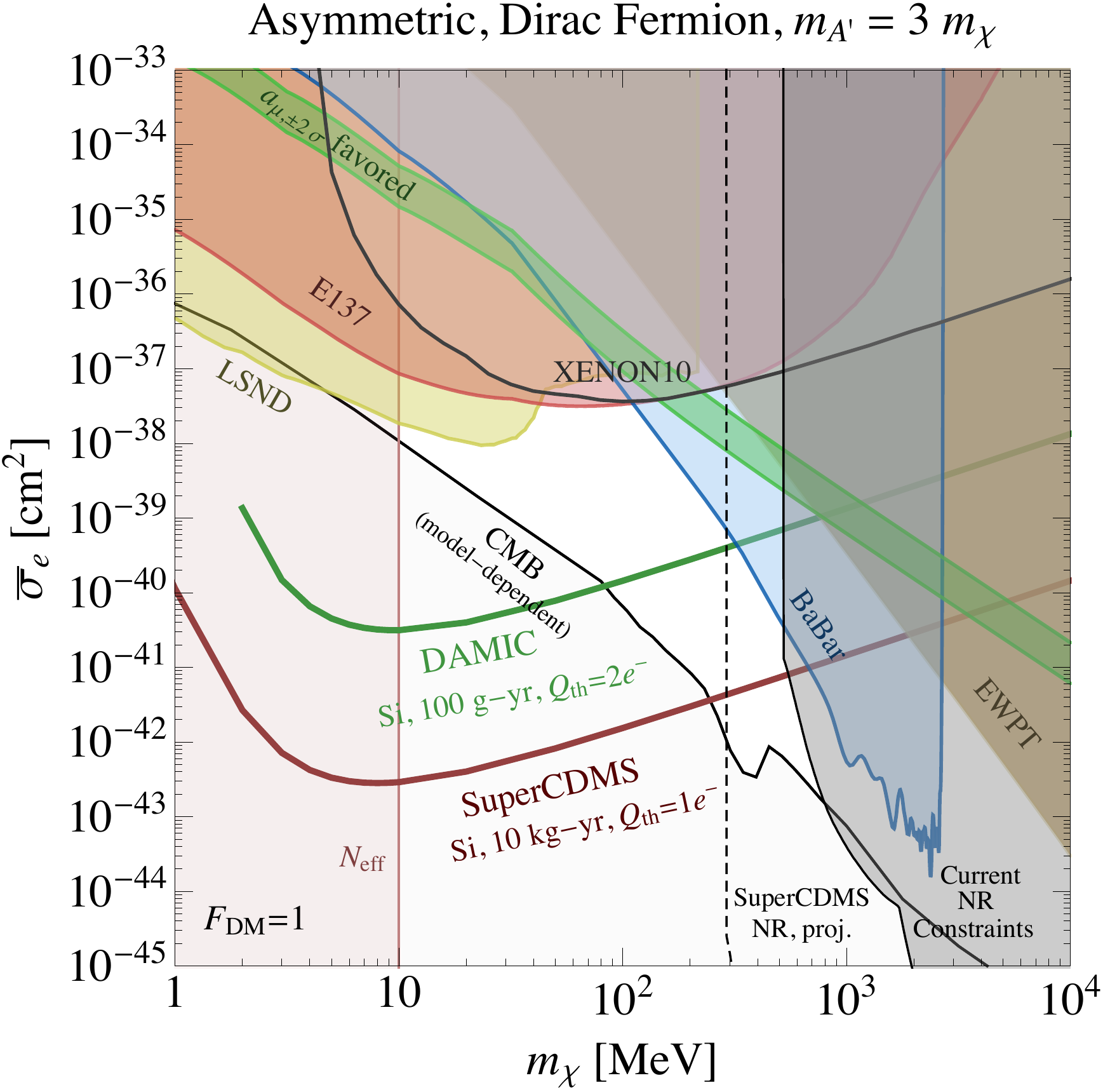}
\vspace{2mm}
\\
\begin{minipage}[t]{0.5\textwidth}
\mbox{}\\[-1.5\baselineskip]
\includegraphics[width=\textwidth]{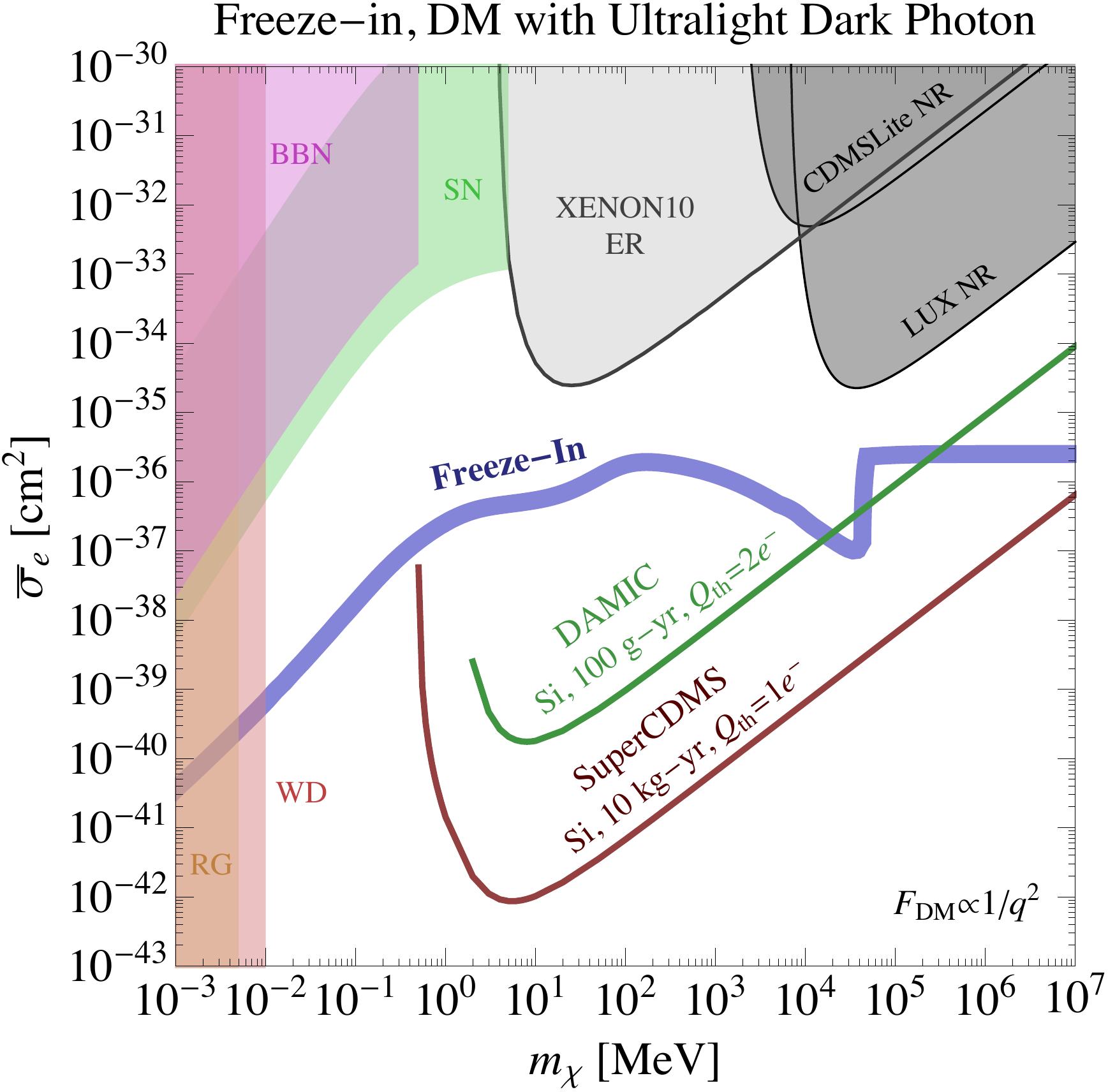}
\end{minipage}
\hfill
\begin{minipage}[t]{0.48\textwidth}
\mbox{}\\[-1.5\baselineskip]
\vspace{-23.5pt}
\caption{\footnotesize {\bf Prospects for Benchmark Models:}
Selected 95\%~C.L.~exclusion reach for the DAMIC (green curves) and SuperCDMS-silicon (dark red curves) experiments, 
compared with other constraints for the benchmark models discussed in \S\ref{sec:models}.  
White regions are unconstrained, while thick blue curves illustrate possible predictive mechanisms for generating the DM abundance. 
\textbf{Top:} DM interacting via a massive dark photon ($F_{\rm DM}(q) = 1$), for 
complex-scalar DM with freeze-out abundance (\textbf{left}), 
and Dirac-fermion DM with asymmetric abundance (\textbf{right}).
\textbf{Bottom:} DM interacting via an ultralight dark photon ($F_{\rm DM}(q) = (\alpha m_e/q)^2$), with an abundance generated by freeze-in. 
The DAMIC and SuperCDMS projections assume 100 g-year and 10 kg-years background-free exposures, with 2- and 1-electron thresholds, respectively, in a silicon target. 
See text for details. 
}
\label{fig:A'-models}
\end{minipage}
%\end{mdframed}
\end{figure}

In Fig.~\ref{fig:A'-models}, we illustrate the parameter spaces of both the $m_{A'} \gg \alpha m_e$ and $m_{A'} \ll \alpha m_e$ regimes, taking the fermionic and complex-scalar cases separately for the former.  
We study three cases, which highlight different possible production mechanisms, and show the interplay between different experimental probes.

\begin{itemize}

\item[(i)] {\bf Freeze-out via the vector portal: complex scalar LDM}\\
We consider the phenomenologically interesting and predictive region $m_{A'} > 2 m_\chi$,  
corresponding to $F_{\rm DM}(q) = 1$.
Annihilation to SM particles occurs via an off-shell $A'$ ($\chi\chi^* \to A'^* \to {\rm SM}$). 
This process is $p$-wave suppressed, allowing the DM abundance to be set by thermal freeze-out while evading constraints from the cosmic microwave background (CMB), e.g.~\cite{Madhavacheril:2013cna, Ade:2015xua}, and from gamma-rays in the Milky-Way halo~\cite{Essig:2013goa}. 
We show the parameter space for this scenario in Fig.~\ref{fig:A'-models} ({\it top left}), taking $m_{A'} = 3m_\chi$ for concreteness. 
The thick blue curve shows the cross-section for which the correct relic abundance is obtained from freeze-out~\cite{Ade:2015xua} (this is largely insensitive to the specific choice of $m_{A'}$).  
Above this line, an asymmetric DM component may complete the DM abundance. 
Below it, the abundance is naively too large, but this region may be viable with alternate hidden-sector freeze-out channels. 
We also show various constraints on this model.  
The black curve labelled ``XENON10'' shows the electron-recoil DM constraint set with XENON10 data~\cite{Essig:2012yx}. 
The black curve labelled ``Current NR Constraints'' shows constraints from conventional nuclear-recoil searches 
from~\cite{Akerib:2013tjd,Agnese:2015nto,Angloher:2015ewa}.
Some measurements only constrain $\epsilon$ as a function of $m_{A'}$.  Among these, we only show the strongest 
constraints, which are a BaBar search for $e^+e^-\to \gamma + invisible$~\cite{Aubert:2008as,Essig:2013vha,Izaguirre:2013uxa} 
as well as electroweak precision tests (EWPT)~\cite{Curtin:2014cca,Hook:2010tw}; 
however, to guide the eye, we also show 
the ``favored'' $2\sigma$-region for which the $A'$ can explain the discrepancy between the measurement and 
SM prediction for the muon anomalous magnetic moment, $a_\mu$~\cite{Pospelov:2008zw}.  
We translate these into the $\overline\sigma_e$ versus $m_\chi$ plane by using the 
constraint on $\alpha_D$ from either perturbativity~\cite{Davoudiasl:2015hxa} 
or $\chi$ self-interactions~\cite{Tulin:2013teo}. 
For these we require that $\alpha_D$ is less than 1.0 and small enough so that 
$\sigma_{\rm self-int}/m_\chi \lesssim 1$~cm$^2$/g for clusters~\cite{Randall:2007ph}. 
A second set of constraints bound some combination of $\epsilon$, $\alpha_D$, and $m_{A'}$:   
the electron beam-dump E137~\cite{Bjorken:1988as,Batell:2014mga} and the proton beam-dump 
LSND~\cite{deNiverville:2011it,Batell:2009di,Kahn:2014sra}. 
We again use the constraint on $\alpha_D$ from self-interactions and perturbativity to translate these into 
the $\overline\sigma_e$ versus $m_\chi$ plane.  
We also show a rough bound on $N_{\rm eff}$, see~\cite{Boehm:2013jpa,Nollett:2013pwa,Ade:2015xua}; 
the presence of additional relativistic degrees of freedom could allow this bound to be evaded.  
For a complementary representation of this parameter space see~\cite{Izaguirre:2015yja}.

\item[(ii)] {\bf Freeze-out via the vector portal: Dirac fermion LDM}\\
In Fig.~\ref{fig:A'-models} ({\it top right}), we consider the same scenario as in (i) but take $\chi$ to be a Dirac fermion.  
This also corresponds to $F_{\rm DM}(q) = 1$.  
The main difference between this scenario and (i) is that the annihilation cross section is now $s$-wave, so that constraints from the CMB preclude the abundance being set by freeze out.  
Instead, we assume the abundance to be asymmetric~\cite{Nussinov:1985xr, Kaplan:1991ah, Kaplan:2009ag}, 
and require the symmetric component to be small enough after freeze-out to avoid the CMB bounds~\cite{Lin:2011gj}.  
This provides a lower bound on the annihilation cross-section and thus on $\overline \sigma_e$, shown with a black solid line.   As before, this lower bound is model-dependent and can be evaded with additional annihilation 
channels. The other constraints are similar. 

\item[(iii)] {\bf Freeze-in via the vector portal}\\
In Fig.~\ref{fig:A'-models} ({\it bottom}), we consider an ultra-light $A'$ mediator ($m_{A'}\ll \alpha m_e$), corresponding to $F_{\rm DM}(q) =(\alpha m_e/q)^2$.  
Here the couplings are so small that the DM would never have thermalized with the SM sector. 
The $\chi$ abundance can receive an irreducible ``freeze-in''~\cite{Hall:2009bx} contribution from $2\to2$ annihilation of SM particles to 
$\chi\bar\chi$ as well as $Z$-boson decays to $\chi\bar\chi$, computed in~\cite{Essig:2011nj} (see also~\cite{Chu:2011be}).  
The parameters required for the abundance again uniquely constrain $\overline\sigma_e$ versus $m_\chi$, as shown by the thick blue curve.  
In addition to the XENON10 electron-recoil constraint~\cite{Essig:2012yx}, 
we also show the bounds from conventional nuclear-recoil searches. 
The nuclear recoil cross-section, $\sigma_{\rm NR}$, can be related to the electron recoil cross-section by 
\begin{equation}
\frac{d \langle \sigma_{\rm NR} v \rangle}{d E_{\rm NR}} = \frac{Z^2 \overline \sigma_e (\alpha m_e)^4}{8 \mu_{\chi e}^2 m_N E_{\rm NR}^2} \eta(v_{\rm min, NR}) \, ,
\end{equation}
where the target nucleus has mass $m_N$ and atomic number $Z$, $E_{\rm NR}$ is the nuclear recoil energy, and $v_{\rm min, NR}=\sqrt{2 m_N E_{\rm NR}}/(2 \mu_{\chi N})$, $v$ is relative velocity of the DM, and $\eta$ is the inverse mean speed defined in Appendix \ref{sec:etavmin}. 
Since this recoil spectrum is peaked towards low energies more than for a contact interaction, determining accurate DM constraints requires a careful analysis of the experimental data. We place approximate bounds from ``CDMSLite''~\cite{Agnese:2013jaa} and LUX~\cite{Akerib:2013tjd} results, taking the former to have 6.2\,kg-days germanium exposure, a 0.84\,keV threshold, 100\% signal efficiency and 10 observed events, and taking the latter to have 10\,tonne-days xenon exposure, a 5\,keV threshold, 50\% signal efficiency and 0 observed events. 
Due to the smallness of the couplings, the other constraints seen in the previous scenarios are absent in this one. Instead, we include various astrophysical constraints on millicharged particles, which are also applicable for DM coupled to an ultralight $A'$~\cite{Davidson:2000hf}.

\end{itemize}

In each of the figures in Fig.~\ref{fig:A'-models} we show the prospects for DAMIC (100 g-years, silicon target, 2-electron threshold) 
and SuperCDMS (10 kg-years, silicon, 1-electron threshold), discussed in \S\ref{subsec:upcoming-prospects}.  
We note that a magnetic-dipole-moment interaction would also give $F_{\rm DM}(q) = 1$, 
while an electric-dipole-moment interaction would give $F_{\rm DM}(q) = \alpha m_e/q$.  
We see that these models above all have concrete predictions that the upcoming generation of direct detection experiments can test.

%%%%%%%%%%%%%%%%%%%%%%%%%%%%%%%%%%%%%%%%%%%%%%%%%
%%%%%%%%%%%%%%%%%%%%%%%%%%%%%%%%%%%%%%%%%%%%%%%%%
%%%%%%%%%%%%%%%%%%%%%%%%%%%%%%%%%%%%%%%%%%%%%%%%%
%%%%%%%%%%%%%%%%%%%%%%%%%%%%%%%%%%%%%%%%%%%%%%%%%
\section{Direct detection of dark matter by electron scattering in semiconductors}
\label{sec:calculating-rates}
%%%%%%%%%%%%%%%%%%%%%%%%%%%%%%%%%%%%%%%%%%%%%%%%%

In this section, we review the theory of DM scattering with bound electrons.
We begin in \S\ref{sec:scattering-kinematics} by considering the simple kinematics of LDM scattering with both nucleons and electrons. 
This makes clear the motivation for using electron recoils to probe LDM.
The discussion also shows that the DM-electron scattering rate is expected to be sensitive to the details of electron binding in the target, \emph{especially for higher energy/ionization thresholds}.
A consequence of this is that to calculate accurate scattering rates, detailed modeling of the electronic structure of the target material is required, involving knowledge of the wavefunctions of all accessible occupied and unoccupied electron levels.

In \S\ref{sec:rate calculation}, we summarize how this scattering-rate calculation is formulated, with a focus on the case of semiconductor targets. 
The key results are Eqs.~(\ref{eq:diff-crystal-rate}) and~(\ref{eq:crystal-form-factor}). 
The former gives the differential scattering rate in terms of the DM model, the DM velocity profile, and crystal form factor.
The latter gives the  \emph{crystal form-factor}, which encodes all the relevant electron binding effects for a given target material. 
This reviews and extends the discussion from Ref.~\cite{Essig:2011nj}. 
In Appendix~\ref{sec:full-derivation}, we provide a full derivation of all the results given here. 
For the interested reader, in Appendix~\ref{sec:atomic-ionization-rate-derivation} we present a derivation of the ionization rate in free atomic targets, as is relevant for xenon targets and which was used in Refs.~\cite{Essig:2011nj,Essig:2012yx}.

%%%%%%%%%%%%%%%%%%%%%%%%%%%%%%
\subsection{Kinematics of dark matter-electron scattering}
\label{sec:scattering-kinematics}

Conventional DM direct detection experiments assume that the DM particle scatters elastically off a target nucleus. This recoiling nucleus then collides with the surrounding matter within the detector, giving off energy in the form of heat, phonons, ionized electrons, scintillation photons, etc, depending on the detector material. However, if the DM particle is light, the momentum transfer, $\vec q$, between the DM and the target nucleus is small and may not provide enough energy for the recoil of the nucleus to be detected. We can see this through the following argument. The energy of the recoiling nucleus in nuclear scattering is 
\begin{equation}
E_{ \rm NR}=\frac{q^2}{2 m_N} \le \frac{2\muchiN^2 v^2}{m_N} \simeq 1 \mathrm{~eV} \times\left(\frac{m_\chi}{100\mathrm{~MeV}}\right)^2\left(\frac{20\mathrm{~GeV}}{m_N}\right) \,.
\label{eq:E_NR}
\end{equation}
For the scaling in the last step of this equation, we have taken the typical DM speed to be $300$ km/s $\approx 10^{-3}c$, and assumed 
$m_\chi \ll m_N$. For $m_\chi=30$ GeV, we find $E_{\rm NR}\sim2$ keV. 
However, if we consider lighter DM masses, such as $m_\chi=100$ MeV, the recoil energy drops to $E_{\rm NR}\sim$ eV, which is well 
below the detection thresholds of current direct detection experiments (e.g.~$\sim 840$~eV$_{\rm NR}$ for CDMSlite~\cite{Agnese:2013jaa} 
and $\sim 4$~keV$_{\rm NR}$ for LUX~\cite{Akerib:2013tjd}).  
Note that the energy of the recoiling nucleus is also not efficiently transferred to electrons, and so is not nearly large enough to ionize 
or excite even a single electron; it is also well below current phonon detection thresholds.
As a result, DM masses below a few hundred MeV escape detection no matter how large their cross section. 

\begin{figure}[!htbp]
\begin{center}
\includegraphics[scale=0.6]{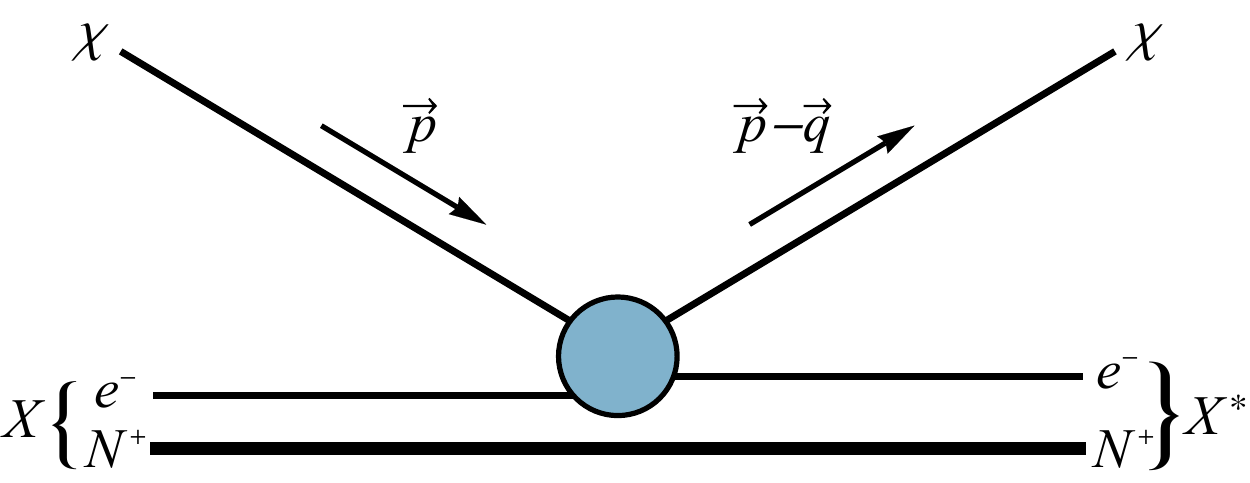}
\caption{\footnotesize The scattering of a DM particle with a bound electron. The DM transfers momentum $\vec q$ to the target, exciting it from the ground state $X$ to an excited state $X^*$, which can be either a higher-energy bound state or an ionized state. }
\label{fig:momenta}
\end{center}
\end{figure}

Now consider a DM particle colliding directly with a bound electron, exciting it to a higher energy level or an unbound state, as illustrated in Fig.~\ref{fig:momenta}. The kinematics are very different from those of a nuclear recoil.  
Firstly, being in a bound state, the electron does not have definite momentum -- in fact it may have arbitrarily high momentum (albeit with low probability).
This breaks the direct relation between recoil energy and momentum transfer given in Eq.~(\ref{eq:E_NR}). 
The energy transferred to the electron, $\Delta E_e$, can still be related to the momentum lost by the DM, $\vec q$, via energy conservation:
\begin{equation}
\Delta E_e = - \Delta E_\chi - \Delta E_N 
= - \frac{|m_\chi \vec v - \vec q|^2}{2 m_\chi} + \frac{1}{2} m_\chi v^2 -\frac{q^2}{2 m_N}
=  \vec q \cdot \vec v - \frac{q^2}{2 \muchiN} \, .
\label{eq:delta-E-electron}
\end{equation}
Here the $\Delta E_N$ term accounts for the fact that the whole atom also recoils. In practice this term is small, which also allows 
us to replace $\mu_{\chi N}$ with $m_\chi$.  We thus define
\bea
E_e \equiv \Delta E_e = -\Delta E_\chi
\eea
as the energy transferred to the electron.\footnote{We emphasize that $E_e$ is the energy transferred to the electron,  
not its kinetic energy. 
Some of this energy goes to overcoming the  binding energy.  
As we will discuss further in \S\ref{sec:conversion-to-ionization-size}, in semiconductors the remaining energy is rapidly redistributed by secondary scattering processes, which can produce further electron-hole pairs.}
Since an arbitrary-size momentum transfer is now possible, the largest allowed energy transfer is found by maximizing $\Delta E_e$ with respect to $\vec q$, giving 
\begin{equation}
\Delta E_e \leq \frac{1}{2} \muchiN v^2 \simeq \frac{1}{2} \mathrm{~eV} \times \left(\frac{m_\chi}{{\rm MeV}}\right)\, .
\label{eq:delta-E-max}
\end{equation}
This shows that \emph{all} the kinetic energy in the DM-atom collision is (in principle) available to excite the electron. 
For a semiconductor with an $\mathcal{O}$(eV) bandgap, ionization can be caused by DM as light as $\mathcal{O}$(MeV).

What is the likelihood of actually obtaining a large enough $q$ to excite the electron? This brings us to the second major difference compared to DM-nuclear scattering: the electron is both the lightest and fastest particle in the problem. 
The typical velocity of a bound electron is $v_e \sim Z_{\rm eff} \alpha$, where $Z_{\rm eff}$ is 1 for outer shell electrons and larger for inner shells. 
This is much greater than the typical DM velocity  of $v\sim10^{-3}$. 
The typical size of the momentum transfer is therefore set by the electron's momentum, 
\begin{equation}
q_{\rm typ} \simeq \mu_{\chi e} v_{\rm rel} \simeq m_e v_e \sim Z_{\rm eff} \alpha m_e \simeq Z_{\rm eff} \times 4\,{\rm keV} 
\label{eq:typical-q}
\end{equation}
where $v_{\rm{rel}}$ is the relative velocity between the DM and electron.

Returning to Eq.~(\ref{eq:delta-E-electron}), the first term on the right dominates as long as $m_\chi$ is well above the bound in Eq.~(\ref{eq:delta-E-max}). This gives a formula for the minimum momentum transfer required 
to obtain an energy $\Delta E_e$: 
\begin{equation}
	q\gtrsim \frac{\Delta E_e}{v} \sim \frac{\Delta E_e}{4 \, Z_{\rm eff} \, \text{eV}} \times q_{\rm typ} \, .
\label{eq:q-min-approx}
\end{equation}
This scaling suggests that the typical available momentum is enough to cause a transition of just a few eV, such as for an electron being excited just across the germanium or silicon bandgap.
Exciting a more energetic transition will require a momentum out on the tail of the electron's momentum-space wavefunction (or probing the tail of the DM velocity distribution), and its probability will be correspondingly suppressed 
(as can be seen clearly in Fig.~\ref{fig:crystalFF} below, which we will discuss in \S\ref{subsec:ff}). 
Ionization of a xenon atom, requiring $\sim$10\,eV energy, falls into the second category, as do most possible transitions to the conduction band in germanium or silicon.

From this argument we expect the rate of DM-electron scattering to be sensitive to the precise forms of the electron energy levels and wavefunctions in the target.
The computation we present below is designed to address this sensitivity by modeling in detail the electronic structure in germanium 
and silicon crystals.
A corollary of this argument is that, given the $v$-dependence in Eq.~(\ref{eq:q-min-approx}), the rate should also be sensitive to the DM velocity profile. 
As this varies over the year, we expect a significant annual modulation in the signal size, a potentially crucial test of the DM origin of a signal.
We discuss the expected annual modulation in \S\ref{sec:annual-mod}.

%%%%%%%%%%%%%%%%%%%%%%%%%%%%%%
\subsection{Calculating excitation rates}
\label{sec:rate calculation}

\subsubsection{General formulation for dark matter-induced electron transitions}

If a DM particle scatters with an electron in a stationary bound state such as in an atom, it can excite the electron from an initial energy level 1 to an excited energy level 2 by transferring energy $\Delta E_{1\to2}$ and momentum $\vec q$. 
The cross section for this process takes quite a different form to the free elastic scattering cross section.

If $\mathcal M_{\rm free}(\vec q\,)$ is the matrix element for free elastic scattering of a DM particle and an electron, then we parametrize the underlying DM--electron coupling using the following definitions~\cite{Essig:2011nj}:
\begin{gather}
\overline{|\mathcal M_{\rm free}(\vec q\,)|^2} \equiv \overline{|\mathcal M_{\rm free}(\alpha m_e)|^2} \times |F_{\rm DM}(q)|^2
\label{eq:DM-form-factor}
\\
\overline \sigma_e \equiv \frac{\mu_{\chi e}^2 \overline{|\mathcal M_{\rm free}(\alpha m_e)|^2}}{16 \pi m_\chi^2 m_e^2} \, ,
\label{eq:sigma-bar-e}
\end{gather}
where $\overline{|\mathcal M|^2}$ is the absolute square of $\mathcal M$, {\it averaged over initial and summed over final particle spins}.
The \emph{DM form factor}, $F_{\rm DM}(q)$, gives the momentum-transfer dependence of the interaction -- for example, $F_{\rm DM}(q)=1$ results from a point-like interaction induced by the exchange of a heavy vector mediator or magnetic dipole moment coupling, $F_{\rm DM}(q)=(\alpha m_e / q)$ for an electric dipole moment coupling, and $F_{\rm DM}(q)=(\alpha m_e / q)^2$ for exchange of a massless or ultra-light vector mediator (see \S\ref{sec:models}). 
$\overline \sigma_e$ parameterizes the strength of the interaction, and in the case of $F_{\rm DM}(q)=1$ is equal to the cross section for free elastic scattering. 
All sensitivity estimates or constraints on LDM will be given for $\overline \sigma_e$, which plays the 
analogous role to $\sigma_{\chi N}$, the DM-nucleon scattering cross section, in (WIMP) DM scattering with nuclei.   
 
With these definitions, the cross section for a DM particle to excite an electron from level 1 to level 2 can be written as (see Appendix~\ref{sec:general-excitation-derivation}) 
\begin{equation}
\sigma v_{1\to 2} = 
\frac{\overline \sigma_e}{\mu_{\chi e}^2} \int \frac{d^3 q}{4 \pi} \,\delta \Big(\Delta E_{1\to2} + \frac{q^2}{2 m_\chi} - \vec q \cdot \vec v \Big)  \times |F_{\rm DM}(q)|^2 | f_{1\to 2}(\vec q \,)|^2 \,,
\label{eq:1to2-sigmav}
\end{equation}
where $f_{1\to2}(\vec q\,)$ is the \emph{atomic form factor} for the excitation. It is given by
\begin{equation}
f_{1\to 2}(\vec q \,) = \int d^3 x \, \psi_2^*(\vec x) \psi_1(\vec x) e^{i \vec q \cdot \vec x} \,,
\label{eq:1to2-form-factor}
\end{equation}
where $\psi_1$ and $\psi_2$ are the normalized wavefunctions of the initial and final electron levels.
We now apply this general result to the special case of electrons in a periodic crystal lattice, such as a semiconductor.

%%%%%%%%%%%%%%%%%%%%%%%%%%%%%%%%%%%%%
\subsubsection{Excitation rate in a semiconductor crystal}
\label{sec:excitation-in-crystal}

\begin{figure}[!htbp]

\hspace{-6mm}
\includegraphics[trim={0 0 47.45 0}, clip, height=2.8in]{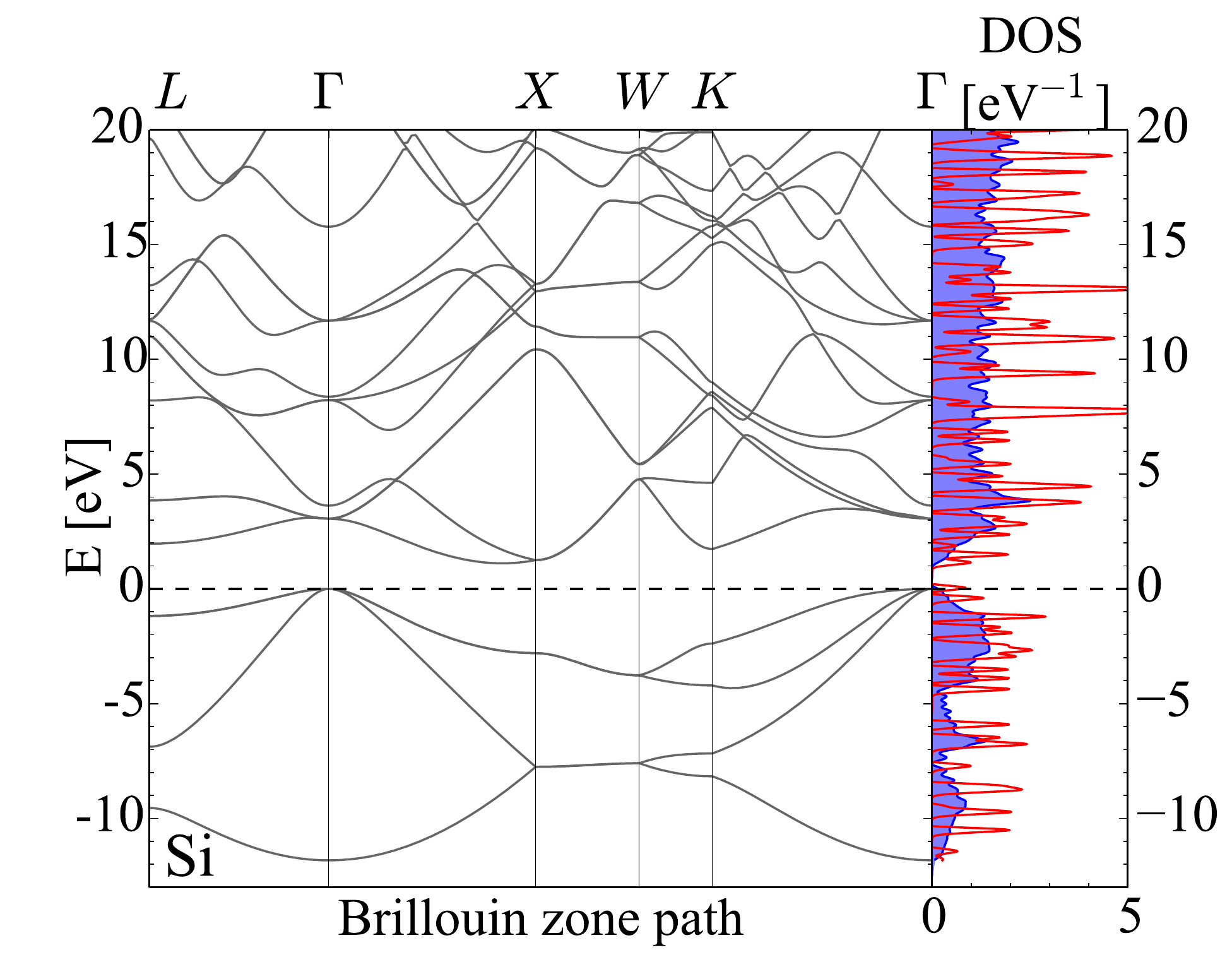}
\hspace{-0.14625cm}
\includegraphics[trim={11.8175 0 0 0}, clip, height=2.8in]{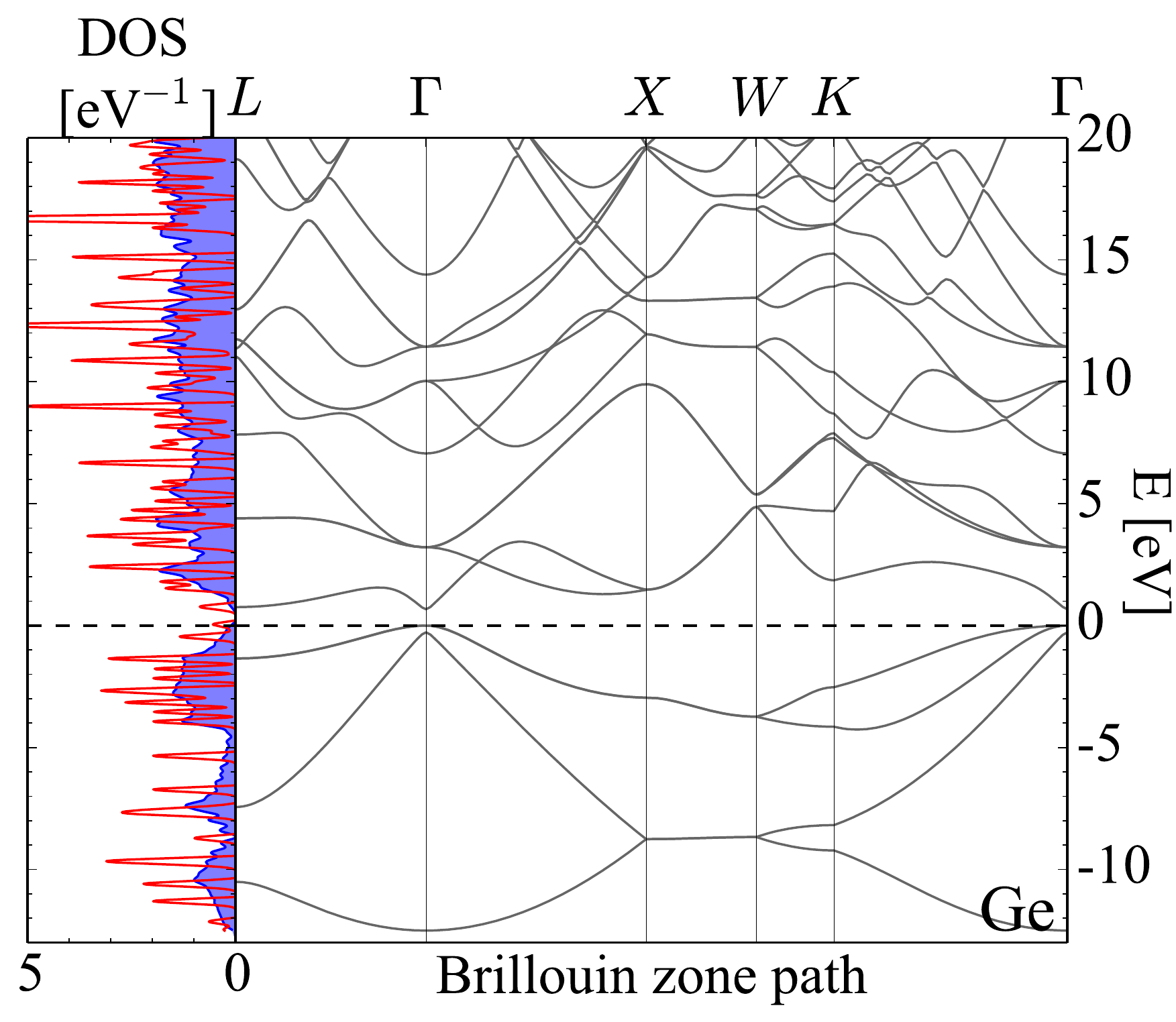}
\caption{\footnotesize Scissor corrected band structure for silicon ({\bf left}) and germanium ({\bf right}) as calculated with \texttt{Quantum ESPRESSO} \cite{QE-2009} with a very fine k-point mesh. The horizontal dashed line indicates the top of the highest valence band. The four bands below the horizontal dashed line are the valence bands while the bands above the dashed line are the conduction bands.  We also show the density-of-states (DOS) as a function of the energy for a very fine k-point mesh (blue) and for our 243 k-point mesh (red). A Gaussian smearing of $0.15$ eV was used to generate a smooth function.}
\label{fig:bandstructure}
\end{figure}

The periodic lattice of a semiconductor crystal has a continuum of electron energy levels, forming a complicated band structure (see Fig.~\ref{fig:bandstructure}). 
A small energy gap separates the occupied valence bands from the unoccupied conduction bands; 
exciting electrons across this bandgap creates mobile electron-hole pairs, which can be manipulated and detected. 
In order to perform practical calculations for this system, the true multi-body electron wavefunction must be replaced with a product of single-particle wavefunctions (this is a well-understood procedure, which we discuss further in \S\ref{sec:numerical}). 
Once found, these single-particle wavefunctions can be used in Eqs.~(\ref{eq:1to2-sigmav}) and (\ref{eq:1to2-form-factor}), giving the cross-section to excite an electron between specific energy levels. 
To find the total rate, these cross sections are integrated over initial and final electron levels, and over the DM velocity distribution.

\paragraph{DM halo dependence.}
Neither the electron band structure, nor the electron wavefunctions, nor the DM velocity distribution are spherically symmetric.  
As noted in~\cite{Essig:2011nj}, the excitation rate will therefore depend on the orientation of the crystal with respect to the galaxy, an effect which may be extremely useful in verifying the DM origin of a signal. 
Here, however, we sidestep this complication by approximating the DM velocity distribution as being a spherically symmetric function $g_\chi(v)$. 
All the relevant information about the DM velocity profile can then be encoded in the function $\eta(v_{\rm min})$ (see Appendix~\ref{sec:velocity-averaging}), defined as
\begin{equation}
\eta(v_{\rm min}) = \int \frac{d^3 v}{v} \, g_\chi(v) \,\Theta (v- v_{\rm min}) 
\label{eq:eta}
\end{equation}
where $\Theta$ is the Heaviside step function. 

When calculating rates, we assume a Maxwell-Boltzmann distribution with a sharp cutoff 
(we describe this in more detail, and give analytic formulas for $\eta(v_{\rm min})$, in Appendix~\ref{sec:etavmin}).  
The requirement of energy conservation is captured by $v_{\rm min}(q, E_e)$, the minimum speed a DM particle requires in 
order for the electron to gain an energy $E_e$ with momentum transfer $q$ (note that $E_e$ was also denoted as $\Delta E_e$ 
in \S\ref{sec:scattering-kinematics}). This is given by
\begin{equation}
v_{\rm min}(q, E_e) = \frac{E_e}{q} + \frac{q}{2 m_\chi} \, .
\label{eq:vmin}
\end{equation}

\paragraph{Differential rate.}
As we show in Appendix~\ref{sec:crystal-excitation-rate-derivation}, the differential electron scattering rate in a semiconductor target (with the approximation of a spherically symmetric DM velocity distribution) can be written as
\begin{align}
\frac{d R_{\rm crystal}}{d \ln E_e} =& 
\frac{\rho_\chi}{m_\chi}\ N_{\rm cell}\ \overline{\sigma}_e\ \alpha
\nonumber \\ 
& \quad \times \frac{m_e^2}{\mu_{\chi e}^2}
\int\! d \ln q \, \bigg(\frac{E_e}{q} \eta \big( v_{\rm min}(q, E_e) \big)\bigg) 
F_{\rm DM}(q)^2 \big| f_{\rm crystal}(q, E_e) \big|^2 \, ,
\label{eq:diff-crystal-rate}
\end{align}
where $\rho_\chi\simeq 0.4$~GeV/cm$^3$ is the local DM density, $E_e$ is the total energy deposited, 
and $N_{\rm cell} = M_{\rm target}/M_{\rm cell}$ is the number of unit cells in the crystal target. 
($M_{\rm cell}=2\times m_{\rm Ge}=145.28~{\rm amu}=135.33~{\rm GeV}$ for germanium, and $M_{\rm cell}=2\times m_{\rm Si} = 56.18~{\rm amu}=52.33~{\rm GeV}$ for silicon.) 
We have written this in such a way that the first line gives a rough estimate of the rate, about 29~(11)~events/kg/day for silicon 
(germanium) for $\rho_\chi =0.4$~GeV/cm$^3$, $m_\chi = 100$~MeV, and $\overline\sigma_e \simeq 3.6\times 10^{-37}$~cm$^2$ 
(the current limit from XENON10~\cite{Essig:2012yx}), while every factor in the second line is a roughly $O(1)$ number for the preferred values of $q$ and $E_e$. 

All the necessary details of the target's electronic structure are contained in the dimensionless \emph{crystal form factor}, $f_{\rm crystal}(q, E_e)$, which is a property purely of the target material and is independent of any DM physics. 
The computation of this form factor is one of the main results of this paper.

\paragraph{Crystal form factor.}
In the periodic lattice of a semiconductor crystal, each electron energy level is labelled by a continuous wavevector $\vec k$ in the first Brillouin Zone (BZ), and by a discrete band index $i$. 
The wavefunctions of these states can be written in Bloch form,
\begin{equation}
\psi_{i \vec k}(\vec x) = \frac{1}{\sqrt V} \sum_{\vec G} u_i(\vec k + \vec G) e^{i (\vec k + \vec G)\cdot \vec x} \, ,
\label{eq:Bloch-wavefunctions}
\end{equation}
where the $\vec G$'s are the reciprocal lattice vectors. 
Here $V$ is the volume of the crystal, and the wavefunctions are taken to be unit-normalized, so that
\begin{gather}
\sum_{\vec G} \big| u_i (\vec k + \vec G) \big|^2 = 1 \, .
\end{gather} 

Using this form for the wavefunctions, we can define the form factor for excitation from valence level $\{i \, \vec k \}$ to conduction level $\{i' \, \vec k' \}$,
\begin{equation}
f_{[i \vec k , i' \vec k', \vec G']} = \sum_{\vec G} u_{i'}^*(\vec k' + \vec G + \vec G') u_i (\vec k + \vec G) \, .
\label{eq:crystal-form-factor-basis}
\end{equation}
The \emph{crystal form factor} required in Eq.~(\ref{eq:diff-crystal-rate}) is then given by
\begin{align}
\big| f_{\rm crystal}(q, E_e) \big|^2 =& 
\frac{2\pi^2 (\alpha m_e^2 V_{\rm cell})^{-1}}{E_e}  \sum_{i\,  i'} \! \int_{\rm BZ} \frac{V_{\rm cell}\ d^3 k }{(2 \pi)^3} \  \frac{V_{\rm cell} \ d^3 k'}{(2 \pi)^3} \times
\nonumber \\
& E_e \ \delta(E_e - E_{i' \vec k'} + E_{i \vec k}) 
\sum_{\vec G'} q \ \delta(q - |\vec k' - \vec k + \vec G'|)\
\big| f_{[i \vec k , i' \vec k', \vec G']} \big|^2 \, .
\label{eq:crystal-form-factor}
\end{align}
 (See Appendix~\ref{sec:crystal-excitation-rate-derivation} for the derivation.)
The band index $i$ is summed over the filled energy bands, while $i'$ is summed over unfilled bands, and the momentum 
integrals are over the 1st BZ. 
$E_{i \vec k}$ is the energy of level $\{i \, \vec k \}$, and $V_{\rm cell}$ is the volume of the unit cell. The numerator in the first factor has units of energy, with value $2\pi^2 (\alpha m_e^2 V_{\rm cell})^{-1}=1.8\,$eV for germanium and 2.0\,eV for silicon. 
The crystal form factor can be computed numerically using established solid-state computational techniques. 
Once it is known, it can be used to find event rates for any DM model and halo profile, using Eq.~(\ref{eq:diff-crystal-rate}), along with Eqs.~(\ref{eq:DM-form-factor}), (\ref{eq:sigma-bar-e}), (\ref{eq:eta}), and (\ref{eq:vmin}). 
We now turn to our own numerical evaluation of the crystal form factor.

%%%%%%%%%%%%%%%%%%%%%%%%%%%%%%%%%%%%%%%%%%%%%%%%%%%%%%
%%%%%%%%%%%%%%%%%%%%%%%%%%%%%%%%%%%%%%%%%%%%%%%%%%%%%%
\section{Numerical computation of the form factor} 
\label{sec:numerical}
%%%%%%%%%%%%%%%%%%%%%%%%%%%%%%%%%%%%%%%%%%%%%%%%%%%%%%

Our aim is to compute the crystal form factor, given by Eq.~(\ref{eq:crystal-form-factor}), for silicon and germanium targets with low energy thresholds ($\simlt 30$\,eV). 
Once these are found, it is possible to calculate scattering rates for any DM model. 
Calculating the form factor requires knowledge of the electron wavefunction coefficients $u_{i}(\vec k + \vec G)$ for all energetically accessible electron levels. 
To calculate these coefficients, we utilize the ``plane wave self-consistent field'' (\texttt{PWscf}) 
code within the \texttt{Quantum ESPRESSO}~\cite{QE-2009} package, based on the formalism of DFT. 
We then input these into our own postprocessing code, {\tt QEdark}, to calculate the form factors. 
In this section, we summarize the key conceptual and numerical details of our computation.  
We provide a review of DFT in Appendix~\ref{sec:DFT-review}, detail the approximations used in the computation of the wavefunctions, and lay out the numerical methods.  
In Appendix~\ref{sec:convergence}, we discuss the convergence of our computation.

%%%%%%%%%%%%%%%%%%%%%%%%%%%%%%%%%%%%%
\subsection{Computational framework}

It is impossible in practice to obtain the exact many-electron wavefunctions that describe interacting electrons in a many-body system such as a crystal.  
Instead, several methods exist to obtain excellent numerical approximations to these wavefunctions.  
We use DFT, which reformulates the interacting quantum many-body problem in terms of functionals of the particle density $n(\vec r)$.  
For the case of electrons, the Hohenberg-Kohn theorems~\cite{Hohenberg:1964zz} imply that all properties of the interacting system are determined once the ground-state electron density is known.  
In order to obtain the ground-state density, we use the Kohn-Sham method~\cite{Kohn:1965zzb} to map the system of interacting electrons into a system of independent electrons under the presence of an auxiliary potential that produces the same ground-state density.  
After this mapping, one has to solve the much simpler non-interacting electron system    in order to obtain the ground-state energy and electron density.  

The mapping from an interacting to a non-interacting many-body system comes at the expense of having to use an approximate auxiliary potential. Typically this potential is split into the mean-field Hartree potential and an exchange-correlation potential.  The latter captures the quantum mechanical effect of having identical electrons and also attempts to capture the correlation energy 
among the interacting electrons.  
The exchange-correlation potential is not known exactly and needs to be approximated.  
We use the Perdew-Burke-Ernzerhof (PBE) functional~\cite{PhysRevLett.77.3865}, 
which belongs to the class of the Generalized Gradient Approximations (GGA).  
We discuss this further in Appendix~\ref{sec:DFT-review}. 

Both silicon and germanium have a diamond lattice structure that contains two atoms in the unit cell. There are two s-shell and two p-shell 
valence electrons per atom (3s and 3p (2s and 2p) for germanium (silicon)), which makes a total of 8 electrons per cell. 
This translates to 4 valence bands, since each band is doubly degenerate in electron spin. 
In silicon, the core electrons have binding energies of $\sim$100\,eV and above, and so are irrelevant for the energies we consider here.  
One must take more care with germanium, since the $3d$ electrons have binding energies of $\sim$30\,eV, and so can be relevant for the higher energy thresholds we consider here.\footnote{We thank the authors of~\cite{Lee:2015qva} for 
discussions regarding this point.}  
In the computation, energetically-inaccessible core electrons can be replaced with a pseudopotential, which increases the computational efficiency by reducing the number of initial states required, and by reducing the resolution needed to describe the wavefunctions. 
We use ultrasoft pseudopotentials~\cite{PhysRevB.31.2163}  in place of all but the outer 
two s-shell and two p-shell valence electrons. 
For germanium, we also use a pseudopotential that allows us to treat the $3d$ electrons as valence states. 
As a result, the computational cost for germanium is slightly higher than that of silicon. 
We use an empirical ``scissor correction" approach~\cite{PhysRevLett.63.1719,PhysRevB.43.4187} to set the band gap to 
$1.11$~eV for silicon and $0.67$~eV for germanium~\cite{ExptGaps}.\footnote{These values are measured at 300~K, and change by 
5--10\% as the temperature approaches 0~K~\cite{PhysRevB.31.2163}.  The effect of this on our results is a few percent and therefore negligible.}

%%%%%%%%%%%%%%%%%%%%%%%%%%%%%%%%%%%%%
\subsection{Discretization procedure and cutoff choices}
\label{sec:discretization}

In order to obtain the crystal form factor with a finite computation, several modifications must be made to Eq.~(\ref{eq:crystal-form-factor}):

\begin{itemize}

\item \textbf{Binning in $q$ and $E_e$}.\quad
The form factor must be evaluated for finite grid of $q$- and $E_e$-values. We do this by averaging over bins of equal width in $q$ and $E_e$:
\begin{equation}
\big| f_{\rm crystal}^{\rm (binned)}(q_n, E_m) \big|^2 \equiv 
\int_{q_n - \frac{1}{2}\Delta q}^{q_n + \frac{1}{2}\Delta q} \frac{d q'}{\Delta q}  
\int_{E_m - \frac{1}{2}\Delta E}^{E_m + \frac{1}{2}\Delta E} \frac{d E'}{\Delta E} 
\big| f_{\rm crystal}(q', E') \big|^2 \, .
\end{equation}
Here $q_n$ is the central value of the n$^{\rm th}$ $q$ bin, and $E_m$ is the central value of the m$^{\rm th}$ energy bin, and $\Delta q$ and $\Delta E$ are the widths of the bins. 
We use 500 $E_e$-bins with $\Delta E=0.1$ eV and 900 $q$-bins with 
$\Delta q = 0.02 \,\alpha m_e$. 

\item
\textbf{Discretization in $\vec k$}.\quad
The continuum of $k$-values in each energy band must be replaced with a discrete mesh of representative $k$-points. 
The $k$-integrals in Eq.~(\ref{eq:crystal-form-factor}) are then replaced with finite sums:
\begin{equation}
\int_{\rm BZ} \frac{d^3 k}{(2 \pi)^3} (...) \longrightarrow \frac{V_{\rm BZ}}{(2\pi)^3} \sum_{\vec k} \frac{1}{2} w_{\vec k} \, (...)
= \frac{1}{V_{\rm cell}} \sum_{\vec k} \frac{1}{2} w_{\vec k} \, (...) \, .
\end{equation}
Here $V_{\rm BZ}$ is the volume of the Brillouin Zone, $V_{\rm cell}$ is the volume of the crystal's unit cell, and $w_{\vec k}$ are the weightings of the k-points, with $\sum w_{\vec k} = 2$ (following the convention of \texttt{Quantum ESPRESSO}). We use a uniform $243~k$-point mesh.

\item
\textbf{Cutoff in $\vec G$, $\vec G'$}.\quad
The wavefunctions are expanded in a finite size plane-wave basis whose reciprocal lattice 
vectors satisfy the ``kinetic energy" cutoff (really a cutoff in the space of $\vec G$-vectors)
\begin{equation}\label{eq:ecut}
\frac{|\vec k+\vec G|^2}{2m_e} \le E_{\rm cut}\,.
\end{equation}
Note that since $q = |\vec{k}' - \vec{k} + \vec G'|$, and since $| \vec G _{\rm max} | \gg | \vec k|$ and $|\vec k'| $, the momentum transfer $q$ essentially has a cutoff of $\sqrt{2m_e E_{\rm cut}}$. We choose a value of $E_{\rm{cut}}=70$~Ry, which allows us to sample a large enough $q$ space to obtain $\mathcal{O}$(1\%) accuracy for our rate calculations. 

\item
\textbf{Energy bands}\quad
As discussed above, we consider initial electron states in the 4 valence bands for silicon and the 4 valence bands + 10 outer core bands (corresponding to the 3d-shell electrons) for germanium. 
We include final-state energy bands up to the $52^{\rm nd}$ conduction band in both germanium and silicon. 
The lowest conduction states \emph{not} included are about 57 eV above the band gap, while the highest energy core states \emph{not} included are more than 60\,eV below the band gap. 
Our choice of bands therefore fully covers any energy transition below $\sim$57\,eV.

\end{itemize}

We can now write the form factor in the form that is implemented in our numerical code:\begin{align}
\begin{aligned}
\big| f_{\rm crystal}^{\rm (numerical)}(q_n, E_m) \big|^2 &=
\frac{2\pi^2 (\alpha m_e^2 V_{\rm cell})^{-1}}{E_m} \sum_{i\,  i'} \! \sum_{\vec k, \vec k'} \sum_{\vec G'} 
\frac{E_m}{\Delta E} \frac{q_n}{\Delta q} \frac{w_{\vec k}}{2} \frac{w_{\vec k'}}{2} 
\big| f_{[i \vec k , i' \vec k', \vec G']} \big|^2  \times
\\
& \Theta \bigg(1 - \frac{|E_{i' \vec k'} \!-\! E_{i \vec k}-E_m|}{ \frac{1}{2}\Delta E} \bigg) 
\, \Theta \bigg(1- \frac{\big||\vec k' - \vec k + \vec G'|-q_n\big|} { \frac{1}{2} \Delta q} \bigg)  \, .
\label{eq:crystal-excitation-rate-discrete}
\end{aligned}
\end{align}
Note that the first line here represents summing over bands, $k$-points, and reciprocal lattice vectors, and calculating the contribution to the form factor from each. The sums are all over finite ranges as discussed above. The second line represents adding each contribution to the appropriate $\{q, \, E_e \}$ bin. 
We present the results of our computation, including prospects for upcoming experiments, in \S\ref{sec:results}.
In Appendix~\ref{sec:convergence} we discuss convergence with respect to the choice of $k$-point mesh and $E_{\rm cut}$.

%%%%%%%%%%%%%%%%%%%%%%%%%%%%%%%%%%%%%%%%%%%%%%%%%%%%
%%%%%%%%%%%%%%%%%%%%%%%%%%%%%%%%%%%%%%%%%%%%%%%%%%%%
\section{Conversion from energy to ionization}
\label{sec:conversion-to-ionization-size}
%%%%%%%%%%%%%%%%%%%%%%%%%%%%%%%%%%%%%%%%%%%%%%%%%%%%

The calculation described in the previous two sections gives the DM--electron scattering rate in a semiconductor crystals as a function of the \emph{total energy} deposited by the dark matter, $E_e$. 

However, experiments will not directly measure the deposited energy itself, but rather the \emph{ionization} signal $Q$ -- \textit{i.e.},~the number of electron-hole pairs produced in an event. 
Linking the two is a complicated chain of secondary scattering processes, which rapidly redistribute the energy deposited in the initial scattering. 

A realistic treatment of the conversion from energy to ionization is a crucial step in calculating the sensitivity of experiments. 
Unfortunately, exact modeling of the secondary scattering processes is extremely challenging and is beyond the scope of this paper. 
Instead, we assume a linear response, which we believe does a reasonable job of capturing the true behavior.  Specifically we assume that, in addition to the primary electron-hole pair produced by the initial scattering, one extra electron-hole pair is produced for every extra $\varepsilon$ of energy deposited above the band-gap energy. 
Here $\varepsilon$ is the mean energy per electron-hole pair as measured in high-energy recoils. 
The ionization $Q$ is then given by
\begin{equation}
Q(E_e) = 1 + \lfloor (E_e - E_{\rm gap})/\varepsilon \rfloor \, ,
\label{eq:linear-ionization-response}
\end{equation}
where $\lfloor x \rfloor$ rounds $x$ down to the nearest integer. 
$\varepsilon$  and the band-gap energy $E_{\rm gap}$ are measured to be~\cite{ExptGaps, Klein:1968}

\begin{equation}
\label{eq:epsilon}
\varepsilon =
\begin{cases} 3.6~\text{eV} & \text{(silicon)} \\ 2.9~\text{eV} & \text{(germanium)} \end{cases} 
\qquad , \qquad
E_{\rm gap} =
\begin{cases} 1.11~\text{eV} & \text{(silicon)} \\ 0.67~\text{eV} & \text{(germanium)} \end{cases} .
\end{equation}

We devote \S\ref{sec:secondary-scattering-discussion} and Appendix~\ref{sec:MC-model} to a discussion motivating this simple treatment. 
We emphasize that, while our treatment is approximate, it
(\emph{a}) is quite separate from the systematic, first-principles calculation of $d R/ d E_e$ described in \S\ref{sec:calculating-rates} 
and \S\ref{sec:numerical}, and does not affect that calculation's accuracy;
(\emph{b}) is probably conservative, since it does not account for fluctuations that could push a low-energy event above the ionization threshold; and 
(\emph{c}) should be possible to improve upon in the future, both with better theoretical modeling and with experimental calibration.

%%%%%%%%%%%%%%%%%%%%%%%%%%%%%%%%%%%%%%
\subsection{Understanding the secondary scattering processes}
\label{sec:secondary-scattering-discussion}

It is experimentally well-established that for high energy electron recoils ($\simgt$ keV), the ionization signal is directly proportional to the deposited energy, 
with a constant average energy $\varepsilon$ deposited per electron-hole pair created,
\begin{equation}
\langle Q \rangle \simeq \frac{E_e}{\varepsilon} \, .
\end{equation}
$\varepsilon$ is several times the bandgap energy, accounting for the fact that only a fraction of the energy deposited goes directly into  pair production. 
Fluctuations around the average ionization are quite small, with the Fano factor (defined as the ratio of the variance to the mean) measured to be~\cite{Lowe1997, Lepy2000}
\begin{equation}
F \equiv \frac{\sigma_Q^2}{\langle Q \rangle} \approx 0.1\text{--}0.15 \, .
\end{equation}

At the low energies we are interested in, $\mathcal O$(1--50\,eV), the energy--ionization relationship has not been directly measured. 
Fortunately, there is reason to expect that the high-energy response can be extrapolated to lower energies. 
It has long been understood (see e.g.~\cite{Klein:1968, Bloom:1980}) that following a high-energy electron recoil, an electronic cascade occurs that rapidly redistributes the energy between many low-energy electrons and holes. 
Roughly speaking, any electron or hole is expected to re-scatter and create an additional electron-hole pair, so long as it has sufficient energy to do so. 
This repeats, distributing the energy over an exponentially increasing number of electron-hole pairs, until all electrons and holes have energy below the pair-creation threshold. 
Note that this threshold is larger than the band gap energy due to the constraints of momentum conservation~\cite{Klein:1968}. 
The excess energy carried by the electrons and holes after the cascade is slowly lost to phonons, as is a fraction of the energy during the cascade.
As a result of the cascade, the vast majority of secondary scatterings that occur after the initial electron recoil are \emph{low} energy scatterings.
This means that, for example, a single 10\,keV electron recoil is approximately equivalent to 100 recoils with 100\,eV each, or 1000 recoils with 10\,eV each. 
This justifies the extrapolation of the high-energy behavior to low energies.

The linear response described by Eq.~\eqref{eq:linear-ionization-response} is not the only tractable approach. 
Other, less simplistic approaches can be taken without resorting to a full first-principles treatment. 
For comparison, in Appendix~\ref{sec:MC-model} we construct a phenomenological Monte Carlo model of the secondary scattering cascade, following~\cite{Bloom:1980}. 
The model is intended to capture the general features of the cascade, without knowledge of the specific microscopic structure of the target material. 
The model reproduces the known high-energy behavior well with only a single tunable parameter, and can be used instead of Eq.~\eqref{eq:linear-ionization-response} when calculating DM scattering rates. 
Unlike the linear treatment, the Monte Carlo model predicts fluctuations about the mean, which can have an important effect for DM masses that are right on the edge of detectability. 
For typical masses, however, we find that the two approaches agree to within a few 10's of percent (see Fig.~\ref{fig:MC-rate-comparison}). 
We conclude that the linear treatment of Eq.~\eqref{eq:linear-ionization-response} is a reasonably realistic approximation, and postpone a more careful treatment to future work.

%%%%%%%%%%%%%%%%%%%%%%%%%%%%%%%%%%%%%%%%%%%%%%
\section{Results}
\label{sec:results}
%%%%%%%%%%%%%%%%%%%%%%%%%%%%%%%%%%%%%%%%%%%%%%

In this section, we present the results of our calculation of the DM--electron scattering rates in silicon and germanium detectors.  
We show the potential reach for single-electron-sensitive experiments, as well as the effect of higher experimental thresholds. 
We also give the full recoil spectra and the annual modulation fraction, which may be crucial for discriminating a possible signal from background. 
Lastly we discuss near-term prospects, focussing on upcoming searches expected from the SuperCDMS and DAMIC collaborations.

Experimental thresholds are set in terms of the ionization signal $Q$ (the number of electron-hole pairs produced in an event) rather 
than the deposited energy $E_e$. 
In the following results, we model the conversion of deposited energy to ionization with the linear treatment described in~\S\ref{sec:conversion-to-ionization-size}. 
We take the DM halo to have a local density of $\rho_{DM}=0.4~\mathrm{GeV/cm}^3$~\cite{Catena:2009mf,Salucci:2010qr}, 
and a Maxwell-Boltzmann velocity distribution with a mean velocity $v_0=230$~km/s and escape velocity $v_{\rm esc}=600$~km/s, 
and we take the average Earth velocity to be $v_{\rm E}=240$~km/s 
(see Appendix~\ref{sec:etavmin} for explicit formulae). 
In Appendix~\ref{sec:convergence} we discuss the numerical convergence of our results. 

Event rates as a function of $Q$, for an extensive range of DM masses, are 
available online at \href{http://ddldm.physics.sunysb.edu}{this link}.  
The crystal form-factor, as a function of $q$ and $E_e$, is also available there. 
Using Eq.~(\ref{eq:diff-crystal-rate}), the information online can be used to re-derive rates using a different DM form-factor 
or velocity profile, or using a different treatment of the energy-to-ionization conversion.

\begin{figure*}[t!]
\centering
\includegraphics[width=0.435\textwidth]{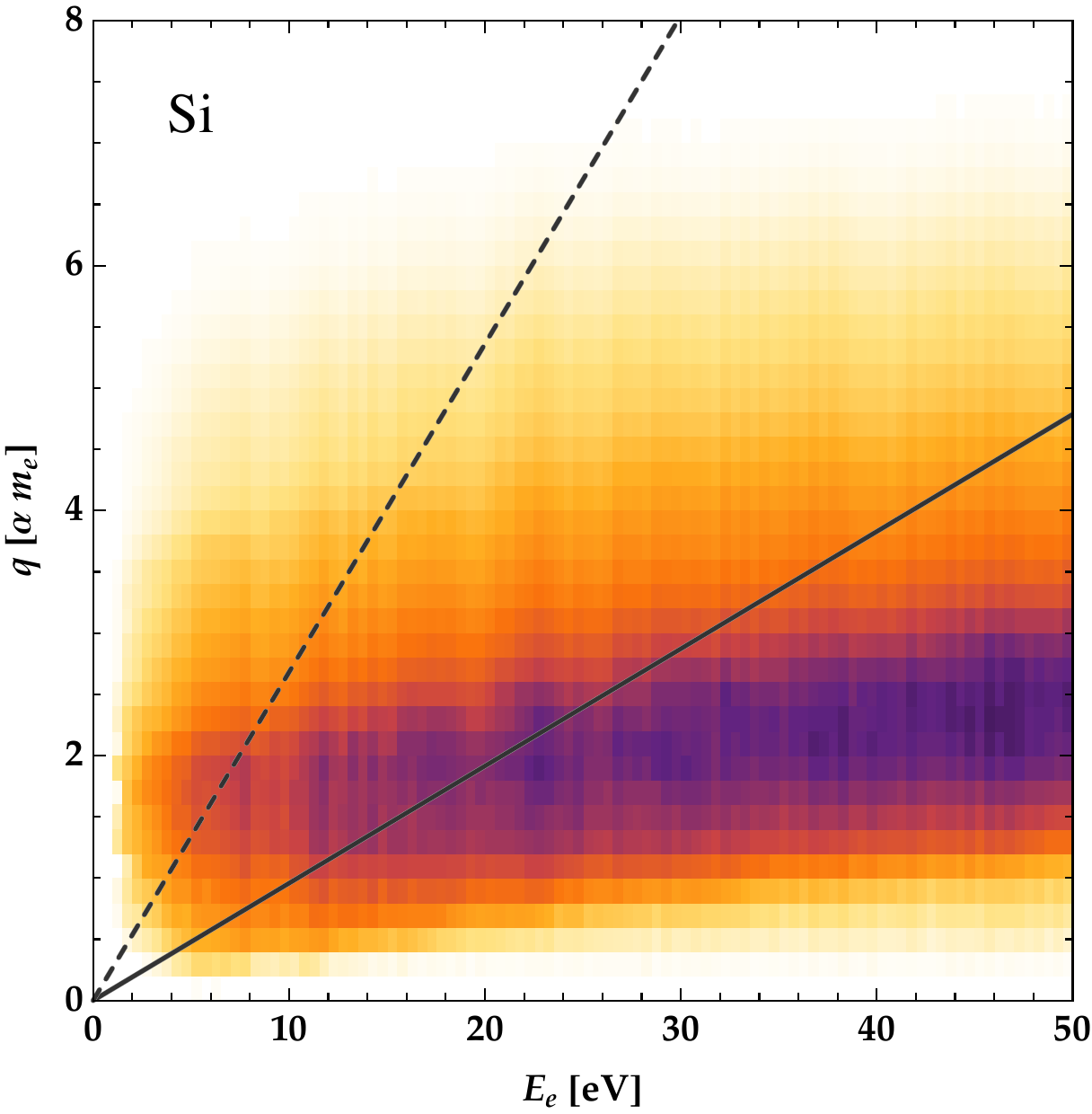}
\hfill
\includegraphics[width=0.435\textwidth]{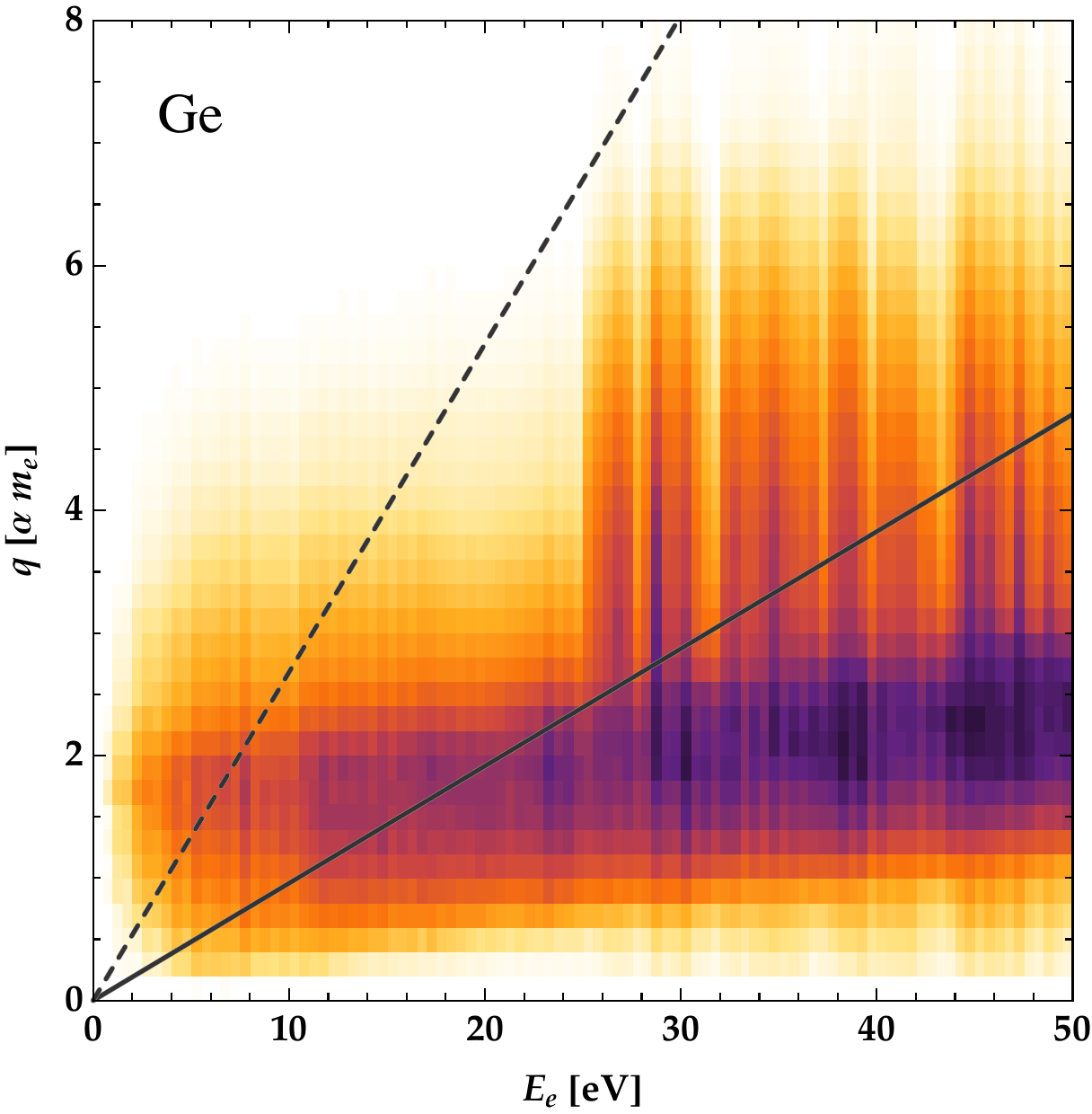}
\hfill
\includegraphics[height=0.48\textwidth]{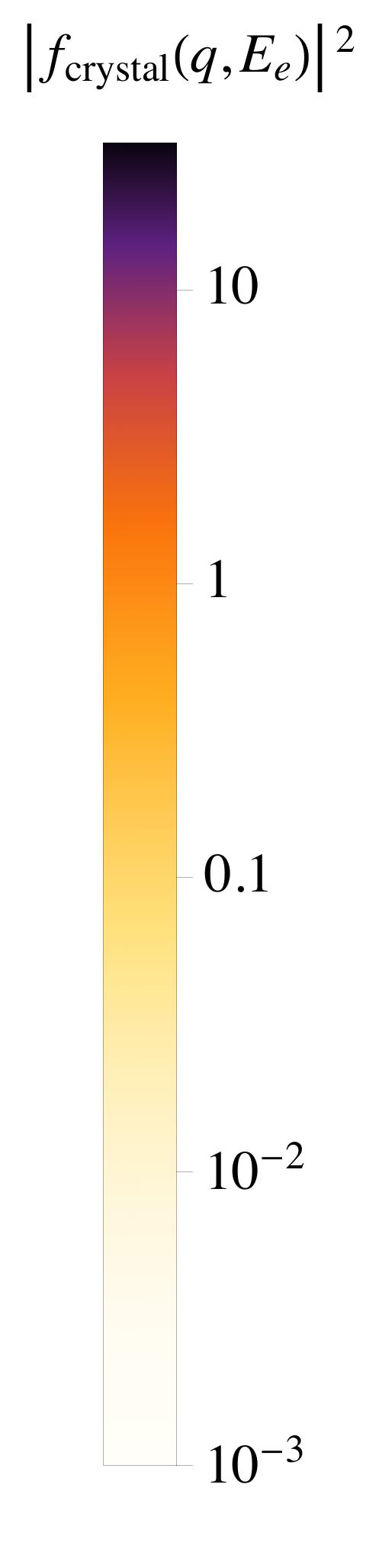}
\caption{\footnotesize 
The crystal form factor $|f_{\rm crystal}(q, E_e)|^2$ as a function of $q$ and $E_e$ for silicon ({\bf left}) and germanium ({\bf right}) (see Eq.~(\ref{eq:crystal-form-factor})). 
In the region below the solid line, $v_{\rm min}>v_{esc}+v_E$ for any DM mass, and electron scattering is thus kinematically inaccessible. 
The dashed line corresponds to $v_{\rm min}=300$~km/s (a typical DM halo velocity) in the heavy DM limit; the region below this line is only kinematically accessible to DM particles with velocities larger than the average velocity. 
For energies above $\sim$\,10\,eV, the scattering rate is suppressed by both the form factor and DM velocity distribution. We see that the 3d electrons in germanium give a sizable contribution to $|f_{\rm crystal}(q, E_e)|^2$ for $E_e>25$ eV.
}
\label{fig:crystalFF}
\end{figure*}

%%%%%%%%%%%%%%%%%%%%%%%%%%%%%%%%%%%%%%%%%%%%%%
\subsection{The crystal form factor}\label{subsec:ff}

Much of the behavior of the scattering rates can be understood from the behavior of the crystal form factor, $|f_{\rm crystal}(q, E_e)|^2$,  
in Eq.~(\ref{eq:crystal-form-factor}).  We show the crystal form factor in Fig.~\ref{fig:crystalFF} as a function of $q$ and $E_e$, for both silicon and germanium.  The rapid fall-off as $q$ increases is clearly visible. 
The solid line in the figure corresponds to $v_{\rm min}=v_{esc}+v_E$ from Eq.~(\ref{eq:vmin}) as $m_\chi \to \infty$.  
The region below this line is kinematically inaccessible for any DM mass.  The dashed line uses the velocity of a typical DM particle 
in the halo, i.e.~$v_{\rm min}=300$~km/s.
We see that larger recoil energies require larger $q$, for which the crystal form factor is suppressed.  
{\it The implication of this is that the DM-electron scattering rates increase dramatically for smaller recoil energies, resulting in 
a dramatic increase in sensitivity as detector thresholds are lowered.}

%%%%%%%%%%%%%%%%%%%%%%%%%%%%%%%%%%%%%%%%%%%%%%
\subsection{Cross-section reach versus detector threshold}\label{subsec:prospects}

\begin{figure}[t]
%\begin{mdframed}
%\begin{center} {\bf Dark Matter--Electron Cross-Section Sensitivity}\end{center}
\vspace{4mm}
\includegraphics[width=0.5\textwidth]{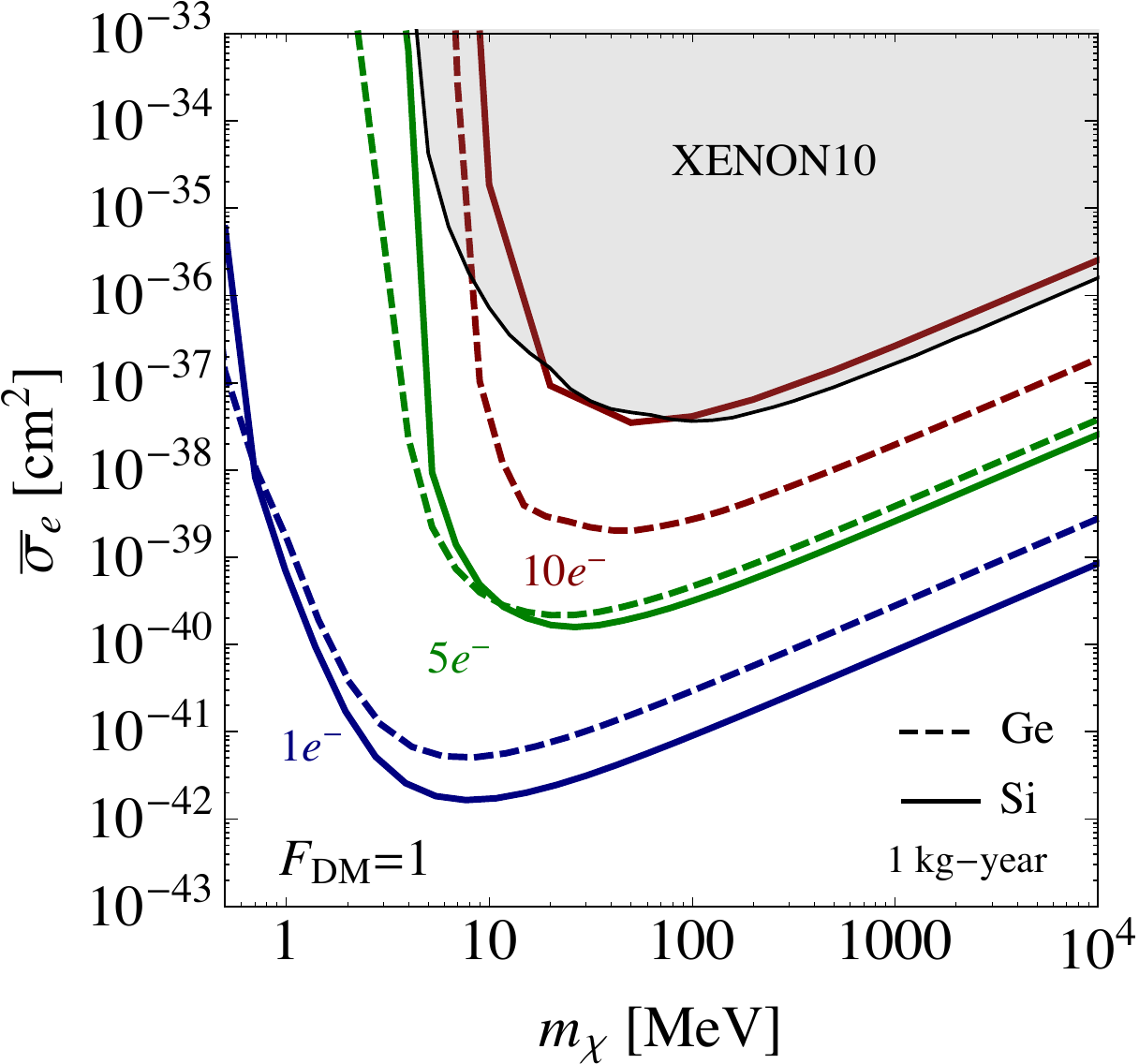}
~\includegraphics[width=0.5\textwidth]{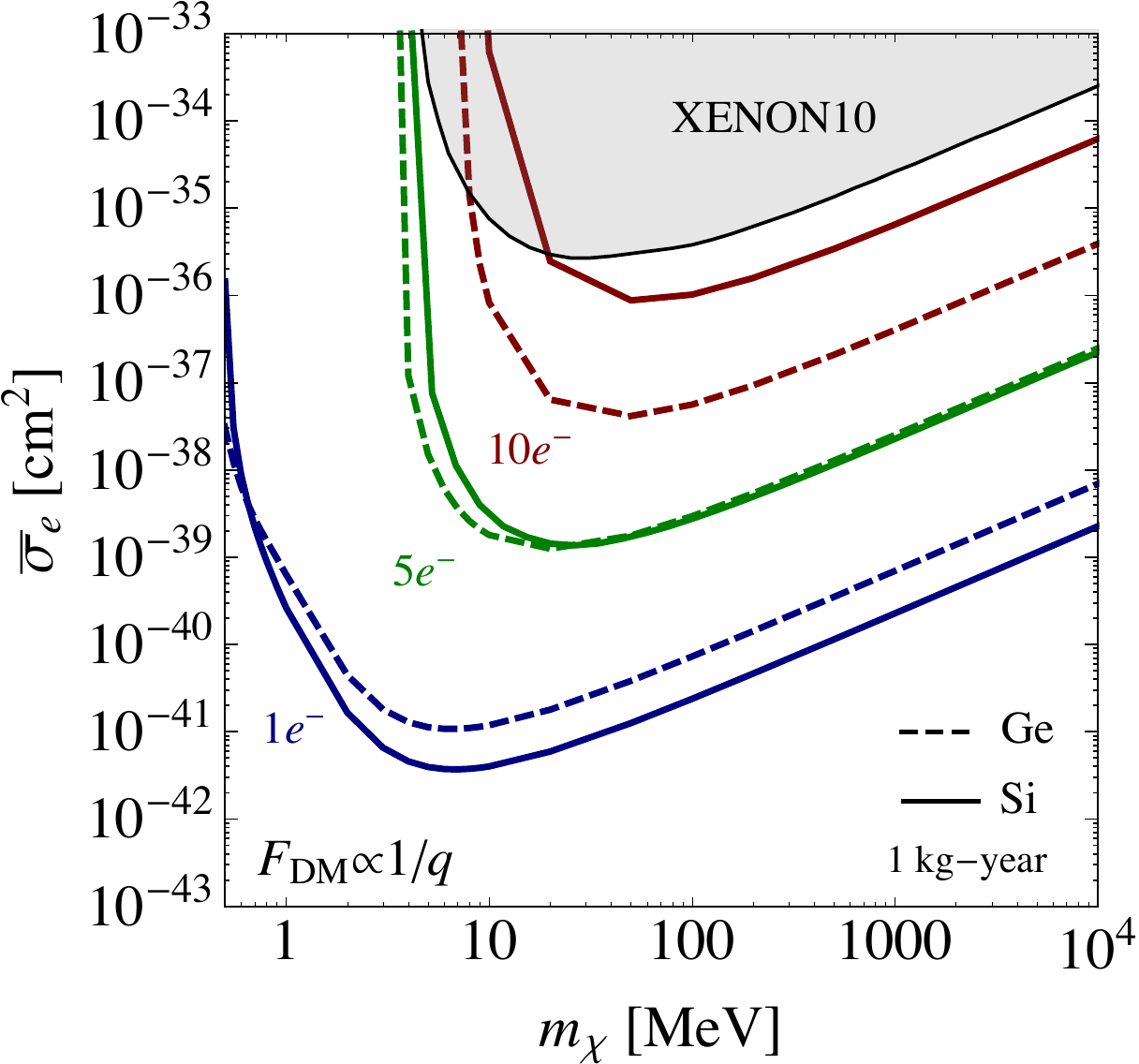}
\vspace{-2mm}
\\
\begin{minipage}[t]{0.5\textwidth}
\mbox{}\\[-\baselineskip]
\includegraphics[width=\textwidth]{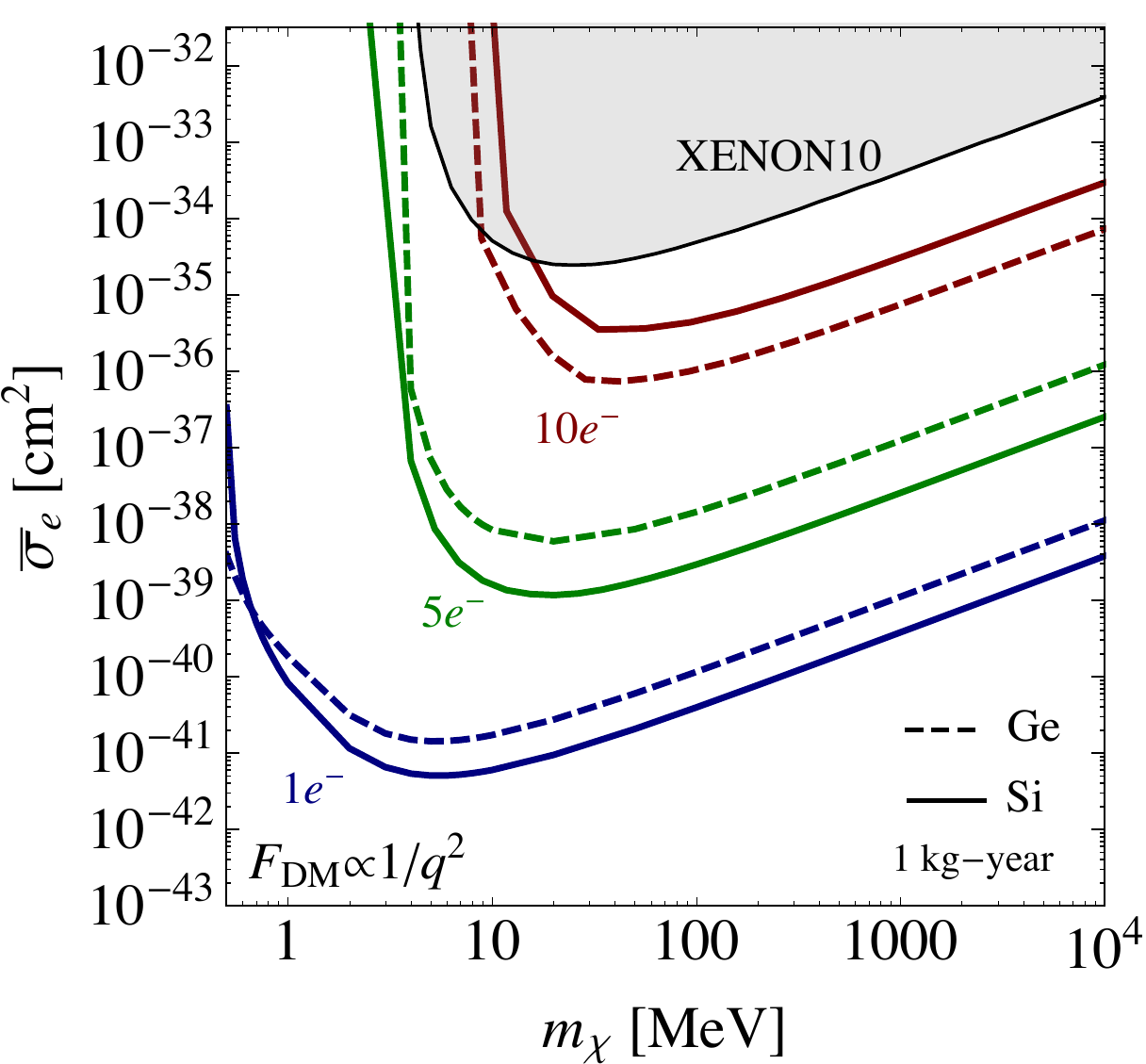}
\end{minipage}
\hfill
\begin{minipage}[t]{0.48\textwidth}
\mbox{}\\[-\baselineskip]
\vspace{-25pt}
\caption{\footnotesize {\bf Dark Matter--Electron Cross-Section Sensitivity:} The 95\%~C.L.~exclusion reach in the DM--electron scattering cross-section, $\overline \sigma_e$, of an experiment with 1 kg-year exposure and zero background events, for different experimental thresholds.
Solid (dashed) lines show the reach for silicon (germanium) targets. 
Ionization thresholds of 1, 5, and 10 electron-hole pairs are shown with blue, green, and red lines, respectively. 
The corresponding energy thresholds are 0, 11.6, and 26.1\,eV  in germanium, and 0, 14.4, and 32.4\,eV in silicon. 
The gray shaded region shows the existing constraint from XENON10 data~\cite{Essig:2012yx}.  
The three plots assume different DM form factors, $\FDM(q)=1$, $\alpha m_e/q$, $(\alpha m_e/q)^2$, corresponding to different DM models.
}  
\label{fig:rates_ERcuts}
\end{minipage}
%\end{mdframed}
\end{figure}

In Fig.~\ref{fig:rates_ERcuts}, we show the sensitivity to the DM--electron scattering cross section, $\overline \sigma_e$, 
versus the DM mass, $m_\chi$, for hypothetical silicon- and germanium-based experiments with a 1 kg-year exposure and zero background, and with various detector thresholds.
The curves show 95\% C.L. limits, \emph{i.e.} 3.6 signal events. 
The blue, green, and red lines show ionization thresholds, $Q_{\rm th}$, of 1, 5, and 10 detected electron-hole 
pairs, respectively, which correspond to 
deposited energies, $E_e$, of 0.67, 12.3, and 26.8\,eV  in germanium, and 1.1, 15.5, and 33.5\,eV in silicon 
(to get the corresponding ionization energy thresholds, subtract 0.67~eV for germanium and 1.1~eV for silicon from these numbers, 
see Eq.~(\ref{eq:linear-ionization-response})). 
The three plots show results for different DM form factors, corresponding to different classes of DM models: 
$\FDM(q)=1$ ({\it top left}), $\FDM(q)=\alpha m_e/q$ ({\it top right}), and $\FDM(q)=(\alpha m_e/q)^2$ ({\it bottom}), 
see \S\ref{sec:models} for details.

As expected, the reach dramatically improves when the threshold is lowered, since the crystal form factor 
strongly suppresses the rate for high electron recoil energies. 
This improvement is most pronounced for $F_{\rm DM}(q) = (\alpha m_e/q)^2$, since lower $q$ tends to correspond to lower recoil energies. 
With a single-electron threshold, the difference in sensitivity for silicon and germanium targets can be accounted for by 
the fact that germanium is 2.6 times heavier, and so has correspondingly fewer valence electrons per kg. 
However, germanium targets are sensitive to slightly lower DM masses due to their lower band-gap. 
In addition, germanium targets become comparably more sensitive than silicon 
targets for ionization thresholds of $Q_{\rm th} \gtrsim 9$ due to the additional contribution from the 3d-shell electrons (see below). 

\begin{figure}[!t]
\includegraphics[width=0.49\textwidth]{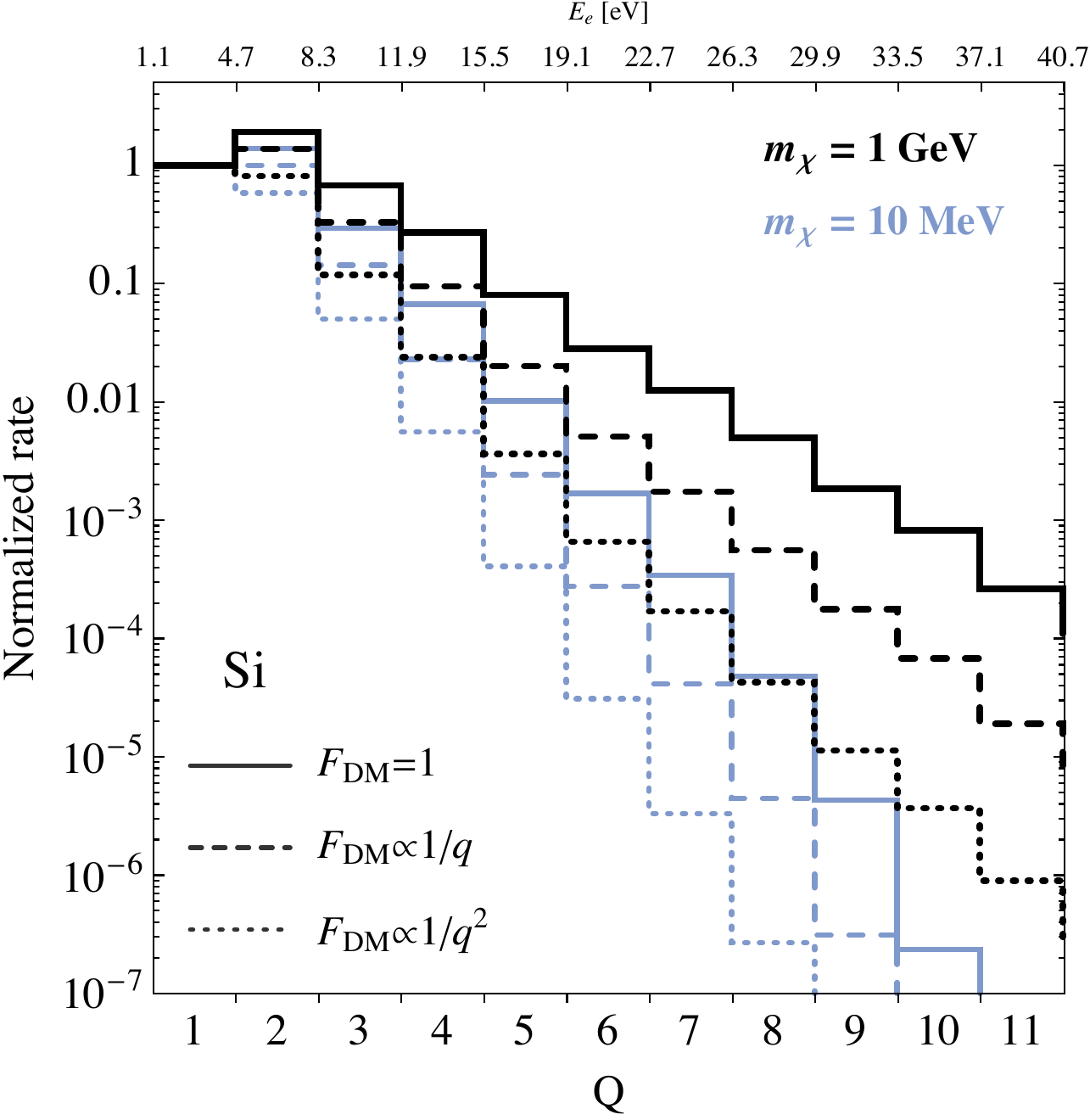}
\hfill
\includegraphics[width=0.49\textwidth]{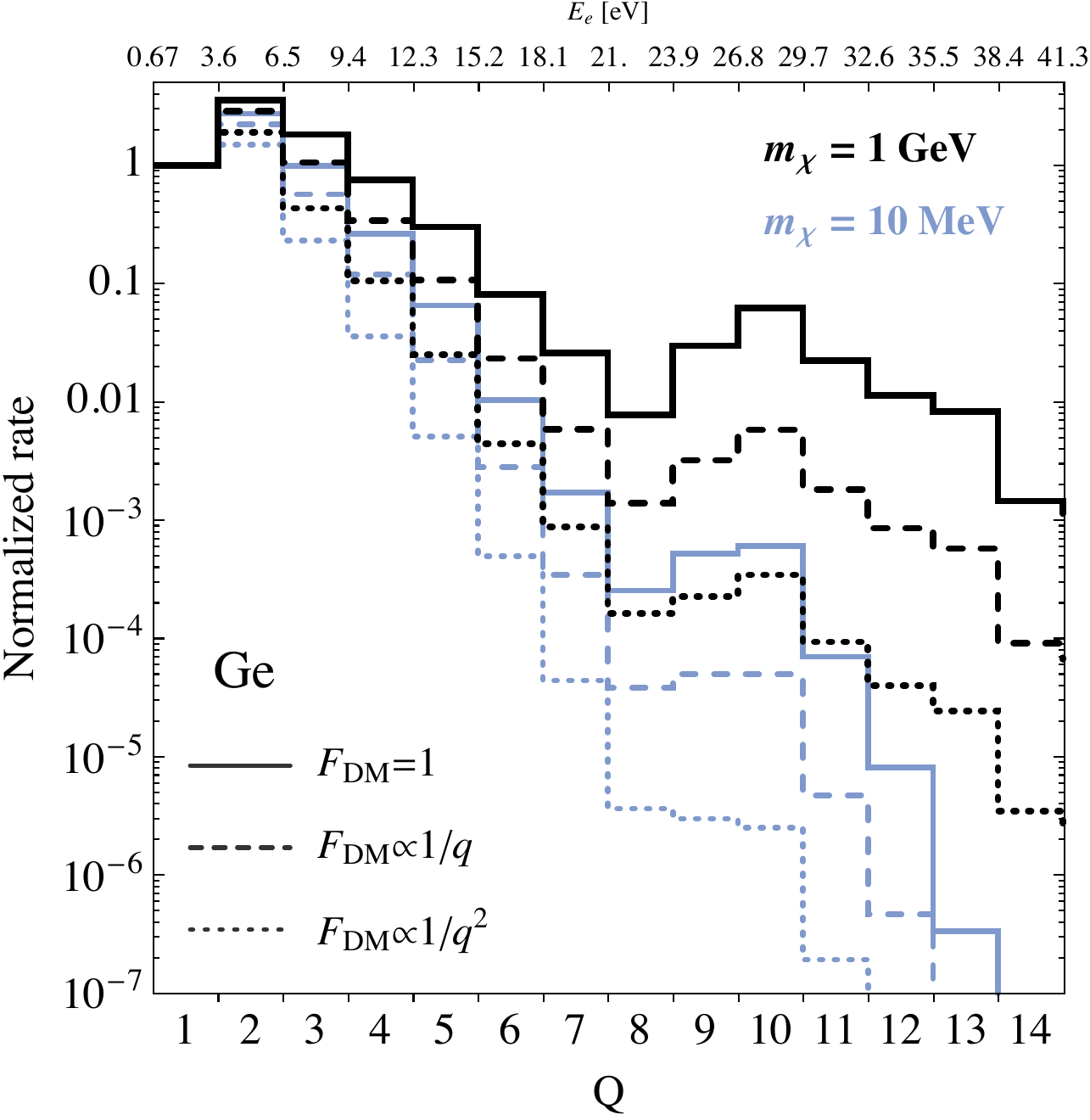}
\caption{\footnotesize Spectrum of events as function of the ionization signal $Q$. The different lines show different DM masses and form-factors, as indicated, in a silicon ({\bf left}) or germanium ({\bf right}) target.
The spectra are normalized to 1 in the first bin. 
The top axes show the values of the deposited energy $E_e$ corresponding to the edges of the bins. Since the typical $E_{e}$ is around a few eV, the distributions peak at $Q=2$ (see \S\ref{sec:scattering-kinematics}). 
}
\label{fig:FDM_ER}
\end{figure}

Fig.~\ref{fig:FDM_ER} shows the spectrum of events as a function of the ionization signal $Q$, for different DM form-factors and masses, in silicon ({\it left}) and germanium ({\it right}) targets.  
The fast fall-off with $Q$ shows the large gain to be made from lowering experimental thresholds towards a 1- or 2-electron threshold, especially for the steeper DM form-factors and for lower DM masses. 
The shape of the spectra may be useful in discriminating a signal from background.

In germanium, the 3d-shell electrons dominate the rate for $E_e\gtrsim 24$~eV, corresponding to $Q \gtrsim 9$ electron-hole pairs 
(the 3d-shell electrons lie about $\sim$15~eV below the bottom of the valence band and $\sim$24~eV below the bottom of the 
conduction band).  
The intuitive reason for the 3d-shell electrons dominating over the valence electrons at large $E_e$ can be seen from 
Eqs.~(\ref{eq:typical-q}) and (\ref{eq:q-min-approx}). 
The typical velocity of the 3d-shell electrons, and hence the typical momentum transferred from the DM, is larger than for the valence electrons, so that the 3d-shell electrons can dominate if they are kinematically accessible.\footnote{For even larger deposited energies, 
$E_e \gtrsim 100$~eV, it is likely that the deeper shells will dominate the rate, for both silicon and germanium.}
We show the effect of neglecting the 3d-shell electrons in Fig.~\ref{fig:FDM_ER-core-valence} in Appendix~\ref{sec:3d-shell}, where we compare the cross-section reach and the recoil spectrum generated by DM scattering with and without the inclusion of the 3d-shell electrons.  
The effect is  significant for ionization thresholds above $Q_{\rm th} \approx 7$ or 8, but not important at lower thresholds. 

We note that there are some differences between our results and those in~\cite{Graham:2012su,Essig:2011nj}.  
In~\cite{Essig:2011nj}, only the case $Q_{\rm th}=1$ was considered and we find that the new computation predicts a 
somewhat lower rate.  
We find that the shape of the recoil spectra in~\cite{Graham:2012su} is noticeably different from ours, which gives rise 
to several differences in the expected limits for the different $Q$ thresholds. Furthermore, for germanium, we also include 
the 3d-shell electrons, which can be important as discussed above.

%%%%%%%%%%%%%%%%%%%%%%%%%%%%%%%%%%%%%%%%%%%%%%
\subsection{Comparison with existing XENON10 limit and discussion of background}

We see from Fig.~\ref{fig:rates_ERcuts} that to surpass the existing limits obtained with XENON10 data~\cite{Essig:2012yx}, a 
germanium- or silicon-based experiment with an ionization threshold of 10 electrons would require a background-free exposure 
of less than 1\,kg-year.  
However, with a single-electron threshold, such an experiment would surpass the XENON10 limit at all masses with a background-free 
exposure of around 1\,kg-day for $F_{\rm DM}(q)=1$, 10\,g-days for $F_{\rm DM}(q) = \alpha m_e/q$, or just a 1\,g-day for 
$F_{\rm DM}(q) = (\alpha m_e/q)^2$. 
In addition, with any exposure such an experiment would place the first ever bounds in the $\sim$1-5\,MeV mass range, below the threshold of the XENON10 search.
The XENON10 detector had a threshold of one electron with an $\mathcal{O}(1)$ detection efficiency, but to obtain one electron required an energy of at least 12.4~eV to overcome the binding energy of an electron in the outer shell.\footnote{Note that in~\cite{Essig:2012yx}, the electrons were   treated as bounded   inside free (xenon) atoms, unlike the electrons in the semiconductor targets here.} 
Moreover, the background in the XENON10 data was much larger than conventional nuclear-recoil background, so that the number of DM events leading to single 
(two, and three) electrons was only limited, at 90\% C.L., to be less than 8,550 (1,550, and 330) counts/kg/year, respectively.  

While the single-electron background in the XENON10 data was rather large, its origin is likely specific to its dual-phase detector setup.  
Many of the single electron backgrounds likely had one, or a combination, of the following origins~\cite{Essig:2012yx}:
(i) electrons, trapped in the potential barrier at the liquid-gas interface, were randomly drawn into the gas phase
(these transiently trapped electrons likely originated from other background events that caused xenon atoms to ionize); 
(ii) photo-dissociation of a negatively charged $O_2$-ion, which received its negative charge from a drifting electron 
that originated from another event; 
(iii) field emission from the cathode. 
XENON100 and LUX may face similar challenges, although an analysis is still in progress. 

The semi-conductor targets will not suffer from these same detector-specific backgrounds.  
They will, of course, have their own unique experimental challenges to deal with, including detector noise and 
dark current, as we will discuss in more detail in \S\ref{subsec:upcoming-prospects} for DAMIC and SuperCDMS. 
These will likely be the limiting instrumental factors in setting the threshold for a particular experiment.  
Once these challenges are overcome, one needs to deal with the physics backgrounds. 
As argued in~\cite{Essig:2011nj}, 
neutrinos are not an important source of background even for the largest exposures considered in this 
paper ($\mathcal{O}$(20 kg-years) for SuperCDMS, see~\S\ref{subsec:upcoming-prospects}).  
Compton scattering or other events that produce recoiling electrons 
will usually lead to a much larger energy deposition and most of them could thus be vetoed, although some backgrounds will  
persist to the lowest energies.  
The size of this background will depend on the shielding and purity of the materials around the detector; 
for SuperCDMS at SNOLAB, the Compton background is estimated to be 
$\mathcal{O}(6\times 10^{-3})$~events/kg/day/keV~\cite{Figueroa}.  
Assuming that it is flat at low energies, this translates into $\mathcal{O}(0.04)$ events/eV for 20 kg-years, which would be negligible.  
We do not expect there to be backgrounds from neutrons, and (cosmogenic) x-ray lines will lie well above our energies 
of interest. 
Surface events and other, unknown, backgrounds may be present at low energies.  
As experiments reach the required sensitivity to probe the few-electron events expected from LDM scattering off electrons, 
a better understanding of all backgrounds will emerge allowing for an attempt to mitigate them if necessary.  
 It is possible that a spectral analysis of a signal will further allow for the removal of some background events.  
Our assumption of zero background for the plots should be taken as the best-case scenario.

%%%%%%%%%%%%%%%%%%%%%%%%%%%%%%%%%%%%%%%%%%%%%%
\subsection{Annual Modulation}
\label{sec:annual-mod}

Even with a significant background event rate, it may be possible to distinguish a signal from background using annual modulation, as long as the background is stable on year time-scales.
Annual modulation is a distinguishing feature of a LDM scattering signal~\cite{Drukier:1986tm, Essig:2011nj}, occurring due to the change in the earth's velocity through the DM halo as it rotates around the Sun. 
For a standard smooth and isotropic DM velocity distribution, the modulation is approximately sinusoidal with year period and a peak around June 2nd (the presence of DM streams or non-trivial DM structure may complicate this, as may gravitational focusing by the Sun~\cite{Lee:2013wza,Lee:2015qva}, which we do not include).
The modulation fraction, $f_{\rm mod}$, is defined to be the ratio of the amplitude of the modulating signal to the median signal rate, which for our assumed halo profile (see Appendix \ref{sec:etavmin}) corresponds to
\bea\label{eq:fmod}
f_{\rm mod} & = & \frac{R_{\rm June~2} - R_{\rm Dec~2}}{2 R_0},~~~~R_0 \equiv R_{\rm Mar~2} = R_{\rm Sept~2}
\eea
where $R_i$ represents the rate at time of year $i$.

\begin{figure}[!t]
\includegraphics[width=0.325\textwidth]{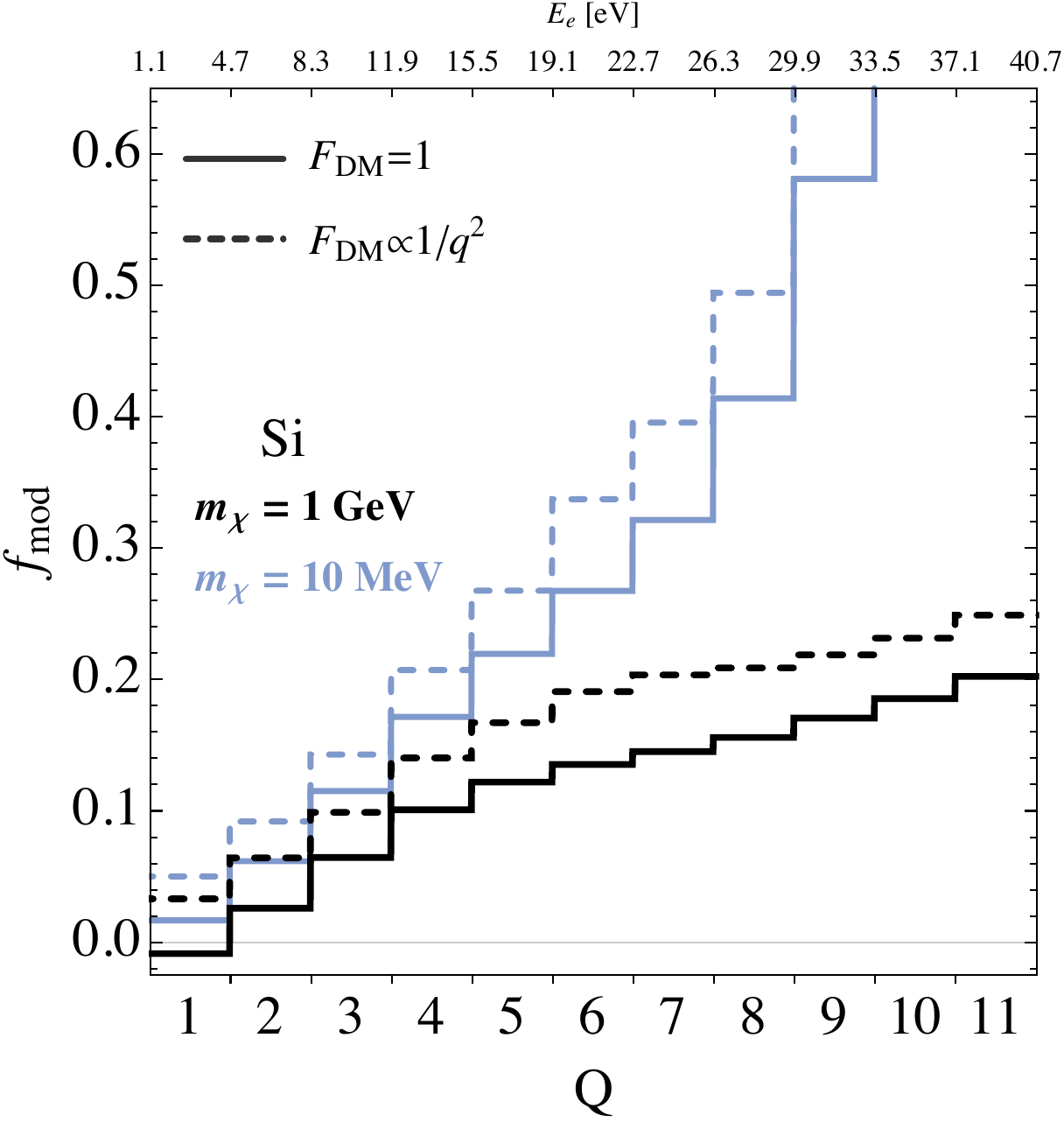}
\hspace{-2pt}
\includegraphics[width=0.325\textwidth]{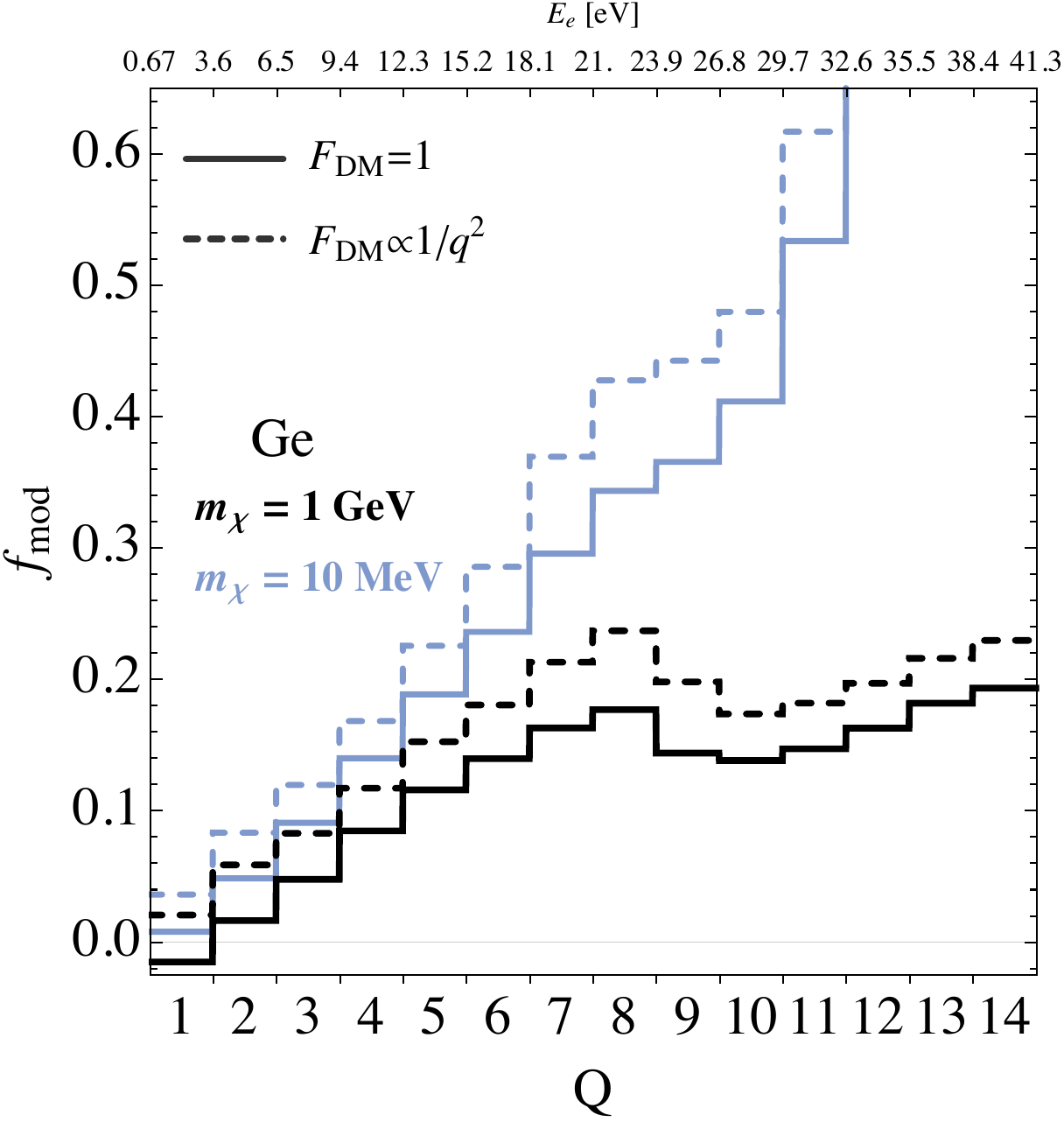}
\hspace{-2pt}
\includegraphics[width=0.33\textwidth]{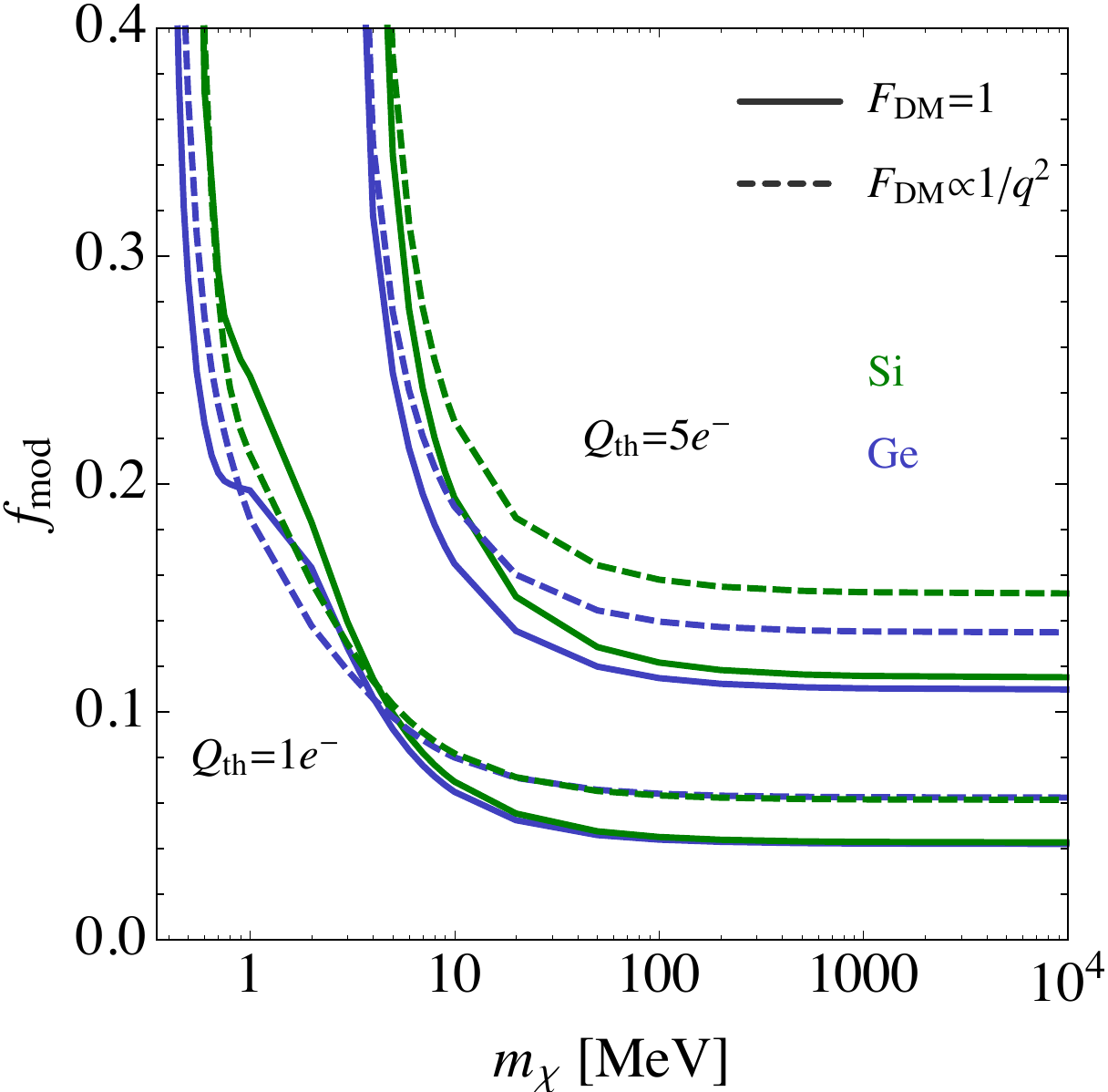}
\vspace{-5pt}
\caption{\footnotesize
Annual modulation fraction of the DM signal, $f_{\rm mod}$. 
{\bf Left:} $f_{\rm mod}$ as a function of the ionization threshold $Q_{\rm th}$, in silicon, for 10~MeV (blue) and 1~GeV (black) DM masses.  
The solid (dashed) lines  correspond to $\FDM \!=\! 1$ ($\FDM \!=\! (\alpha m_e/q)^2$). 
The top axis indicates the energies corresponding to the edges of the bins. 
{\bf Center:}  Same as {\it left} for germanium.
{\bf Right:}  $f_{\rm mod}$ as a function of DM mass, for ionization thresholds of 1 and 5 electron-hole pairs.  
The solid (dashed) lines correspond to $\FDM=1$ ($\FDM = (\alpha m_e/q)^2$), while the blue (green) lines correspond to germanium (silicon).}
\label{fig:fmod}
\end{figure}

For DM scattering off electrons, the modulation fraction can be significantly larger than for the usual elastic scattering 
of (heavy) WIMPs off nuclei. 
As we saw in Fig.~\ref{fig:crystalFF} (see discussion in \S\ref{sec:scattering-kinematics}), DM--electron scattering relies on the tail of the DM velocity distribution, especially for energies above $\sim$\,5\,--10\,eV. 
We plot the modulation fraction in Fig.~\ref{fig:fmod}.
The {\it left} and {\it center} plots show $f_{\rm mod}$ as a function of ionization $Q$ for different masses and DM form-factors for the two elements. 
$f_{\rm mod}$ rises from a few percent for single-electron events to above 10\% for events with more than $\sim$\,3 electrons. 
Comparing with the spectrum in Fig.~\ref{fig:FDM_ER}, we see that there is a trade-off between modulation fraction and event rate. 
Events with several electron-hole pairs provide large modulation without sacrificing too much in the rate, and may give the best prospects for annual modulation searches depending on the background.
The modulation fraction also rises near the mass threshold, as we show on the right of Fig.~\ref{fig:fmod}  
for ionization thresholds of $Q_{\rm th} = 1$ and 5. Note that the high-mass value of $f_{\rm mod}$ for the single-electron threshold, at 4\,--6\%, is larger than the values in the $Q=1$ bin of the {\it left} and {\it center} plots, because the total rate is \emph{not} dominated by the single-electron events.

\begin{figure}[!t]
%\begin{mdframed}
%\begin{center} {\bf Annual Modulation: Discovery Reach}\end{center}
\includegraphics[width=0.49\textwidth]{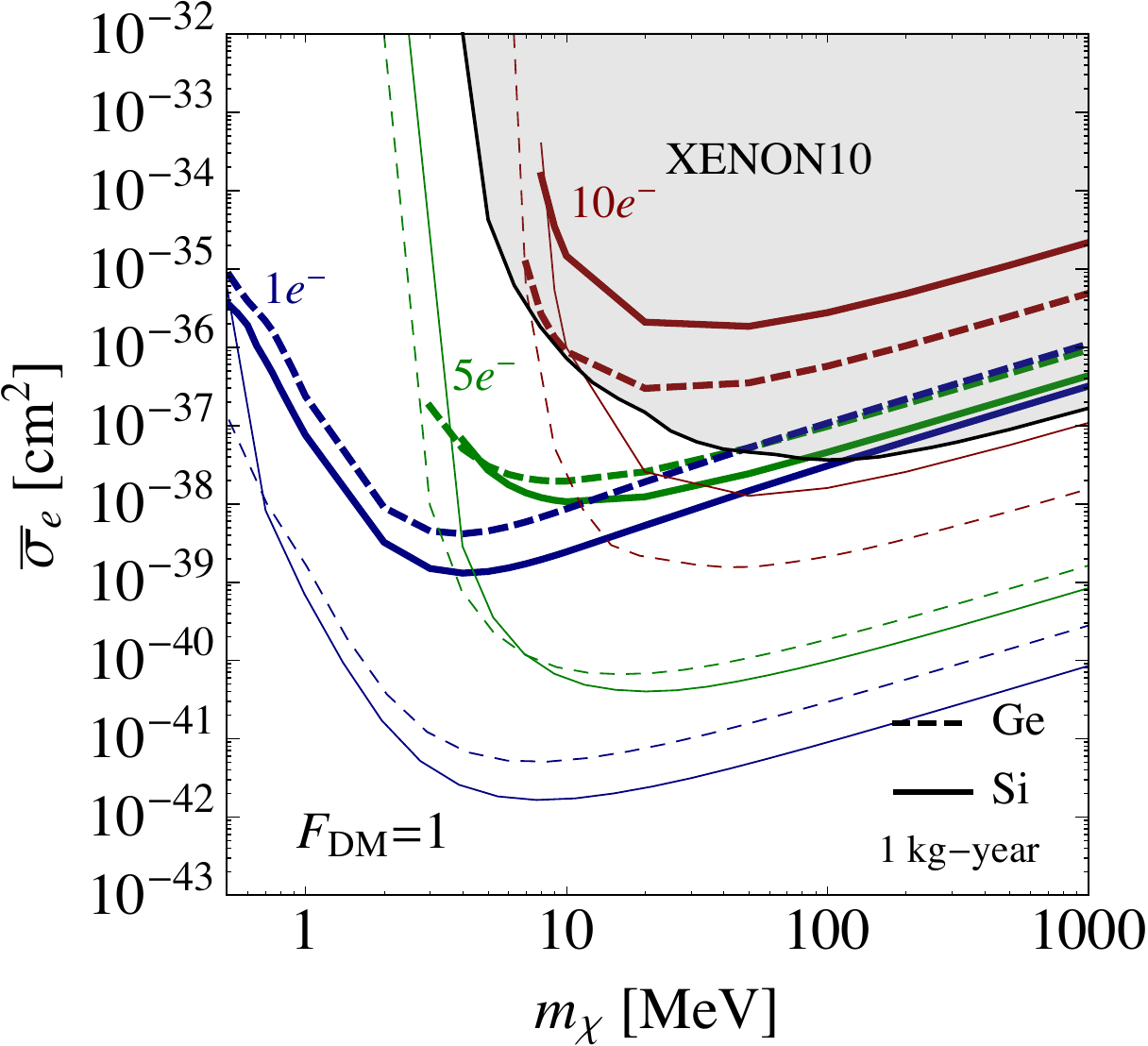}
~\includegraphics[width=0.49\textwidth]{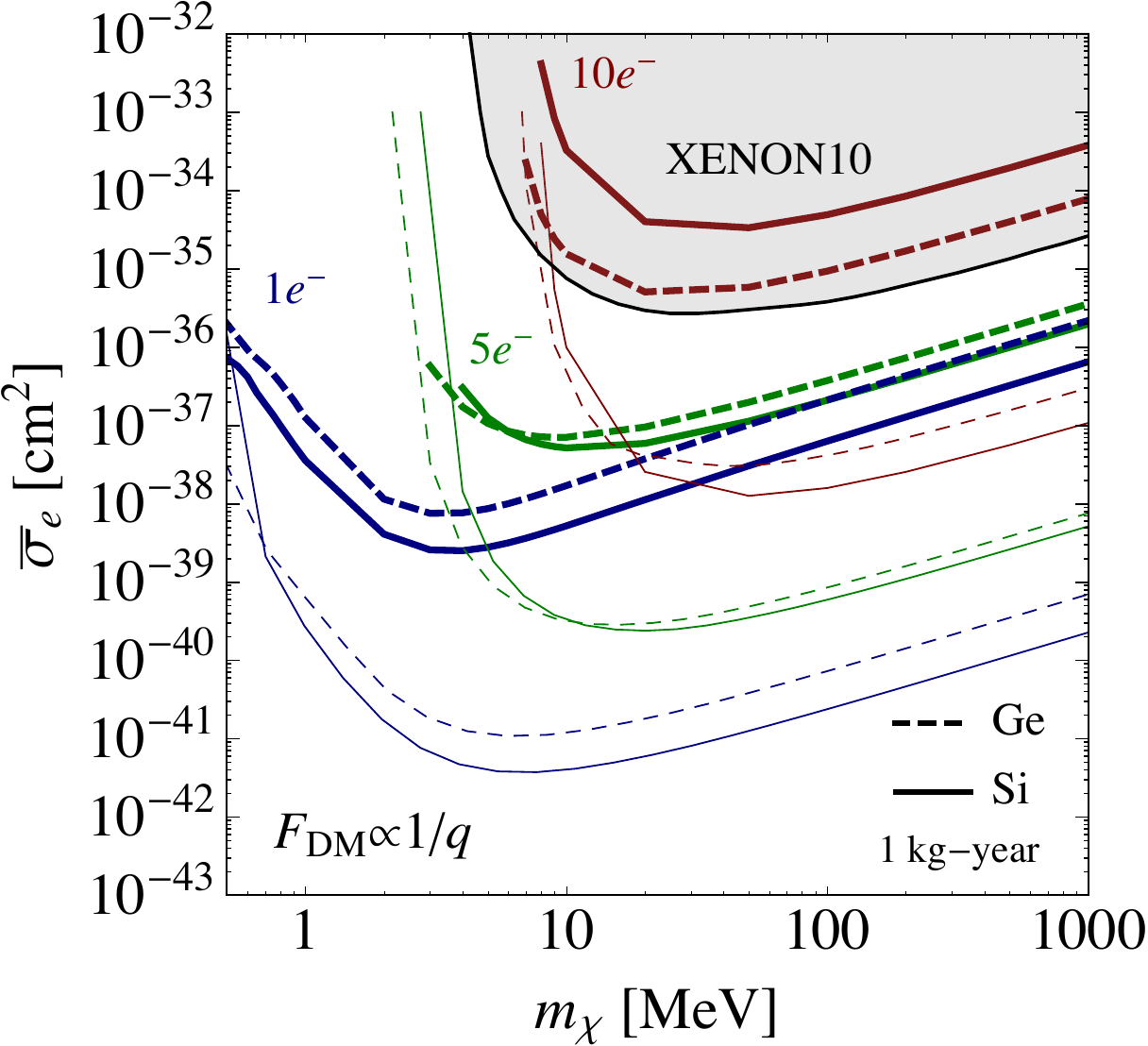}
\vspace{2mm}
\\
\begin{minipage}[t]{0.5\textwidth}
\mbox{}\\[-\baselineskip]
\includegraphics[width=\textwidth]{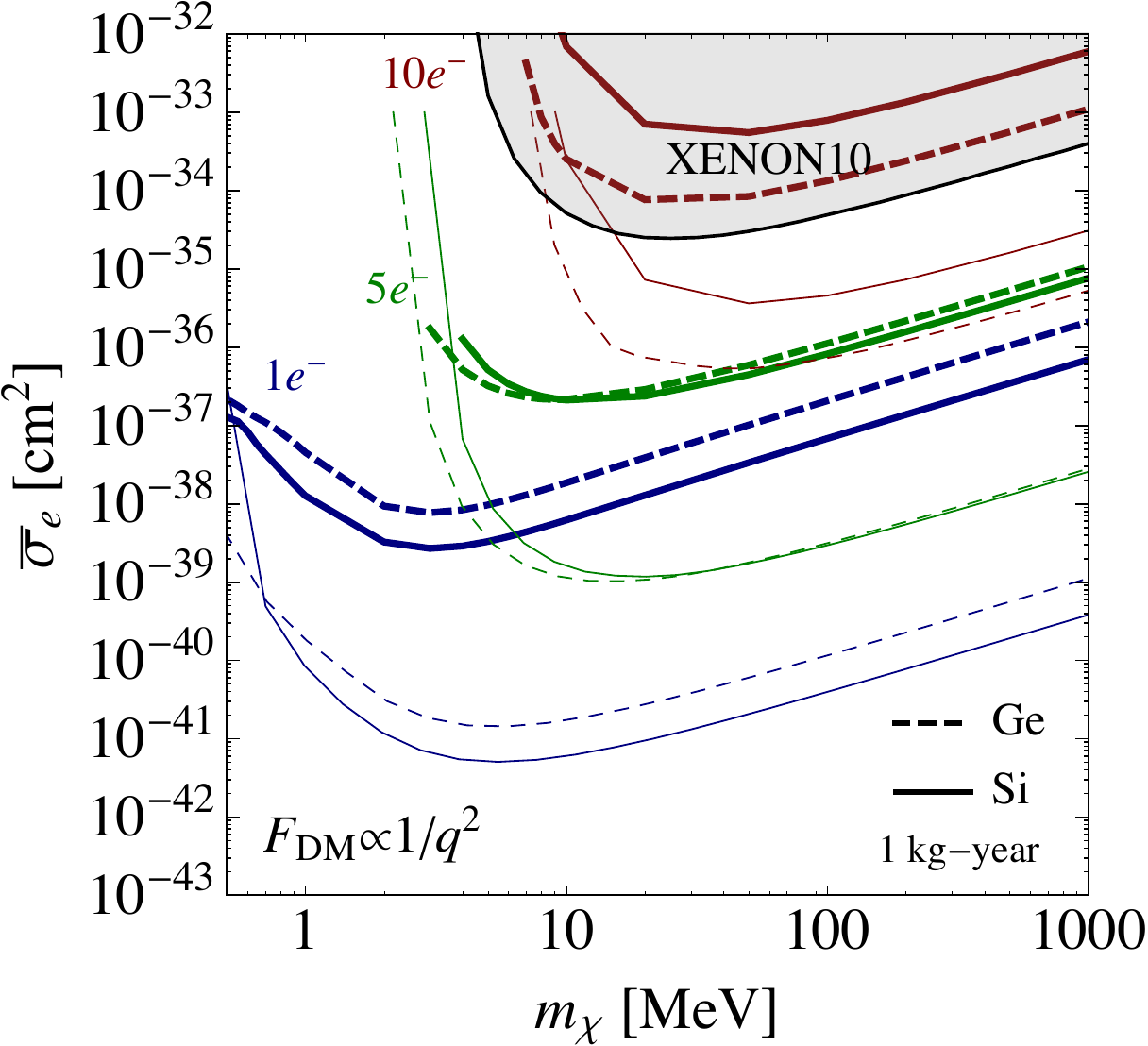}
\end{minipage}
\hfill
\begin{minipage}[t]{0.48\textwidth}
\mbox{}\\[-\baselineskip]
\vspace{-23.5pt}
\caption{\footnotesize {\bf Annual Modulation: Discovery Reach:}
The $5\sigma$ discovery reach in the $m_\chi$--$\overline \sigma_e$ plane of an annual modulation search for DM--electron scattering
%The 95\%~C.L.~discovery reach in the $m_\chi$--$\overline \sigma_e$ plane of an annual modulation search for DM--electron scattering, 
for an experiment with a 1 kg-year background-free exposure.
Solid (dashed) lines show the reach for silicon (germanium) targets. 
Ionization thresholds of 1, 5, and 10 electron-hole pairs are shown with blue, green, and red lines, respectively. 
The corresponding energy thresholds are 0, 12.3, and 26.8\,eV  in germanium, and 0, 15.5, and 33.5\,eV in silicon. 
The gray shaded region shows the existing constraint from XENON10 data~\cite{Essig:2012yx}.  
The three plots assume different DM form factors, as indicated, corresponding to different DM models. Thin lines are from Fig.~\ref{fig:rates_ERcuts}, showing the exclusion reach of a search with the same exposure seeing no events.
} 
\label{fig:ann-mod-discovery-reach}
\end{minipage}
%\end{mdframed}
\end{figure}

Once a signal is found in an electron scattering search, increasing the exposure of the experiment until the annual modulation can be tested will be a crucial step in claiming a DM discovery. 
In Fig.~\ref{fig:ann-mod-discovery-reach} we show the $5\sigma$ discovery reach of an annual modulation search in the mass--cross-section plane. 
We calculate this cross section by requiring
\bea
\Delta S/\sqrt{S_{tot}+B}=5\,,
\eea
where $\Delta S\equiv f_{mod} S_{tot}$ is the modulation amplitude, 
$S_{tot}$ is the total number of signal events, and $B$ is the number of background events. 
The thick curves in Fig.~\ref{fig:ann-mod-discovery-reach} show the discovery reach for different thresholds and DM form-factors, assuming a background-free exposure of 1\,kg-year. 
(A non-zero background will of course weaken the reach, following the equation above.) 
This figure mirrors Fig.~\ref{fig:rates_ERcuts}, which shows the exclusion reach obtained using a simple counting search instead of a modulation search, but otherwise with the same assumptions. 
The curves of Fig.~\ref{fig:rates_ERcuts} are replotted as the thin curves in Fig.~\ref{fig:ann-mod-discovery-reach}, for comparison. 
We see that a substantial discovery reach is possible with a 1~kg-year exposure: at  
low masses for $F_{\rm DM}(q) = 1$, and at all masses for $F_{\rm DM}(q) = \alpha m_e/q $ or  $(\alpha m_e/q)^2$.

Finally, we comment that taking into account the directional (sub-daily) modulation, which is expected in crystalline detectors, will further allow for an improved sensitivity to a DM signal.  As discussed in \S\ref{sec:excitation-in-crystal}, we have averaged-out such directional effects in this work, and we postpone their study to future work.

%%%%%%%%%%%%%%%%%%%%%%%%%%%%%%%%%%%%%%%%%%%%%%
%%%%%%%%%%%%%%%%%%%%%%%%%%%%%%%%%%%%%%%%%%%%%%
\subsection{Prospects for near-term experiment}
\label{subsec:upcoming-prospects}

In this subsection, we discuss the near-term prospects for electron-scattering searches with the DAMIC and SuperCDMS experiments.

%%%%%%%%%%%%%%%%%%%%%%%%%%%%%%%%%%%%%%%%%%%%%%
\subsubsection{DAMIC}

DAMIC~\cite{Estrada:2008zz,Barreto:2011zu,Chavarria:2014ika} uses thick, fully-depleted silicon CCDs for their target material.  
These CCDs are ten times more massive than conventional CCDs, allowing them to be competitive targets for DM direct detection. 
In~\cite{Barreto:2011zu}, DAMIC used one 0.5~g CCD to perform an engineering run, obtaining an exposure of 107~g-days.  
They were able to constrain DM-{\it nuclear} scattering for DM masses almost as low as 1~GeV.  
Work is ongoing to increase the total mass of the detector (by using more CCDs) as well as the detector's sensitivity to low
threshold energies (by using so-called ``Skipper CCDs'')~\cite{EstradaTiffenberg}.  

\paragraph{The first direct detection limit using a semi-conductor target.} Here we investigate the (albeit weak) constraints on DM-electron scattering from their existing result, 
and give reach estimates based on their projected detector improvements.
For the engineering run~\cite{Barreto:2011zu}, DAMIC used a single 0.5~g CCD, for an exposure of 107~g-days.  
They obtained the following values for the read-out noise and the dark current: 
\begin{enumerate}
\item[(i)]
A readout noise of below 2 electrons/pixel, corresponding to $2\times 3.6$~eV = 7.2~eV of r.m.s.~readout noise.  
The CCD has about 4.2 million pixels, so that one requires a threshold of $\sim 13$ electrons ($\sim 47$~eV) 
for the noise to produce a signal above threshold in less than one pixel.  
DAMIC chose a threshold of $\sim 40$~eV ($\sim 11$ electrons) for the search for DM-nuclear scattering; 
we expect $\sim 35$ pixels to reach this threshold.  
In our recast of their data for DM-electron scattering below, we will use the same 40 eV threshold.  
\item[(ii)] 
A dark current of $\sim 1$~electron/CCD/day (at the chosen 120~K operating temperature).  Since the exposure of the CCD is 
a few hours, before being read-out within a few minutes, the threshold is limited by the read-out noise, and not the dark current.
\end{enumerate}
We can use the result in ~\cite{Barreto:2011zu} to constrain DM-electron scattering.  
We will assume that the efficiency to select electron recoil events is the same as selecting nuclear recoils, i.e.~$7 \times 10^{-3}$.  
Fig.~12 in~\cite{Barreto:2011zu} shows the data that was recorded by DAMIC, and we see that {\it zero} events were recorded 
in the first bin above the threshold (40~eV to 100~eV).  
This may be as a result of the efficiency being very low in this bin; nevertheless, it could also be a sign 
that backgrounds may be small at such low energies, boding well for future runs with even lower energy thresholds.  
In any case, this information is sufficient to derive the current DAMIC limit on LDM, which 
we show with a green shaded region (bounded by a green line) in Fig.~\ref{fig:rates_DAMIC}.  
We see that with the current threshold, the form-factor suppression is too large for this constraint to compete with the existing 
XENON10-based limit~\cite{Essig:2012yx}.  Nevertheless, this is the first direct detection limit for sub-GeV DM 
using a semi-conductor target.

%%%%%%%%%%%%%%%%%%%%%%%%%%%%%%%%%%%%%%%%
\paragraph{Projections for future DAMIC runs with improved ``Skipper'' CCDs.} 

There are two main challenges that need to be overcome by DAMIC (and similar experiments) to allow them to push to low thresholds~\cite{EstradaTiffenberg}: 
(i) reduce the noise in reading out the ionization deposited in the detector, and (ii) reduce the dark current.  
The read-out noise can be reduced substantially by taking more time to read the CCD, while the dark current (i.e.~genuine electron-hole pairs produced by thermal excitations in the silicon substrate) can be reduced by lowering the temperature and improving the quality of the silicon. The contribution from the dark current will increase with the readout time, so it will take some optimization to find a way to reduce the readout noise while keeping the contribution from the dark current manageable.
Lowering the temperature also reduces the electron mobility in the substrate, requiring a careful trade-off.
Here we project what future data runs can achieve with the improved DAMIC Skipper CCDs.

The DAMIC Collaboration has been working on so-called ``Skipper CCDs'', which will reduce the r.m.s.~read-out noise down to 
0.2~electrons/pixel/day, with the possibility of going down to 0.1~electrons/pixel/day~\cite{Barreto:2011zu,EstradaTiffenberg}.  
This is done with a new output circuit that enables multiple read-outs.  
The size of the CCDs can be anything up to $4\times 4 = 16$~million pixels (Mpix)~\cite{Tiffenberg}, 
but we will assume 8 Mpix for the projections below.  
The 0.2 (0.1) electrons/pixel/day correspond to read-out noise of 0.72 (0.36)~eV;  
assuming gaussian noise, the 0.1 electron/pixel/day will allow for sensitivity down to single electrons, 
although the 0.2 electrons/pixel/day may require a threshold of two electrons to avoid the noise faking a DM signal.  
However, for such low read-out noise, the dark current becomes the limiting factor in determining the energy threshold.
Significant non-gaussian tails could change this conclusion. 

The dark current has been measured currently at $5\times 10^{-3}$ electrons/pixel/day~\cite{Tiffenberg}.  
As mentioned above, lowering the operating temperature and improving the silicon substrate quality will decrease 
this, and it is reasonable to expect further improvement; the theoretical lower limit is ${\cal O}(10^{-7})$~\cite{Tiffenberg}. 
Below we provide projections under both assumptions.  

\begin{table}[t!]
\begin{center}
\begin{tabular}{|c||c|c|c|c|}
\hline
\multirow{2}{*}{Ionization threshold} & \multicolumn{2}{c|}{dark current: $5 \!\times\! 10^{-3}$elec./pixel/day } & \multicolumn{2}{c|}{dark current: $10^{-7}$elec./pixel/day}\\
\cline{2-5}
& 1 kg-day & 100 g-year & 1 kg-day & 100 g-year \\
\hline
\hline
$Q_{\rm th} = 1\quad$ & $1.6\times 10^{7}$ & $5.8\times 10^{8}$ & 320.0 &  $1.2\times 10^{4}$\\ 
$\phantom{Q_{\rm th} =\;}2\quad$ & $1.7\times 10^3$& $6.1\times 10^4$ & $6.7\times 10^{-7}$ & $\mathbf{2.4\times 10^{-5}}$ ($\star$, $\dagger$)\\ 
$\phantom{Q_{\rm th} =\;}3\quad$ & \bf{0.1} ($\star$) & \bf{4.2} ($\diamond$) & $9.3\times 10^{-16}$ & $3.4\times 10^{-14}$\\ 
$\phantom{Q_{\rm th} =\;}4\quad$ & $6.0\times10^{-6}$ & $2.2\times10^{-4}$ & $9.6\times10^{-25}$ & $3.5\times10^{-23}$ \\ 
\hline
\end{tabular}
\caption{\footnotesize Expected number of events with at least $Q_{\rm th}$ electron-hole pairs under different assumptions for the dark current and exposure ($5\times10^{-3}$electrons/pixel/day and 
$10^{-7}$electrons/pixel/day) and assuming either (i) 4 CCDs (10~g) and an exposure of 1~kg-day; 
or (ii) 40 CCDs (100~g) and an exposure of 100~g-years.
In both cases, we assume that it takes one hour to read the entire CCD, and that it is read continuously.
Projected exclusion reach based on a simple counting experiment are given in Fig.~\ref{fig:rates_DAMIC} for the entries marked with a single 
star ($\star$). 
A projected discovery reach based on seeing the annual modulation of the signal, with negligible background,
is also given in Fig.~\ref{fig:rates_DAMIC} for the entry marked with a dagger ($\dagger$) (we find that the 
prospects from annual modulation for the entry marked with a diamond ($\diamond$) are very similar).  
See text for details. 
\label{tab:dark-currents}
}
\end{center}
\end{table}

The effect of the dark current on the threshold depends on the number of pixels and the exposure length of the CCD.  
The CCD is read pixel-by-pixel, and can be read continuously from one side to the other, before cycling back again to the 
beginning.  
We will assume that the 8~million pixels of the CCD are all read in one hour, so that its 
exposure is one hour for the purposes of calculating the dark current. 
We will consider the following two scenarios for the number of CCDs, the mass, and exposure (we assume an efficiency of 1 for
making our projections below):
\begin{itemize}
\item[(i)]  There are currently four prototype skipper CCDs, each with a mass of 2.5~g, which were produced as part of an R\&D project 
(these will be deployed at the MINOS near site this year).  
For our first set of projections, we will assume that data is taken over 100~days (livetime), for a total exposure of 1kg-day.  
\item[(ii)]
If the testing of the skipper CCDs at MINOS goes well, one can expect that several more of them will be deployed to
search for DM.  
Thus, for our second set of projections, we will assume that 40~CCDs are deployed (for a total mass of 100~g) 
and that data is taken again over 365~days (livetime) for a total exposure of 100 g-years.  
\end{itemize}
Table~\ref{tab:dark-currents} gives the expected number of events with at least $Q$ electron-hole pairs 
%under the two assumptions for the 
%dark currents and for these two scenarios (assuming Poisson statistics).  
for the two scenarios assuming Poisson statistics.
We see that DAMIC could have a threshold of 2 electrons if the dark current can be reduced to below $10^{-7}$electrons/pixel/day, but 
a 3-electron threshold is required for the present dark current rate of $5\times 10^{-3}$electrons/pixel/day.  
In Fig.~\ref{fig:rates_DAMIC}, we show solid green lines that indicate the 95\% C.L.~prospects for the entries marked with a star ($\star$), 
i.e. we show the cross section to obtain 3.6 signal events, assuming zero background events.  We also show the reach of an annual modulation search for the entry marked with a dagger ($\dagger$), i.e. a 2-electron threshold with no background in a 100~g-year exposure. This is shown by the dashed green line in Fig.~\ref{fig:rates_DAMIC}. 
We checked that the prospects from annual modulation for the entry marked with diamond ($\diamond$) are very similar.  
We see from these projections that DAMIC can significantly improve upon the current XENON10 limit, especially at the lowest DM masses.

\subsubsection{(Super)CDMS}

The CDMS experiment uses cryogenic solid state detectors operated at temperatures below $100~$mK.  
In the WIMP search, they distinguish electron from nuclear recoils by measuring the ratio of the ionization versus phonon energy 
deposited into the crystal.  This ratio will be smaller for nuclear recoils than for electron recoils.  
Here, we are interested in their ability to detect electron recoils.  

The signal from a low-energy recoiling electron can be dramatically enhanced by applying a relatively large bias voltage, $V_b \sim \mathcal{O}(50-100~{\rm V})$, across the target material.  
The work done in drifting an electron-hole pair out of the crystal, $e V_b$, is emitted as 
Luke-Neganov~\cite{Luke:1990ir,Neganov:1985,Wang:2010} phonons, which will be picked up by the phonon sensors.  
This was done for ``CDMSlite'', which has yielded electron recoil thresholds with an $\mathcal{O}$(1) detection efficiency of 170\,eV~(i.e.~$\mathcal{O}(50)$~electrons)~\cite{Agnese:2013jaa}.  
Here we discuss the prospects of future versions of SuperCDMS. 

In Fig.~\ref{fig:rates_DAMIC}, we show 3 projections for SuperCDMS, 
two for silicon (with an exposure $\sim$10~kg-years) assuming an electron-hole pair threshold, $Q_{\rm th}$, of either 4 or 1, and a signal detection efficiency of 0.7, and one for an annual modulation search assuming $Q_{\rm th} = 2$ with the same exposure and efficiency. The cross-section reach for germanium is very similar to those of silicon.

The electron-hole-pair thresholds are based on the following assumptions. 
The $Q_{\rm th}=4$ threshold is based on the numbers used by SuperCDMS for Snowmass~\cite{Cushman:2013zza}, while 
the $Q_{\rm th}=1$ threshold is based on an ambitious but achievable best-case scenario.  
For Snowmass, a phonon energy resolution of 50~eV was assumed.  
As there may be non-gaussian tails, a $7\sigma$ threshold was assumed,  corresponding to a threshold of 
$E_e=350$~eV.  Taking the bias voltage to be 100~V, this translates into $Q_{\rm th}=3.5$, which we round up to 4.  
For the second set of projections, we assume that further R\&D can push the noise threshold down to better than $\sim$14\,eV, 
which is ambitious but achievable in principle~\cite{Pyle}.
A $7\sigma$ threshold corresponds to $\sim$100\,eV, so that a bias voltage of 100~V is sufficient to achieve 
sensitivity to $Q_{\rm th}=1$.  
In practice, the bias voltage can be optimized as well.  A larger bias voltage would allow for a reduced threshold in terms 
of the number of electron-hole pairs needed to pass the phonon energy threshold, but could lead to breakdowns of the substrate. However, it has been demonstrated that the bias voltages needed for sensitivity to $Q_{\rm th}=1$ are achievable in both silicon and 
germanium~\cite{PyleFigueroa}. 

As can be seen from Fig.~\ref{fig:rates_DAMIC}, SuperCDMS has the potential to improve drastically upon the existing XENON10 limit, 
especially at low DM masses. 

%%%%%%%%%%%%%%%%%%%%%%%%%%%%%%%%%%%%%%%%%%%%%%%%%%%%%%%%%%%%%
%%%%%%%%%%%%%%%%%%%%%%%%%%%%%%%%%%%%%%%%%%%%%%%%%%%%%%%%%%%%%
%%%%%%%%%%%%%%%%%%%%%%%%%%%%%%%%%%%%%%%%%%%%%%%%%%%%%%%%%%%%%
%%%%%%%%%%%%%%%%%%%%%%%%%%%%%%%%%%%%%%%%%%%%%%%%%%%%%%%%%%%%%
\section{Conclusions}\label{sec:conclusions}
Direct detection experiments have so far primarily focused on searching for WIMPs, and as a result of an intense research effort, 
the path forward in this direction 
is rather well-defined.  Within the next decade,   WIMPs will either be found or become significantly less motivated.  
However, other theoretically motivated candidates exist that could constitute the DM in our Universe.  
In this work, we focused on a class of DM candidates that have a mass between a few-hundred keV to a GeV.  
We showed that tremendous progress can be made in exploring the direct-detection parameter space of these candidates over 
the next few years, by searching for DM-induced electron recoils in experiments with targets that consist of  semiconductor 
materials.    
The technology currently used in WIMP searches can be adapted for such light-DM searches by improving the ionization sensitivity, and this is being actively pursued.
The backgrounds are expected to be quite different in nature to those in WIMP searches, and there is reason to believe that they will be small and controllable.

The calculation of the DM-electron scattering rate and the subsequent electron recoil spectrum in semiconductor targets 
is much more challenging than for DM-nuclei scattering.  
We have provided detailed formulae for the scattering rate and recoil spectrum, expressed in terms of simple DM properties and a target-dependent ``crystal form factor", which encodes the quantum structure of the target electrons. We numerically calculated the crystal form factor for germanium and silicon  with our code {\tt QEdark}, which is based on the software package {\tt Quantum Espresso}  that calculates the crystal wave functions and energy levels. Convergence tests indicate that our results are accurate at the few percent level.   
{\tt QEdark} will be publicly available at \href{http://ddldm.physics.sunysb.edu}{\tt ddldm.physics.sunysb.edu}, 
together with the crystal form factors.  
With these, upcoming experiments  can derive their own sensitivities or limits. 

The crystal form factor is a steeply falling function of the electron recoil energy.
Consequently, even a small improvement in 
an experiment's detector threshold translates into a significant increase in the sensitivity to DM-electron scattering. 
We have provided the projected sensitivity for a variety of experimental thresholds, showing that upcoming experiments including DAMIC and SuperCDMS can probe orders of magnitude of unexplored DM parameter space in the next few years.
In addition to setting limits, sub-GeV dark matter can be discovered via its expected modulation signal.  We showed that in the case of electron-scattering, annual modulation is sizable  and could provide the necessary signal for discovery.  Additional sub-daily modulation is expected due to the orientation-dependent nature of scattering in crystalline detectors. We have ignored directionality in this work, deferring it to future study. 

Calculating the experimentally observable signal requires a conversion from energy deposition to the ultimate ionization signal. This conversion requires a detailed knowledge of the secondary scattering processes in crystals, at energies below the existing experimental sensitivity. We therefore used a phenomenological model for secondary interactions, and studied its possible systematic uncertainties using a Monte Carlo model. We find that our predictions suffer from systematic uncertainties of order a few tens of percent, and is likely conservative. 
Further theoretical and experimental study of secondary interactions would be useful to improve the modeling of this conversion.

To summarize, our work provides the necessary tools for experiments which use semiconductor targets to search for sub-GeV dark matter to derive accurate limits. Technologies adapted from WIMP searches and currently under development can be employed in searches for sub-GeV dark matter. This highly-motivated direction in dark matter searches is a natural progression from the WIMP program, and we expect that it will take a leading role in the search for dark matter.

\section*{Note added:}
While this work was being completed, Ref.~\cite{Lee:2015qva} appeared, which also deals with DM-electron scattering in germanium. Ref.~\cite{Lee:2015qva} is complementary to our work, its main point being the effect of ``gravitational focusing" on the modulation signal of DM-electron scattering. Ref.~\cite{Lee:2015qvar} derives scattering rates using a semi-analytic approach, which builds on the method of Ref.~\cite{Graham:2012su}, but is significantly less detailed than the method we have presented here. We find that their results are comparable to ours within a factor of a few, but with some notable differences. In particular, Ref.~\cite{Lee:2015qva} finds increasingly higher rates than us at increasingly higher recoil energies. Most strikingly, Ref.~\cite{Lee:2015qva} finds that scattering of the 3d shell electrons dominates the total rate when it is kinematically accessible (we find the the 3d shells cause a bump in the spectrum, but with a rate subdominant to lower energy events). We attribute these differences to the inherent sensitivity of the calculation to the tails of the electron wavefunctions for energies above $\mathcal O$(10 eV), as we discussed in \S\ref{sec:scattering-kinematics}. 
%With our detailed, self-consistent computational approach, we believe our results to be more reliable, and hope that this may be verified with future studies.

\section*{Acknowledgments}
We are very grateful to Julien Billard, Enectali Figueroa-Feliciano, Matt Pyle, and Javier Tiffenberg for extensive discussions 
and correspondence on the experimental capabilities of SuperCDMS and DAMIC.  We also thank Philip Allen, Brian Batell, Juan Estrada,  
Eder Izaguirre, Gordan Krnjaic, Samuel Lee, Mariangela Lisanti, Deyu Lu, Aaron Manalaysay, Siddharth Mishra-Sharma, 
and Benjamin Safdi for many useful discussions. 
R.E.~is supported by the DoE Early Career research program 
DESC0008061 and through a Sloan Foundation Research Fellowship.  T.-T.Y.~is supported also by grant DESC0008061. M.~F.-S. 
and A.S.~acknowledge support from DoE grant DE-FG02-09ER16052.  
J.M.~is supported by grant DE-SC0012012. T.V.~is supported in part by a grant from the Israel Science Foundation, 
the US-Israel Binational Science Foundation, the EU-FP7 Marie Curie, CIG fellowship and by the I-CORE Program 
of the Planning and Budgeting Committee, and The Israel Science Foundation (grant NO 1937/12). J.M. and T.-T.Y. wish to thank the hospitality of the Aspen Center for Physics, which is supported by National Science Foundation grant PHY-1066293, where this work was completed. 
This research used resources of the National Energy Research Scientific Computing Center, 
a DOE Office of Science User Facility supported by the Office of Science of the U.S.~Department of Energy under 
Contract No.~DE-AC02-05CH11231 and the \textit{HANDY} computer cluster at the Stony Brook University Institute for Advanced Computational Science.

%%%%%%%%%%%%%%%%%%%%%%%%%%%%%%%%%%%%
%%%%%%%%%%%%%%%%%%%%%%%%%%%%%%%%%%%%
%%%%%%%%%%%%%%%%%%%%%%%%%%%%%%%%%%%%
%%%%%%%%%%%%%%%%%%%%%%%%%%%%%%%%%%%%
%%%%%%%%%%%%%%%%%%%%%%%%%%%%%%%%%%%%
%%%%%%%%%%%%%%%%%%%%%%%%%%%%%%%%%%%%
%%%%%%%%%%%%%%%%%%%%%%%%%%%%%%%%%%%%
%%%%%%%%%%%%%%%%%%%%%%%%%%%%%%%%%%%%
\appendix
%%%%%%%%%%%%%%%%%%%%%%%%%%%%%%%%%%%%

%%%%%%%%%%%%%%%%%%%%%%%%%%%%%%%%%%%%
%%%%%%%%%%%%%%%%%%%%%%%%%%%%%%%%%%%%
%%%%%%%%%%%%%%%%%%%%%%%%%%%%%%%%%%%%
%%%%%%%%%%%%%%%%%%%%%%%%%%%%%%%%%%%%
\section{Derivation of scattering rate formulae}
\label{sec:full-derivation}
%%%%%%%%%%%%%%%%%%%%%%%%%%%%%%%%%%%%%%%%%%%%%%%%%

%%%%%%%%%%%%%%%%%%%%%%%%%%%%%%%
\subsection{General formula for dark matter-induced electronic transitions}
\label{sec:general-excitation-derivation}

If a DM particle scatters with an electron in a stationary bound state such as in an atom, it can excite the electron from an initial energy level 1 to an excited energy level 2, by transferring to it energy $\Delta E_{1\to2}$ and momentum $\vec q$. 
The cross section for this process can be derived in a standard way using non-relativistic quantum mechanics, but here we derive it starting from the usual formula for the cross-section in field-theory, in order to make easier connection with the underlying particle physics. 
We treat the electron as being bound in a static background potential -- in other words we approximate the atoms as being infinitely heavy objects which can absorb momentum without recoiling. 
This is an excellent approximation ($<1\%$ error), since the momentum-transfers we will be interested are typically of order keV.

The cross section for free 2 $\to$ 2 scattering is given by
\begin{equation}
\sigma v_{\rm free} = \frac{1}{4 E_\chi' E_e'} \int \frac{d^3 q}{(2\pi)^3} \frac{d^3 k'}{(2\pi)^3} \frac{1}{4 E_\chi E_e} (2 \pi)^4 \delta(E_i - E_f) \delta^3(\vec k + \vec q - \vec k') \overline{|\mathcal M_{\rm free}(\vec q\,)|^2} \, ,
\label{eq:free-cross-section}
\end{equation}
where $\mathcal{M}_{\rm free}$ is the usual field-theory matrix element, and $\overline{|\mathcal M|^2}$ represents its absolute square averaged over initial spins and summed over final spins. 

If the electron were unbound, the non-relativistic scattering amplitude would be given by
\begin{equation}
\langle \chi_{\vphantom{\vec k}\vec p - \vec q \,}, e_{\vec k'} | H_{\rm int} | \chi_{\vphantom{\vec k}\vec p\,}, e_{\vec k} \rangle
= C \, \mathcal{M}_{\rm free}(\vec q \,) \times (2 \pi)^3 \delta^3(\vec k - \vec q - \vec k') \, ,
\label{eq:free-amplitude}
\end{equation} 
where $| \chi_{\vphantom{\vec k}\vec p\,}, e_{\vec k} \rangle$ is plane-wave state for a DM particle of momentum $\vec p$ and an electron of momentum $\vec k$, $H_{\rm int}$ is the interaction Hamiltonian, and $C$ is an unimportant coefficient. 
However, because the electron is bound it is instead given by
\begin{align}
\langle \chi_{\vphantom{\vec k}\vec p - \vec q \,}, e_2| H_{\rm int} | \chi_{\vphantom{\vec k}\vec p\,}, e_1 \rangle&=
\bigg[ \int \frac{\sqrt V d^3 k'}{(2\pi)^3} \widetilde \psi_2^*(\vec k') \langle \chi_{\vphantom{\vec k}\vec p \, '}, e_{\vec k'} | \bigg]
H_{\rm int}
\bigg[ \int \frac{\sqrt V d^3 k}{(2\pi)^3} \widetilde \psi_1(\vec k) | \chi_{\vphantom{\vec k}\vec p\,}, e_{\vec k} \rangle \bigg]
\nonumber \\ &= 
C \, \mathcal M_{\rm free}(\vec q\,) \times \!\int \frac{V d^3 k}{(2\pi)^3} \widetilde \psi_2^*(\vec k + \vec q\,) \widetilde \psi_1(\vec k)\, ,
\label{eq:bound-amplitude}
\end{align}
where $\widetilde \psi_1$, $\widetilde \psi_2$ are the (unit normalized) momentum-space wavefunctions of the initial and final electron levels. 
We have purposefully used plane-wave normalization for both the free and bound electron states: $\langle e_{\vec k} | e_{\vec k} \rangle = \langle e_1 | e_1 \rangle = (2\pi)^3 \delta^3(\vec 0) \equiv V$, where $V$ is the volume of space (which always cancels in the end).

To find the cross section for this excitation process, we can use the free 2 $\to$ 2 scattering cross section formula but with two replacements: one to account for the modified scattering amplitude, and the other to account for the different final-state phase space.
Squaring Eqs.~(\ref{eq:free-amplitude},~\ref{eq:bound-amplitude}), we see that the bound-state scattering amplitude is accounted for by making the replacement
\begin{equation}
V (2 \pi)^3 \delta^3(\vec k - \vec q - \vec k')  |\mathcal M_{\rm free}|^2
\longrightarrow
|\mathcal M_{\rm free}|^2 \times V^2 | f_{1\to 2}(\vec q \,)|^2 \, ,
\label{eq:matrix-element-replacement}
\end{equation}
where $f_{1\to 2}(\vec q\,)$ is the \emph{atomic form factor},
\begin{equation}
f_{1\to 2}(\vec q \,) = \int \frac{d^3 k}{(2\pi)^3} \widetilde \psi_2^*(\vec k + \vec q\,) \widetilde \psi_1(\vec k) \, .
\label{eq:k-space-1to2-form-factor}
\end{equation} 
Fourier transforming Eq.~(\ref{eq:k-space-1to2-form-factor}) gives the definition given in Eq.~(\ref{eq:1to2-form-factor}).

Since there is only one final electron state being considered, we also need to remove the usual final-state phase space integral:
\begin{equation}
\text{free-electron phase space} = V \int \frac{d^3 k'}{(2 \pi)^3}
\longrightarrow 1 \,.
\label{eq:phase-space-replacement}
\end{equation}

Combining Eqs.~(\ref{eq:free-cross-section},~\ref{eq:matrix-element-replacement},~\ref{eq:phase-space-replacement}), 
we can write the formula for the cross-section for a DM particle to excite an electron from level 1 to level 2:
\begin{equation}
\sigma v_{1\to 2} = 
\frac{1}{4 E_\chi' E_e'} \int \frac{d^3 q}{(2\pi)^3} \frac{1}{4 E_\chi E_e} 2 \pi \delta(E_i - E_f) \overline{|\mathcal M_{\rm free}(\vec q\,)|^2} \times | f_{1\to 2}(\vec q \,)|^2 \, .
\end{equation}
Since we are in the non-relativistic regime, the energies are given by
\begin{align}
E_i &= m_\chi + m_e + \frac{1}{2} m_\chi v^2 + E_{e,1}
\\
E_f &= m_\chi + m_e + \frac{|m_\chi \vec v - \vec q\,|^2}{2 m_\chi} + E_{e,2} \, .
\end{align}
Using the following definitions~\cite{Essig:2011nj} to parametrize the underlying DM--electron coupling
\begin{gather}
\overline{|\mathcal M_{\rm free}(\vec q\,)|^2} \equiv \overline{|\mathcal M_{\rm free}(\alpha m_e)|^2} \times |F_{\rm DM}(q)|^2
\\
\overline \sigma_e \equiv \frac{\mu_{\chi e}^2 \overline{|\mathcal M_{\rm free}(\alpha m_e)|^2}}{16 \pi m_\chi^2 m_e^2} \, ,
\end{gather}
the cross section simplifies to
\begin{equation}
\sigma v_{1\to 2} = 
\frac{\overline \sigma_e}{\mu_{\chi e}^2} \int \frac{d^3 q}{4 \pi} \,\delta \Big(\Delta E_{1\to2} + \frac{q^2}{2 m_\chi} - q v \cos \theta_{q v} \Big)  \times |F_{\rm DM}(q)|^2 | f_{1\to 2}(\vec q \,)|^2 \, .
\label{eq:1to2-cross-section}
\end{equation}

%%%%%%%%%%%%%%%%%%%%%%%%%%%%%%%
\subsection{Average rate in a dark matter halo}
\label{sec:velocity-averaging}

The actual rate of excitation events, for a given transition and a given target electron, is found by multiplying Eq.~(\ref{eq:1to2-cross-section}) by the DM density and averaging over the DM velocity distribution $g_\chi(\vec v)$,
\begin{equation}
R_{1\to2} = \frac{\rho_\chi}{m_\chi} \int d^3 v \, g_\chi(\vec v) \, \sigma v_{1\to 2} \, .
\end{equation}

In general, both the electron wavefunctions and the DM velocity distribution will not be spherically symmetric. 
As noted in~\cite{Essig:2011nj}, the rate will then depend on the orientation of the target with respect to the galaxy.  
Here we ignore this interesting complication, and approximate the velocity distribution as being spherically symmetric. 
We can then use the $d^3 v$ integral to eliminate the $\delta$-function in Eq.~(\ref{eq:1to2-cross-section}), giving
\begin{align}
R_{1\to2} &= \rho_\chi/m_\chi \frac{\overline \sigma_e}{\mu_{\chi e}^2} \int \frac{d^3 q}{4 \pi} \int \frac{v^2 d v d \phi_v}{q v} \, g_\chi(v) \,\Theta \Big(v- v_{\rm min}(q, \Delta E_{1\to2}) \Big)  \times |F_{\rm DM}(q)|^2 | f_{1\to 2}(\vec q \,)|^2 
\nonumber \\
&= \rho_\chi/m_\chi \frac{\overline \sigma_e}{8\pi \mu_{\chi e}^2} \int d^3 q \,\frac{1}{q} \eta\big(v_{\rm min}(q, \Delta E_{1\to2}) \big)  |F_{\rm DM}(q)|^2 | f_{1\to 2}(\vec q \,)|^2 \, .
\label{eq:general-excitation-rate}
\end{align}
Here $\eta(v_{\rm min})$ has its usual definition,
\begin{equation}
\eta(v_{\rm min}) = \int \frac{d^3 v}{v} \, g_\chi(v) \,\Theta (v- v_{\rm min}) \, ,
\end{equation}
and $v_{\rm min}$ is a function of $q$ and the energy transfer given by
\begin{equation}
v_{\rm min}(q, \Delta E_{1\to2}) = \frac{\Delta E_{1\to2}}{q} + \frac{q}{2 m_\chi} \, .
\end{equation}

%%%%%%%%%%%%%%%%%%%%%%%%%%%%%%%
\subsection{Ionizing an isolated atom} 
\label{sec:atomic-ionization-rate-derivation}

For the purposes of connecting with previous work~\cite{Essig:2011nj}, in this subsection we consider ionization of electrons bound in isolated atomic potentials. 
We derive the ionization rate of such a system, assuming a spherical atomic potential and filled shells. 
This approximation was used in~\cite{Essig:2011nj} to model a liquid xenon target material, and the results below reproduce Eqs.~(5) and (6) of that paper. 
The full calculation of event rates in liquid xenon would require knowledge of electron wavefunctions in the dense, disordered xenon liquid. 
This is a more challenging calculation than for a semiconductor crystal, but can in principle be performed with similar methods -- we leave this for future work.
The corrections, however, can be argued to be small, lowering the ionization threshold and increasing the event rate.

An electron ionized from an atom can be treated as being in one of a continuum of positive-energy bound states. 
These states are affected by the potential well of the atom, but can be approximated as free particle states at asymptotically large radii.
We denote their wavefunctions as $\widetilde \psi_{k' l' m'}(\vec x)$, where $l'$ and $m'$ are angular quantum numbers, and $k'$ is the momentum at asymptotically large radius. 
The energy of such a state is therefore $E_R = k'^2/2 m_e$. 

The ionization rate for such an atom is found by taking Eq.~(\ref{eq:general-excitation-rate}), summing over occupied electron shells, and integrating over the phase space of all possible ionized states. Since these are asymptotically free spherical-wave states, the phase space is
\begin{equation}
\text{ionized electron phase space} = \sum_{l' m'} \int \frac{k'^2 d k'}{(2\pi)^3}
 = \frac{1}{2} \sum_{l' m'} \int \frac{k'^3 d \ln E_R}{(2\pi)^3} \, ,
\end{equation}
when the wavefunction normalization is, as in~\cite{Essig:2011nj}, taken to be
\begin{equation}
\langle \widetilde \psi_{k' l' m'} | \widetilde \psi_{k l m} \rangle = (2\pi)^3 \delta_{l' l} \delta_{m' m} \frac{1}{k^2} \delta(k-k') \, .
\end{equation}
Plugging this in, the ionization rate is given by
\bea
R_{\rm ion} & = & \frac{\rho_\chi}{m_\chi} \frac{\overline \sigma_e}{16\pi \mu_{\chi e}^2}  \times \\
& &  \sum_{\substack{\rm occupied\\\rm states}}\sum_{l' m'} \int \frac{k'^3 d \ln E_R \, d^3 q }{(2\pi)^3 q}\, \eta\big(v_{\rm min}(q, E_{B i} + k'^2/2 m_e) \big)  |F_{\rm DM}(q)|^2 | f_{i\to k' l' m'}(\vec q \,)|^2 \, .\nonumber
\eea
where $E_{B i}$ is the binding energy of occupied state $i$.

Since the potential is assumed to be spherically symmetric, and we are ionizing a full atomic shell, we can sum $| f_{1\to k' l' m'}(\vec q \,)|^2$ over initial and final angular momentum variables and the result cannot depend on the direction of $\vec q$. 
This means we can define the dimensionless \emph{ionization form factor},
\begin{equation}
\big| f_{\rm ion}(k', q) \big|^2 = \frac{2 k'^3}{(2\pi)^3} \sum_{\substack{\rm occupied\\\rm states}} \sum_{l' m'} \Big| \int d^3 x \, \widetilde \psi_{k' l' m'}^*(\vec x) \psi_i(\vec x) e^{i \vec q \cdot \vec x} \Big|^2 \, .
\end{equation}
After applying this definition to the previous equation, we can replace the $d^3 q$ integral with $4\pi q^2 dq$, giving
\begin{equation}
\frac{d R_{\rm ion}}{d \ln E_R} = \rho_\chi/m_\chi \frac{\overline \sigma_e}{8 \mu_{\chi e}^2}  \int q dq\,  |F_{\rm DM}(q)|^2 \big| f_{\rm ion}(k', q) \big|^2 \eta\big(v_{\rm min}(q, E_{B i} + k'^2/2 m_e) \big)  \, .
\end{equation}
This reproduces the formulae given in~\cite{Essig:2011nj}.

%%%%%%%%%%%%%%%%%%%%%%%%%%%%%%%
\subsection{Excitations in a semiconductor crystal}
\label{sec:crystal-excitation-rate-derivation}

In the periodic lattice of a semiconductor crystal, the electron energy levels form a complicated band structure, with an energy gap separating the filled valence bands and the unoccupied conduction bands (Fig.~\ref{fig:bandstructure}). 
Each possible electron level is labelled by a band index $i$ and a wavevector $\vec k$ in the first Brillouin Zone (BZ). 
Due to the periodicity of the potential, the wavefunctions of these states are in Bloch form,
\begin{equation}
\psi_{i \vec k}(\vec x) = \frac{1}{\sqrt V} \sum_{\vec G} u_i(\vec k + \vec G) e^{i (\vec k + \vec G)\cdot \vec x} \, ,
\end{equation}
where the $\vec G$'s are the reciprocal lattice vectors. 
Here $V$ is the volume of the crystal, and the wavefunctions are taken to be unit-normalized, so that
\begin{gather}
\sum_{\vec G} \big| u_i (\vec k + \vec G) \big|^2 = 1
\end{gather} 
(We use the relations $\int d^3 x \, e^{i \vec k \cdot \vec x} = (2\pi)^3 \delta^3(\vec k)$ and $(2\pi)^3 \delta^3(\vec 0) = V$.)

With this form for the wavefunctions, the form factor Eq.~(\ref{eq:1to2-form-factor}) to excite from valence level $\{i \, \vec k \}$ to conduction level $\{i' \, \vec k' \}$ becomes
\begin{align}
\big| f_{i \vec k \to i' \vec k'}(\vec q) \big|^2 
&= \Big| \sum_{\vec G \, \vec G'} \frac{(2 \pi)^3 \delta^3(\vec k +\vec q - \vec k' - \vec G')}{V} u_{i'}^*(\vec k' + \vec G + \vec G') u_i (\vec k + \vec G) \Big|^2
\\
&= \sum_{\vec G'} \frac{(2 \pi)^3 \delta^3\big(\vec q - (\vec k' + \vec G' - \vec k)\big)}{V} \Big| \sum_{\vec G} u_{i'}^*(\vec k' + \vec G + \vec G') u_i (\vec k + \vec G) \Big|^2 \, .
\label{eq:crystal-form-factor-step-2}
\end{align}
We define the term in the absolute square in Eq.~(\ref{eq:crystal-form-factor-step-2}) to be $f_{[i \vec k , i' \vec k', \vec G']}$:
\begin{equation}
f_{[i \vec k , i' \vec k', \vec G']} = \sum_{\vec G} u_{i'}^*(\vec k' + \vec G + \vec G') u_i (\vec k + \vec G) \, .
\end{equation}
Inserting this into Eq.~(\ref{eq:general-excitation-rate}), we can use the $\delta$-function to eliminate the $d^3 q$ integral, giving
\begin{align}
R_{i \vec k \to i' \vec k'} = \frac{\rho_\chi}{m_\chi} \frac{\pi^2 \overline \sigma_e}{\mu_{\chi e}^2} \frac{1}{V} 
\sum_{\vec G' }
\frac{1}{q} \eta\big(v_{\rm min}(q, E_{i' \vec k'} - E_{i \vec k}) \big) \, |F_{\rm DM}(q)|^2 \, \big| f_{[i \vec k , i' \vec k', \vec G']} \big|^2 \bigg|_{q = |\vec k' + \vec G' - \vec k|} \, .
\label{eq:crystal-excitation-rate-level-to-level}
\end{align}

The total excitation rate for an electron in level $\{i \, \vec k \}$ is found by summing Eq.~(\ref{eq:crystal-excitation-rate-level-to-level}) over all unfilled final energy levels $i'$,
\begin{equation}
R_{i \vec k \to \rm any} = \sum_{i'} \int_{\rm BZ} \frac{V d^3 k'}{(2 \pi)^3} R_{i \vec k \to i' \vec k'} \, .
\label{eq:crystal-excitation-ik-to-any}
\end{equation}
Note that we do not sum over final electron spins here as that sum has already been included in the definition of $\overline \sigma_e$.

The total rate of excitation events in the crystal, $R_{\rm crystal}$, is given by summing Eq.~(\ref{eq:crystal-excitation-ik-to-any}) over all filled initial levels $i$,
\begin{equation}
R_{\rm crystal} = 2 \sum_i \int_{\rm BZ} \frac{V d^3 k}{(2 \pi)^3} R_{i \vec k \to \rm any} \, .
\end{equation}
Here the extra factor of 2 is the sum over the two degenerate spin states of the filled valence bands.
 
Putting this together gives the total excitation rate in a crystal,
\begin{equation}
R_{\rm crystal} = 
\frac{\rho_\chi}{m_\chi} \frac{2\pi^2 \overline{\sigma}_e}{\mu_{\chi e}^2} V
\sum_{i\,  i'} \! \int_{\rm BZ} \frac{d^3 k \, d^3 k'}{(2 \pi)^6}  
\sum_{\vec G' }
\frac{1}{q}  \eta \big( v_{\rm min}(q, E_{i' \vec k'} - E_{i \vec k}) \big) \, F_{\rm DM}(q)^2 \,
\big| f_{[i \vec k , i' \vec k', \vec G']} \big|^2 \, ,
\end{equation}
where again $q = |\vec k' + \vec G' - \vec k|$.
Note that this is the total event rate for the whole crystal, and so it is appropriate that it is proportional to the volume $V$ of the whole crystal.
Since the dependence on the DM velocity distribution and interaction type are entirely encoded in $\eta$ and $F_{\rm DM}$, which are functions only of the momentum transfer $q$ and energy deposited $E_e$, it is useful to insert delta-functions into the above expression as follows:
\begin{align}
R_{\rm crystal} &= 
\frac{\rho_\chi}{m_\chi} \frac{2\pi^2 \overline{\sigma}_e}{\mu_{\chi e}^2} V
\int d\ln E_e \, d\ln q \frac{1}{q}  \eta \big( v_{\rm min}(q, E_e) \big) \, F_{\rm DM}(q)^2 
\nonumber \\ 
\times &
\sum_{i\,  i'} \! \int_{\rm BZ} \frac{d^3 k \, d^3 k'}{(2 \pi)^6}  
E_e \delta(E_e - E_{i' \vec k'} + E_{i \vec k}) 
\sum_{\vec G' } q \delta(q - |\vec k' + \vec G' - \vec k|)
\big| f_{[i \vec k , i' \vec k', \vec G']} \big|^2 \, .
\end{align}

Using $V = N_{\rm cell} V_{\rm cell}$, where $V_{\rm cell}$ is the volume of the crystal's unit cell and $N_{\rm cell}$ is the number of cells, the differential rate can then be written in the form of Eq.~(\ref{eq:diff-crystal-rate}),
\begin{align}
\frac{d R_{\rm crystal}}{d \ln E_e} =
\frac{\rho_\chi}{m_\chi} N_{\rm cell} \overline{\sigma}_e \alpha
  \times \frac{m_e^2}{\mu_{\chi e}^2}
\int\! d \ln q \, \bigg(\frac{E_e}{q} \eta \big( v_{\rm min}(q, E_e) \big)\bigg) 
F_{\rm DM}(q)^2 \big| f_{\rm crystal}(q, E_e) \big|^2 \, ,
\end{align}
where the \emph{crystal form-factor} is defined as in Eq.~(\ref{eq:crystal-form-factor}),
\begin{align}
\big| f_{\rm crystal}(q, E_e) \big|^2 =& 
\frac{2\pi^2 (\alpha m_e^2 V_{\rm cell})^{-1}}{E_e}  \sum_{i\,  i'} \! \int_{\rm BZ} \frac{V_{\rm cell} d^3 k }{(2 \pi)^3}  \frac{V_{\rm cell} d^3 k'}{(2 \pi)^3} \times
\nonumber \\
& E_e \delta(E_e - E_{i' \vec k'} + E_{i \vec k}) 
\sum_{\vec G'} q \delta(q - |\vec k' - \vec k + \vec G'|)
\big| f_{[i \vec k , i' \vec k', \vec G']} \big|^2 \, .
\end{align}

%%%%%%%%%%%%%%%%%%%%%%%%%%%%%%%%%%%%
%%%%%%%%%%%%%%%%%%%%%%%%%%%%%%%%%%%%
\section{Derivation of inverse mean speed, $\eta(v_{min})$}
\label{sec:etavmin}
%%%%%%%%%%%%%%%%%%%%%%%%%%%%%%%%%%%%

In this section, we will derive analytic expressions for $\eta(v_{min})$. For simplicity we assume that the DM velocity distribution, $g_\chi(\vec v_\chi)$, takes the form of a Maxwell-Boltzmann distribution in the galactic rest frame, with a hard cutoff at the galactic escape velocity. In the Earth's frame the velocity distribution then takes the form
\begin{equation}
g_\chi(\vec v_\chi)
=\frac{1}{K}e^{-\frac{|\vec v_\chi+\vec v_E|^2}{v_0^2}} \Theta(v_{\rm esc} - |\vec v_\chi+\vec v_E|) \, ,
\end{equation}
where $\vec v_\chi$ is the DM velocity in the Earth frame, and $\vec v_E$ is the Earth's velocity in the galactic rest frame.
We take $v_0=230$ km/s for the typical velocity, and $v_{\rm esc}=600$ km/s for the escape velocity. 
We take $v_E=240$ km/s for the average Earth velocity relative to the DM halo, adding (subtracting) 15 km/s for the Earth velocity in June (December). 

The normalization factor $K$ is determined by requiring $\int d^3 v g_\chi(\vec v)=1$, giving
\begin{eqnarray}
K&=& v_0^3\pi\left[ \sqrt{\pi} \textrm{Erf}\left(\frac{v_{\rm esc}}{v_0}\right)-2  \frac{v_{\rm esc}}{v_0} e^{-\left(\frac{v_{\rm esc}}{v_0}\right)^2}\right] \, .
\end{eqnarray}
Using these values, we obtain $K=6.75\times 10^{22}$ [cm/s]$^3$ or $2.50\times 10^{-9}$ in natural units. 

\noindent We then define the function $\eta(v_{\rm min})$,
\begin{align}
\eta(v_{min}) =&\int d^3 v_\chi \,g_\chi(\vec v_\chi) \frac{1}{v_\chi} \Theta(v_\chi - v_{\rm min})
\nonumber\\
=&\frac{1}{K}\int 2\pi d \cos\theta \, dv_\chi \, v_\chi ~e^{-(v_\chi^2+v_E^2-2v v_E c_\theta)/v_0^2}
 \Theta(v_\chi - v_{\rm min})  \Theta(v_{\rm esc} - v_\chi)\,,
\label{eq:etavmin}
\end{align}

\noindent where $c_\theta=\cos\theta$ is the angle between the velocity and the velocity of the Earth. We can explicitly solve Eq.~(\ref{eq:etavmin}), but need to consider two cases:
\begin{enumerate}
\item $v_{\rm min}<v_{\rm esc}-v_E$ 
\item $ v_{\rm esc}-v_E< v_{\rm min}<v_{\rm esc}+v_E$
\end{enumerate}

\noindent where $v_{esc},~v_E,~v_{min}>0.$ 

\noindent This gives us
\begin{eqnarray}
 \eta_1(v_{\rm min})&=&\frac{v_0^2\pi}{2 v_E K}\left(-4e^{-v_{\rm esc}^2/v_0^2}v_E+\sqrt{\pi}v_0\left[\textrm{Erf}\left(\frac{v_{\rm min}+v_E}{v_0}\right)-\textrm{Erf}\left(\frac{v_{\rm min}-v_E}{v_0}\right)\right]\right)\\
  \eta_2(v_{\rm min})&=&\frac{v_0^2\pi}{2 v_E K }\left(-2e^{-v_{\rm esc}^2/v_0^2}(v_{\rm esc}-v_{\rm min}+v_E)+\sqrt{\pi}v_0\left[\textrm{Erf}\left(\frac{v_{\rm esc}}{v_0}\right)-\textrm{Erf}\left(\frac{v_{\rm min}-v_E}{v_0}\right)\right]\right)\nonumber\\
\end{eqnarray}
where the subscript corresponds to the case number. Note that the two cases converge to the same value for $v_{\rm min}=v_{\rm esc}-v_E$.

%%%%%%%%%%%%%%%%%%%%%%%%%%%%%%%%%%%%
%%%%%%%%%%%%%%%%%%%%%%%%%%%%%%%%%%%%
%%%%%%%%%%%%%%%%%%%%%%%%%%%%%%%%%%%%
%%%%%%%%%%%%%%%%%%%%%%%%%%%%%%%%%%%%
\begin{figure}[t]
\begin{center}
\includegraphics[width=0.4\textwidth]{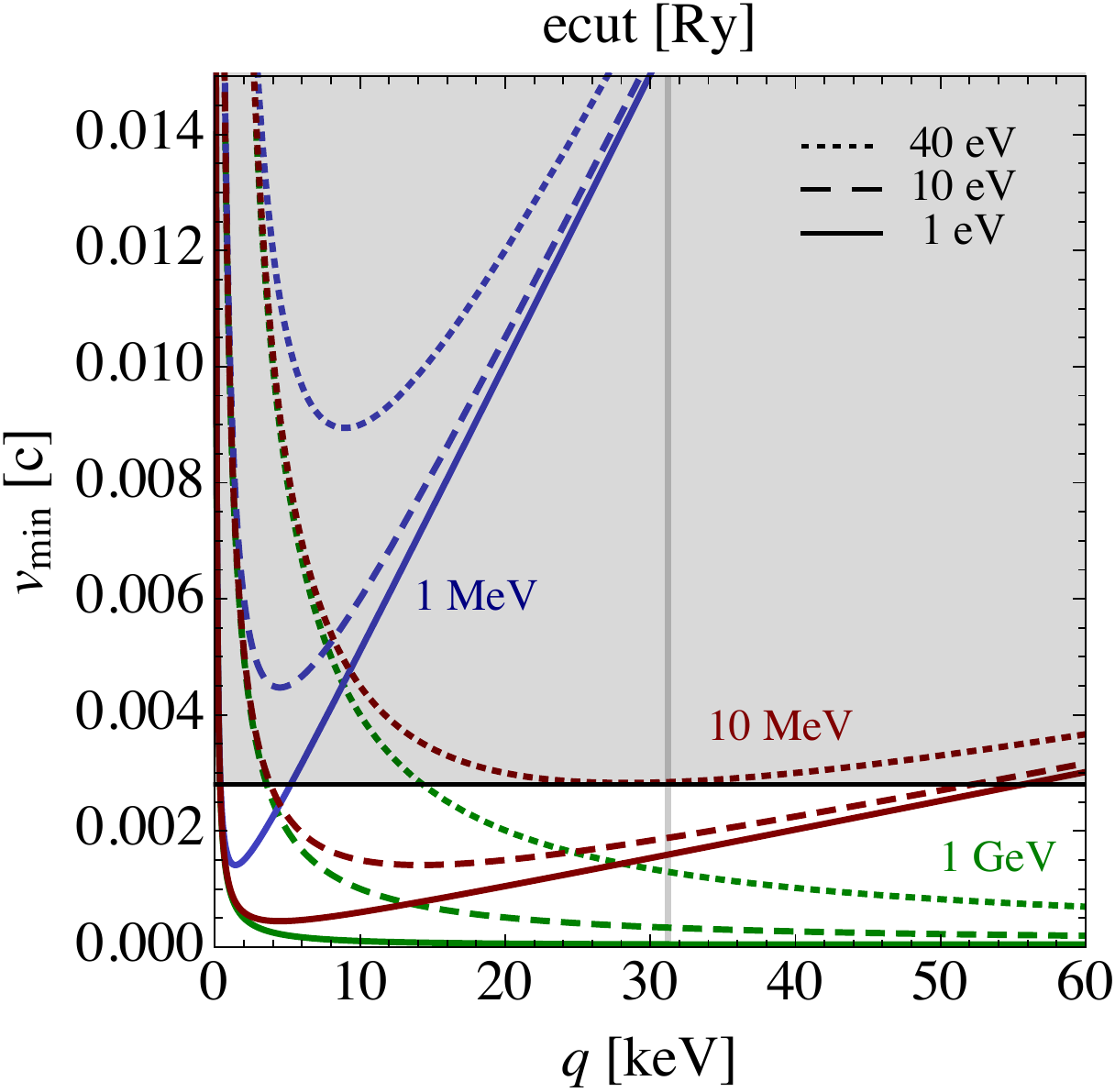}
\caption{\footnotesize The minimum velocity, $v_{min}$ as a function of $q$ for $m_\chi=1$~MeV, 10~MeV, and 1~GeV and $E_e=1$, 10, and 40~eV. 
The grey shaded region indicates where $v_{min}>v_E+v_{esc}$ and is thus prohibited. The vertical grey line indicates the maximum $q$ value for a given $E_{cut}=70$ Ry.}
\label{fig:vmin}
\end{center}
\end{figure}

\section{Convergence of the Numerical Results} 
%%%%
\label{sec:convergence}

\begin{figure}[t]
\center
\includegraphics[width=0.40\textwidth]{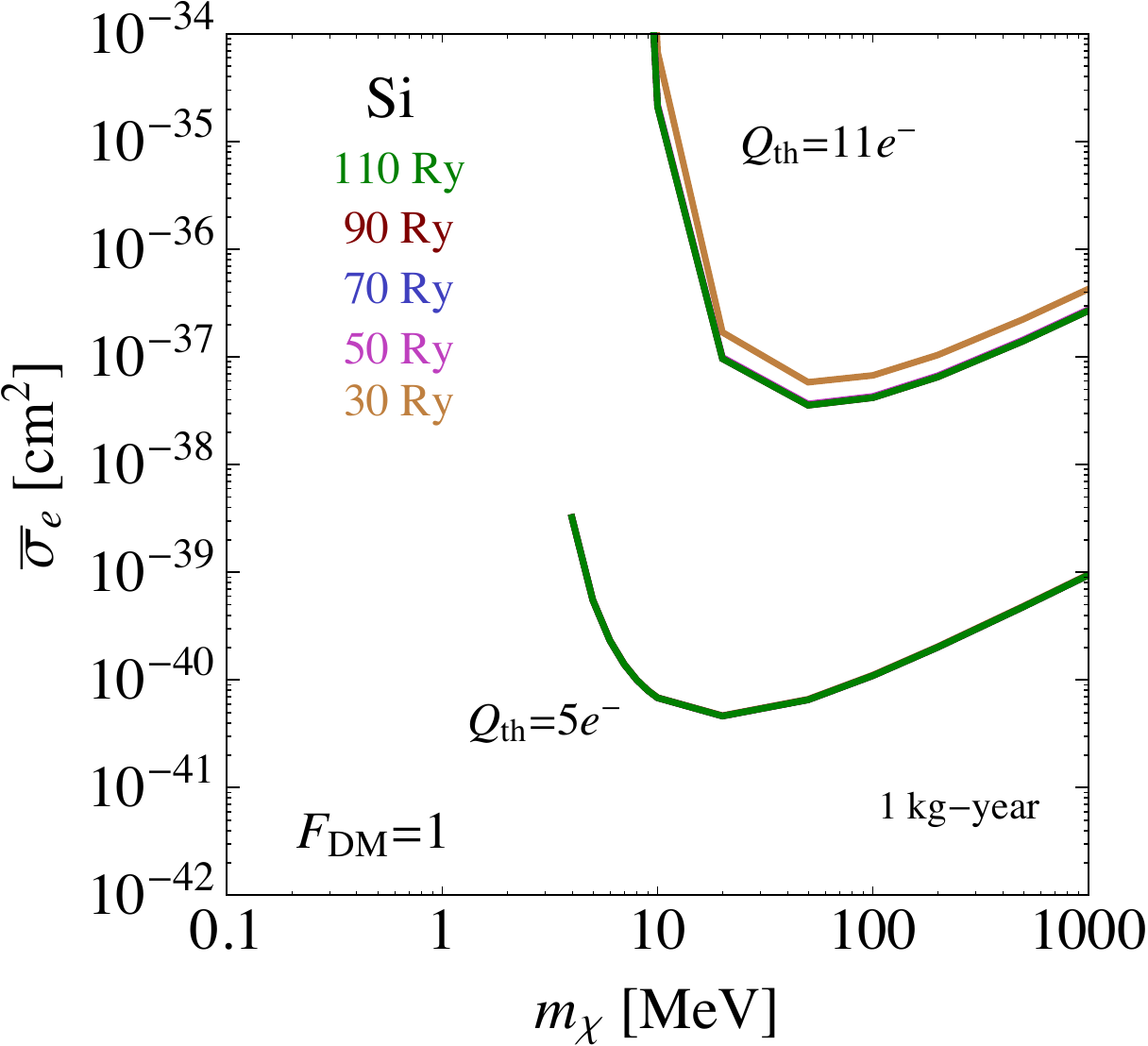}
\includegraphics[width=0.38\textwidth]{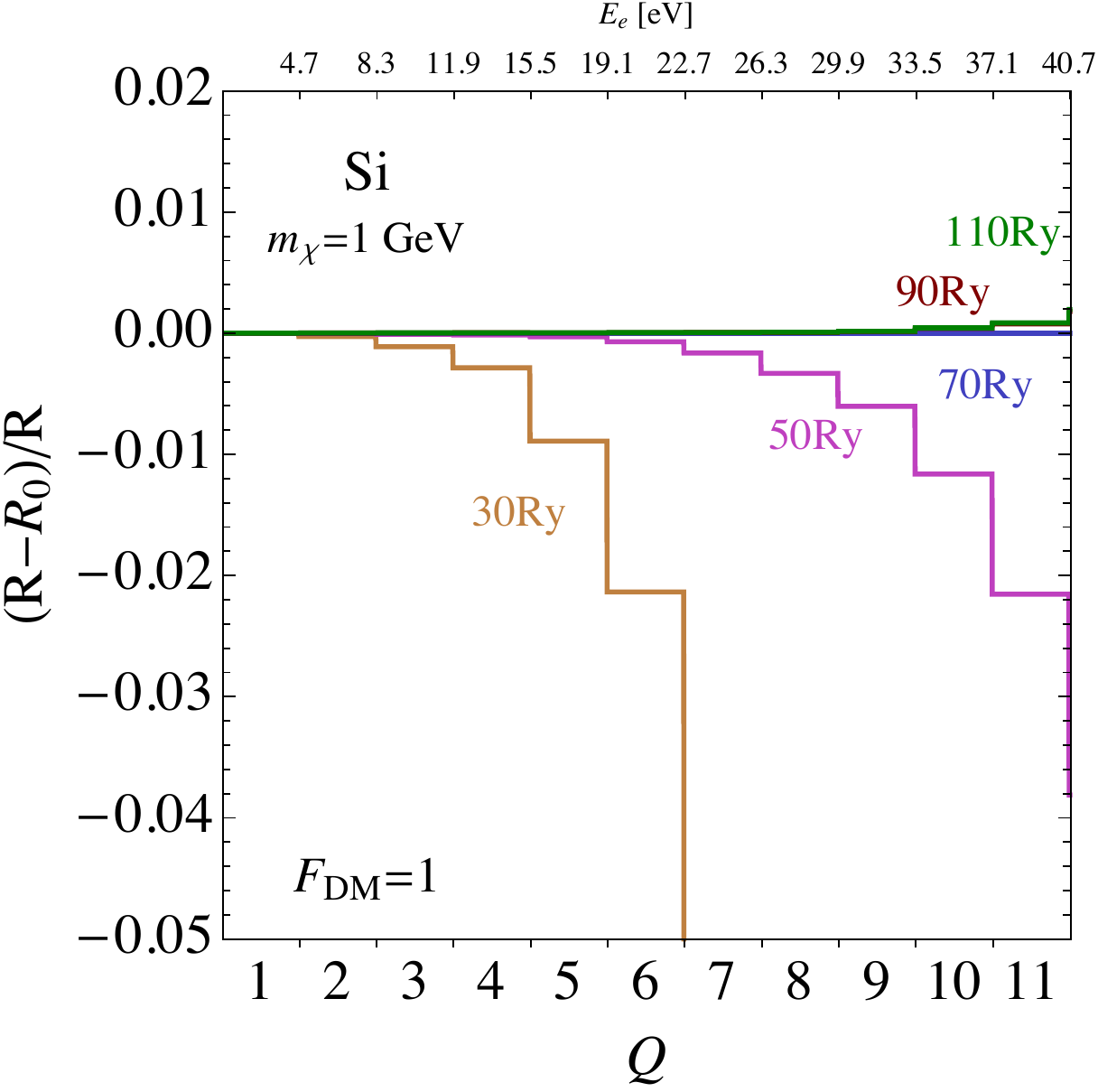}
\vspace{-5pt}
\caption{\footnotesize {\bf Left}: Cross-section sensitivities for ionization thresholds of $Q_{\rm th}=5$ and $Q_{\rm th}=11$ electrons in silicon for $E_{\rm cut}=30, 50, 70, 90$, and 110~Ry (we take a mesh consisting of 27 $k$-points).  Note that most lines are on top of each other, demonstrating the weak dependence of $\overline{\sigma}_e$ on $E_{\rm cut}$.  
{\bf Right}:  Difference in the rate, $R$, for a given $E_{\rm cut}$, to that at $E_{\rm cut} = 70$~Ry, $R_0$, in silicon for $m_\chi=1$ GeV. We see that the error in choosing $E_{\rm cut} = 70$~Ry is $<{\mathcal O}(1\%)$ for the thresholds considered in this paper.}
\label{eq:Ecut-convergence-Si}
\end{figure}

\begin{figure}[t]
\center
\includegraphics[width=0.40\textwidth]{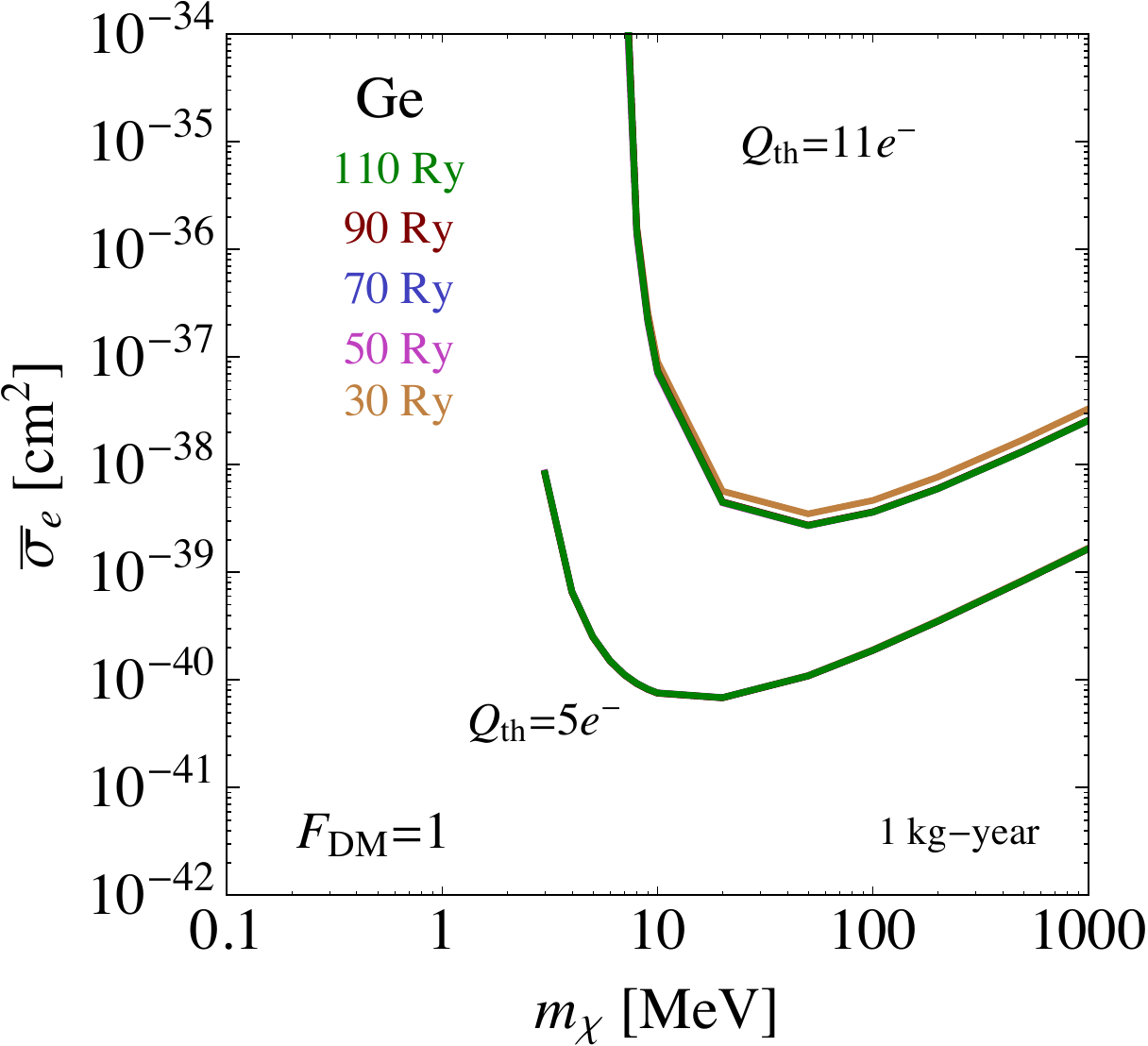}
\includegraphics[width=0.38\textwidth]{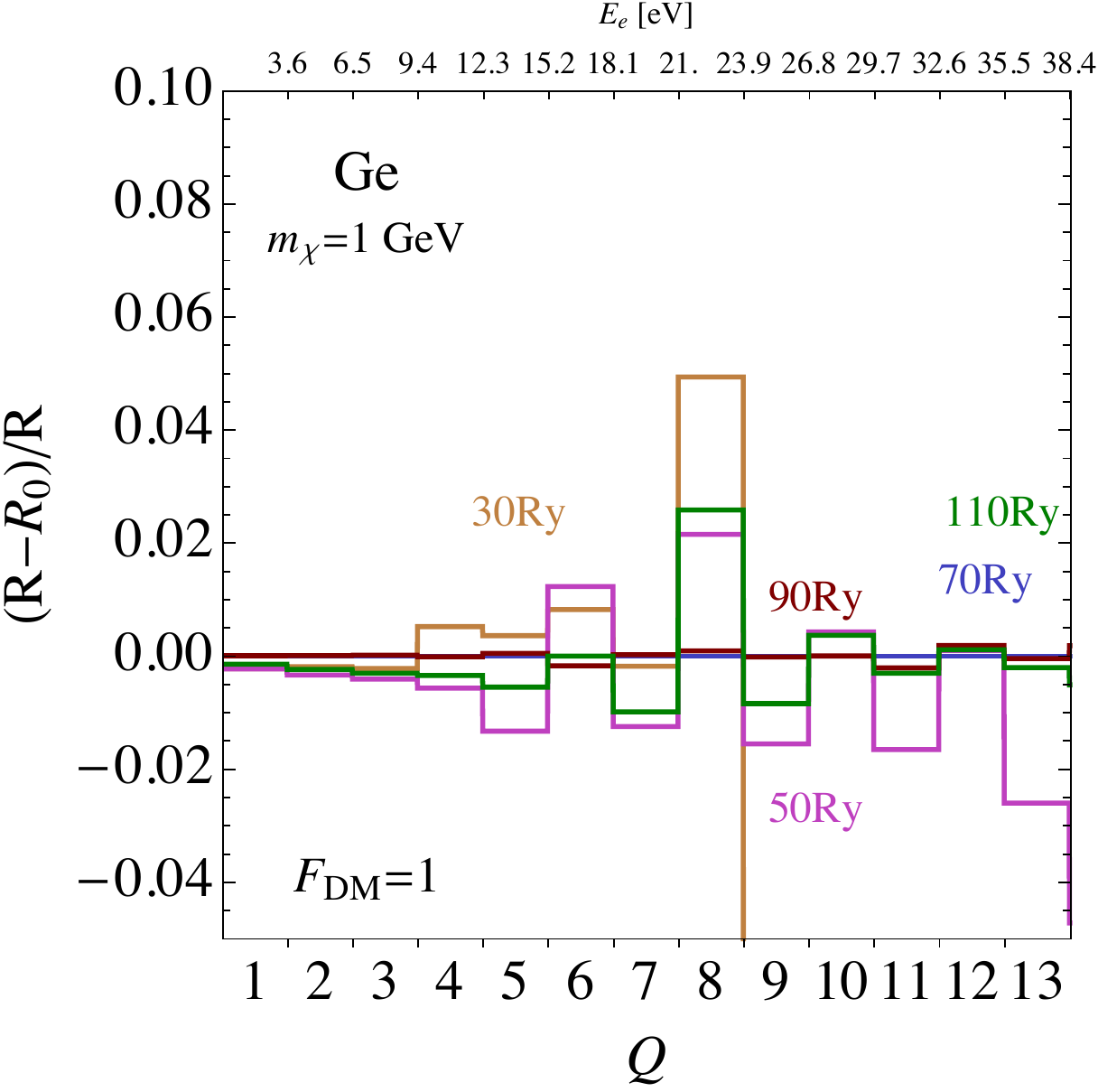}
\vspace{-5pt}
\caption{\footnotesize {\bf Left}: Cross-section sensitivities for ionization thresholds of $Q_{\rm th}=5$ and $Q_{\rm th}=11$ electrons in germanium for $E_{\rm cut}=30, 50, 70, 90$, and 110~Ry (we take a mesh consisting of 27 $k$-points).  Note that most lines are on top of each other, demonstrating the weak dependence of $\overline{\sigma}_e$ on $E_{\rm cut}$.  
{\bf Right}:  Difference in the rate, $R$, for a given $E_{\rm cut}$, to that at $E_{\rm cut} = 70$~Ry, $R_0$, in germanium for $m_\chi=1$ GeV. The structure of the distributions arise from the effect of the $3d$ electrons.}
\label{eq:Ecut-convergence-Ge}
\end{figure}

\begin{figure}[t]
\center
	\includegraphics[width=0.4\textwidth]{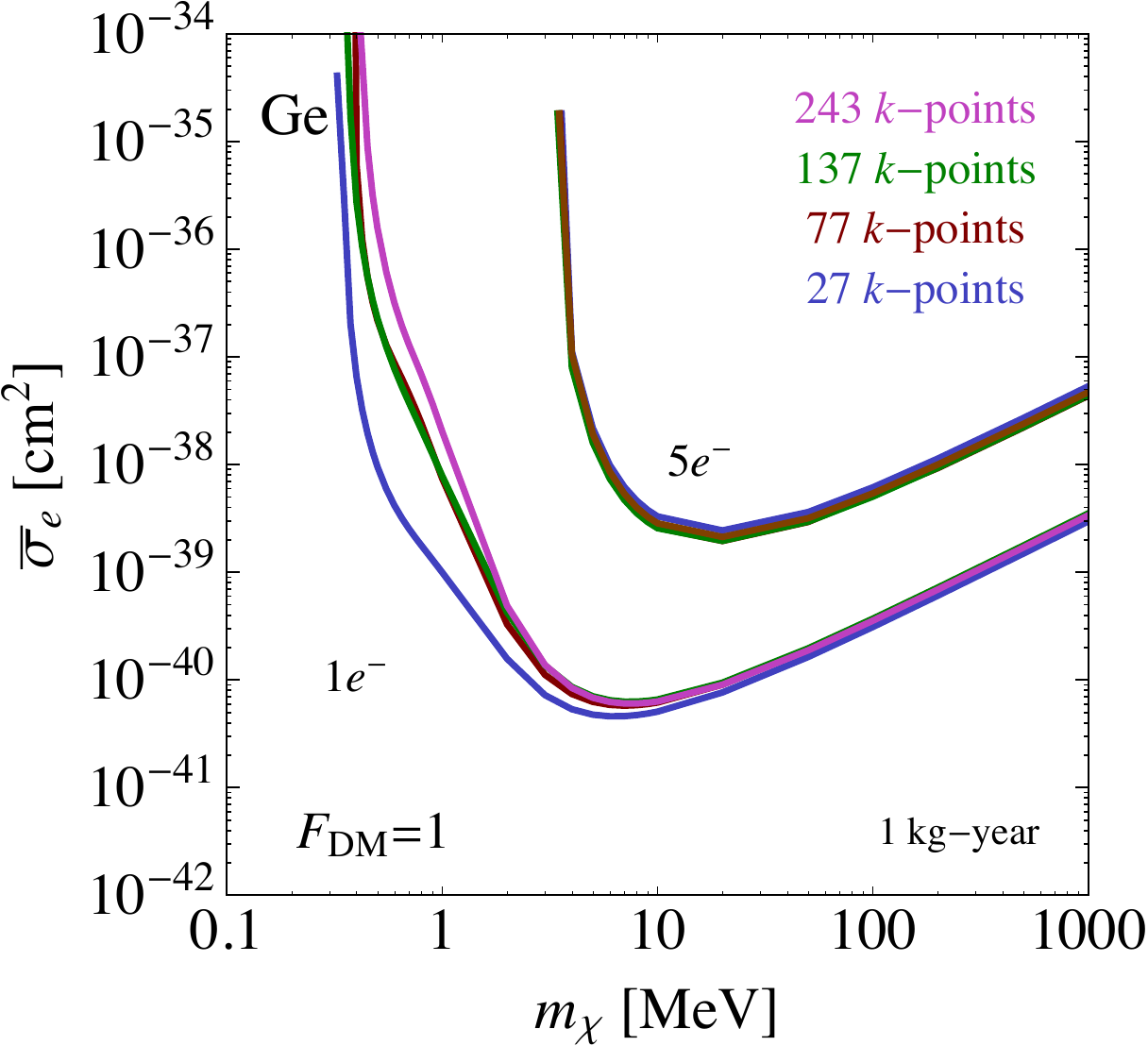}
	\includegraphics[width=0.38\textwidth]{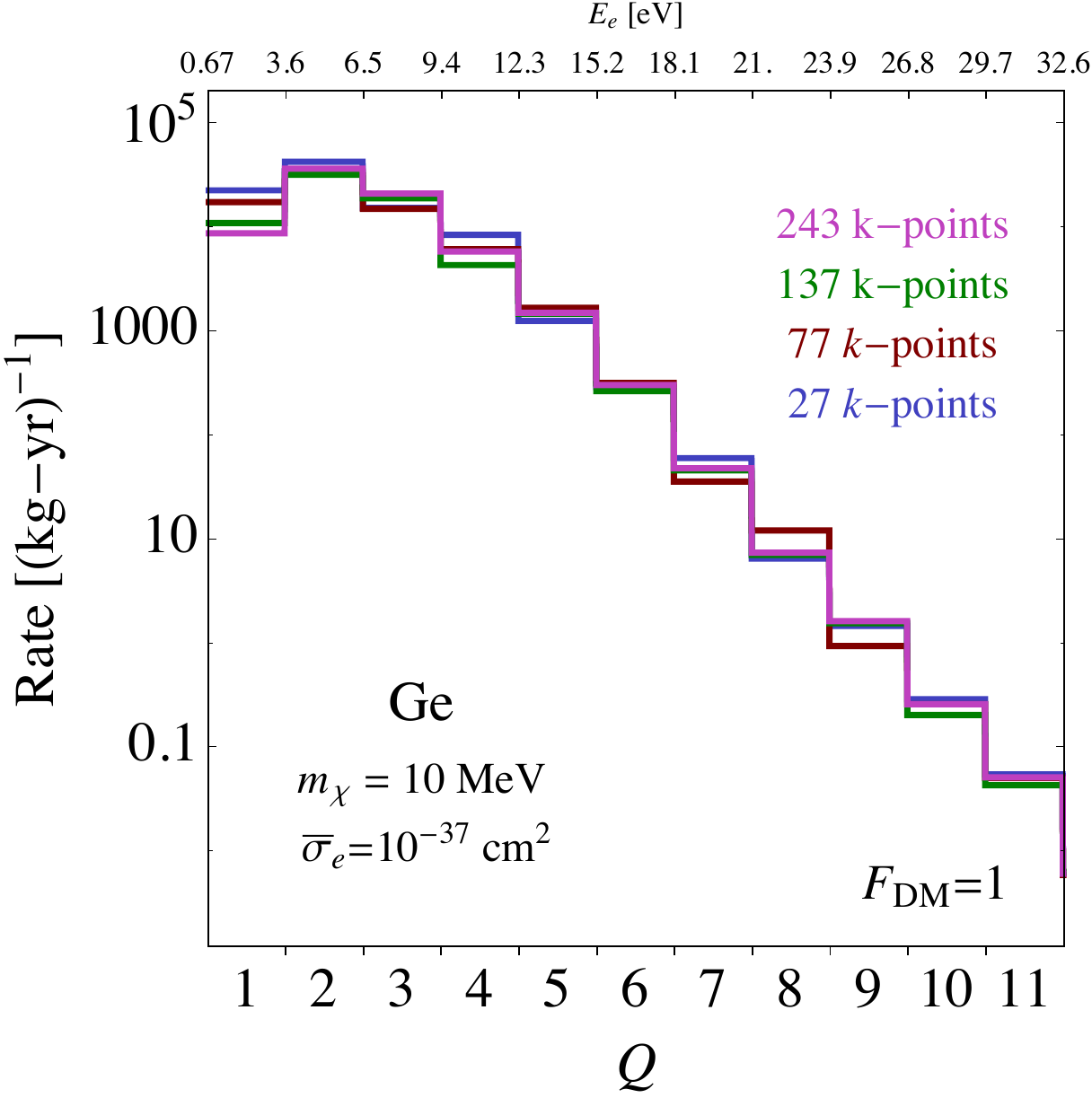}
\caption{\footnotesize {\bf Left}: Cross-section sensitivity for $Q=1$ and $Q=5$ electrons in germanium for 27, 77, 137, and 243 $k$-points.   {\bf Right}: The energy spectra using 27, 77, 137, and 243 $k$-points for $m_\chi=10$ MeV and $\overline\sigma_e=10^{-37} \rm{cm}^2$. We see that the choice of the number of $k$-points used in the mesh has an effect at low DM masses and low $Q$.} 
\label{fig:nks_convergence}
\end{figure}

\begin{figure}[t]
\center
\includegraphics[width=0.44\textwidth]{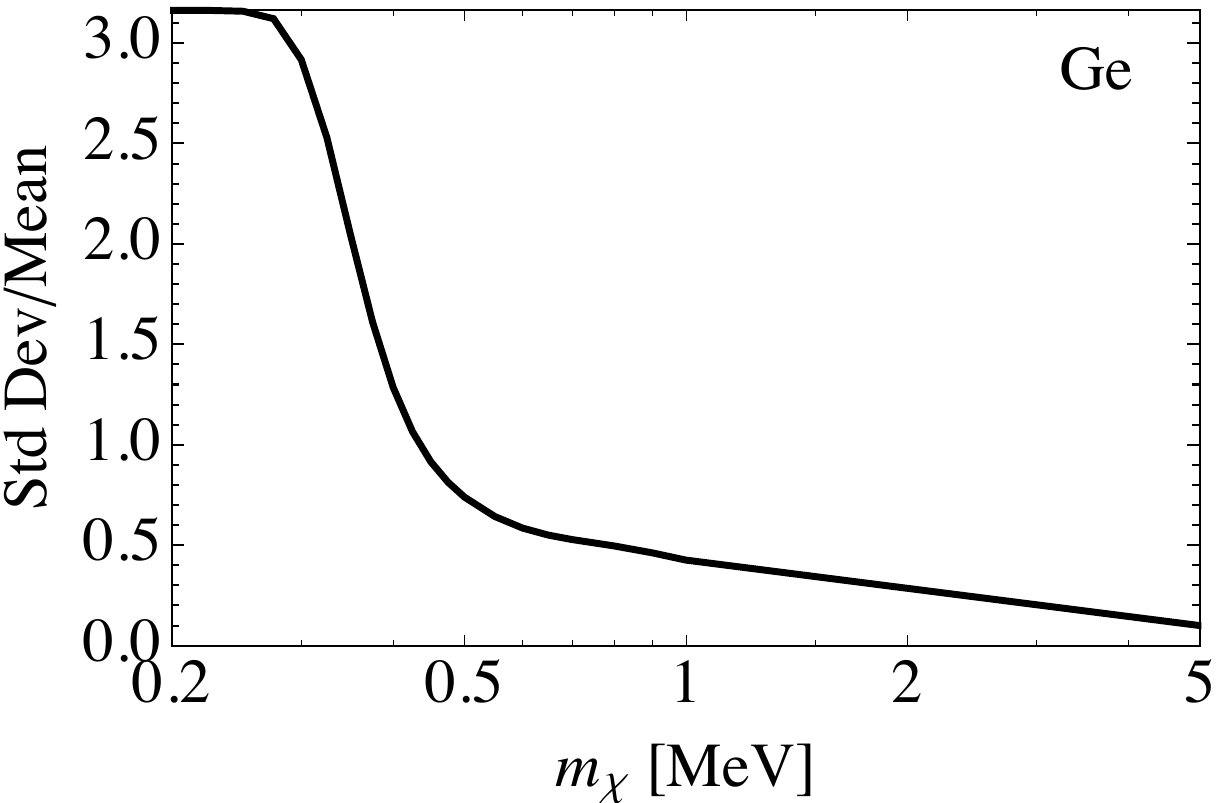}
\includegraphics[width=0.45\textwidth]{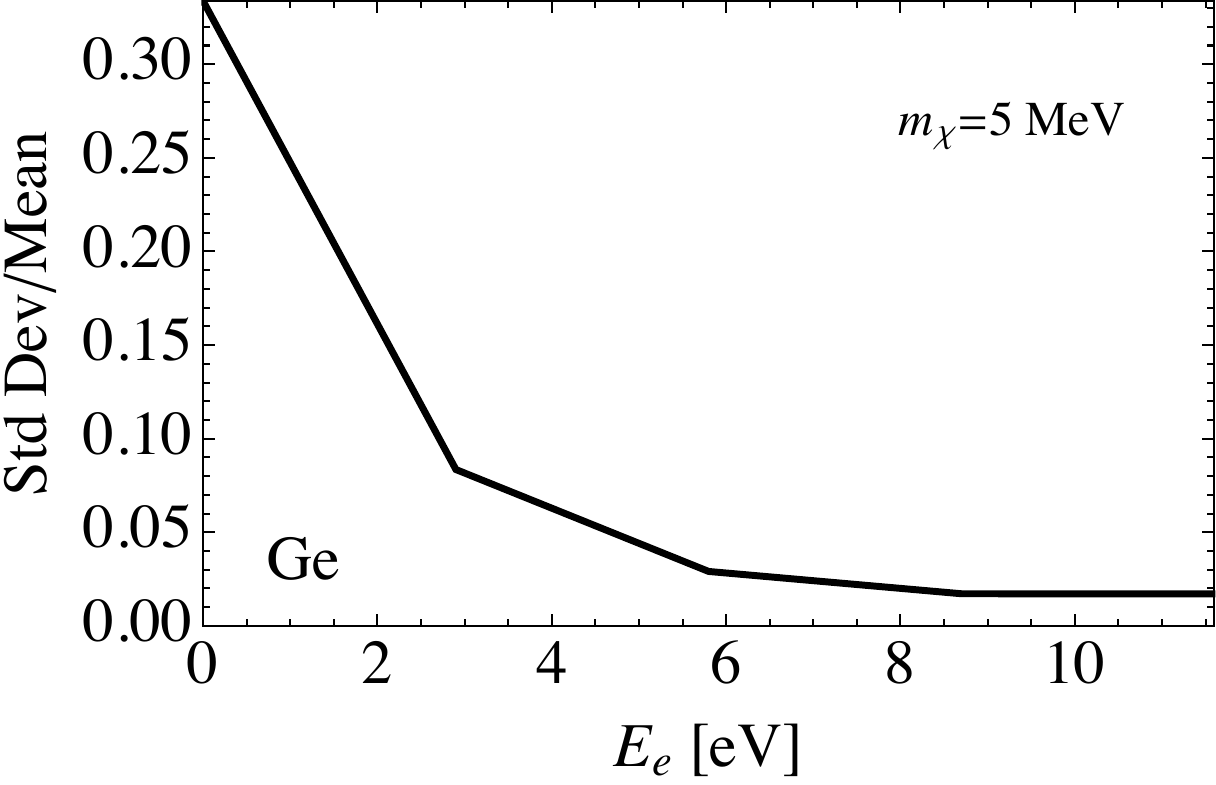}
\caption{\footnotesize The effects of our choice of $k$-point mesh on our germanium results by perturbing the mesh with random shakes of amplitudes up to half the lattice-spacing. On the left, we look at the standard deviation of the shaken runs over the mean value as a function of DM mass. On the right, we look at the standard deviation of the shaken runs over the mean value as a function of $E_e$ for $m_\chi=5$ MeV. We see that our choice of $k$-mesh spacing is accurate to a few tens of percent for masses above 1 MeV.}
\label{fig:nks_shakes}
\end{figure}

In this section, we investigate the dependence of our calculation on the kinetic-energy cutoff, $E_{\rm cut}$, (see Eq.~(\ref{eq:ecut})) 
and on the $k$-point mesh. 
The choice of $E_{\rm cut}$ determines the maximum allowed three-momentum transfer $q$, which impacts the maximum 
$E_e$ that we sample. Truncating the range of $q$ can have more of an effect for high DM masses and high electron thresholds, since these two regimes depend on high values of $q$. 
On the other hand, the $k$-points included in the mesh determine the computationally allowed values of $q$. 
A higher-resolution mesh is particularly important for low $E_e$ transitions to the bottom of the conduction band, and is therefore especially 
important for low DM masses and low electron thresholds. 

In Fig.~\ref{fig:vmin}, we show the dependence of $v_{min}$ in Eq.~(\ref{eq:vmin}) on $q$ for different $m_\chi$ and $E_e$.  The choice of $E_{\rm cut}$ (top axis) determines the range of $q$ (bottom axis).  In Fig.~\ref{eq:Ecut-convergence-Si}, we show the ratio of the rate for different values of $E_{\rm cut}$ to the rate at $E_{\rm cut} = 70$~Ry for 
$m_\chi=1$~GeV for silicon targets. We see that the error in the rate with our choice of $E_{\rm cut} = 70$~Ry is $<{\mathcal O}(1\%)$. The $E_{\rm cut}$ convergence in germanium is slightly worse due to the fact that we are solving for the 3d electrons instead of including them in the pseudopotential. This effect is greatest near the 3d shell energies (a few percent uncertainty), as seen in Fig.~\ref{eq:Ecut-convergence-Ge}. In left plots of both Fig.~\ref{eq:Ecut-convergence-Si} and ~\ref{eq:Ecut-convergence-Ge}, we see a step-like transition between 30 Ry and the other curves for 11e because 30 Ry is not a high enough energy cutoff to accurately calculate the rate for 11e. The irregular behavior in the distributions on the right side of Fig.~\ref{eq:Ecut-convergence-Ge} are from the semicore electrons in Ge. We do not see the same behavior in Fig.~\ref{eq:Ecut-convergence-Si}, which considers Si and no semicore electrons.

We investigate the effects of our choice of $k$-point mesh on our results in two 
ways.\footnote{We do this only for the valence electrons without including the 3d-shell electrons, since energy of the latter 
are nearly constant as a function of $\vec{k}$.  In any case, the 3d-shell electrons are important at large $E_e$, while the choice of $k$-point 
mesh is important only at low $E_e$.}
First, we vary the number of $k$-points in our mesh and find that there is sensitivity to our choice at low masses and thresholds, see Fig.~\ref{fig:nks_convergence}. 
Second, we perturb each point on the mesh with a random shake of amplitude up to half the lattice-spacing so as to cover the entire $k$-space.   We use an energy cutoff of $E_{\rm cut}=70$ Ry and 243 $k$-points. The amplitude of our perturbations is $\Delta k=0.08$ a.u. as the lattice spacing for 243 $k$-points is 0.17 a.u. We run 10 independent simulations and plot the results in Fig.~\ref{fig:nks_shakes}. We find that our choice of $k$-point mesh does not appreciably affect our results for masses above $\sim 1$ MeV.

%%%%%%%%%%%%%%%%%%%%%%%%%%%%%%%%%%%%%%
%%%%%%%%%%%%%%%%%%%%%%%%%%%%%%%%%%%%%%
\section{The importance of the 3d-shell in germanium}
\label{sec:3d-shell}

\begin{figure}[!h]
\center
\includegraphics[trim = 0 -20 0 0pt, clip,width=0.4\textwidth]{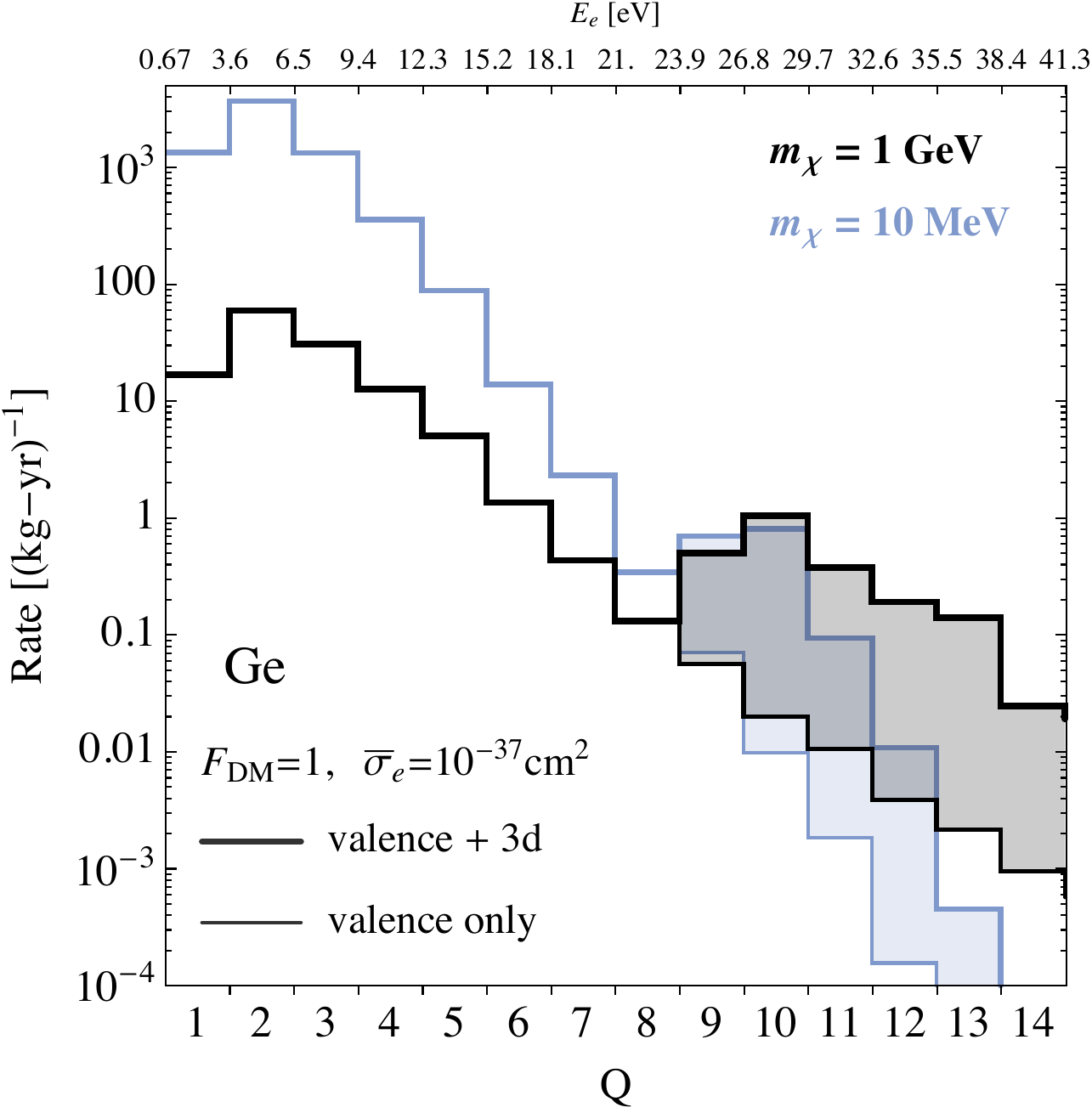}
\hspace{10pt}
\includegraphics[width=0.43\textwidth]{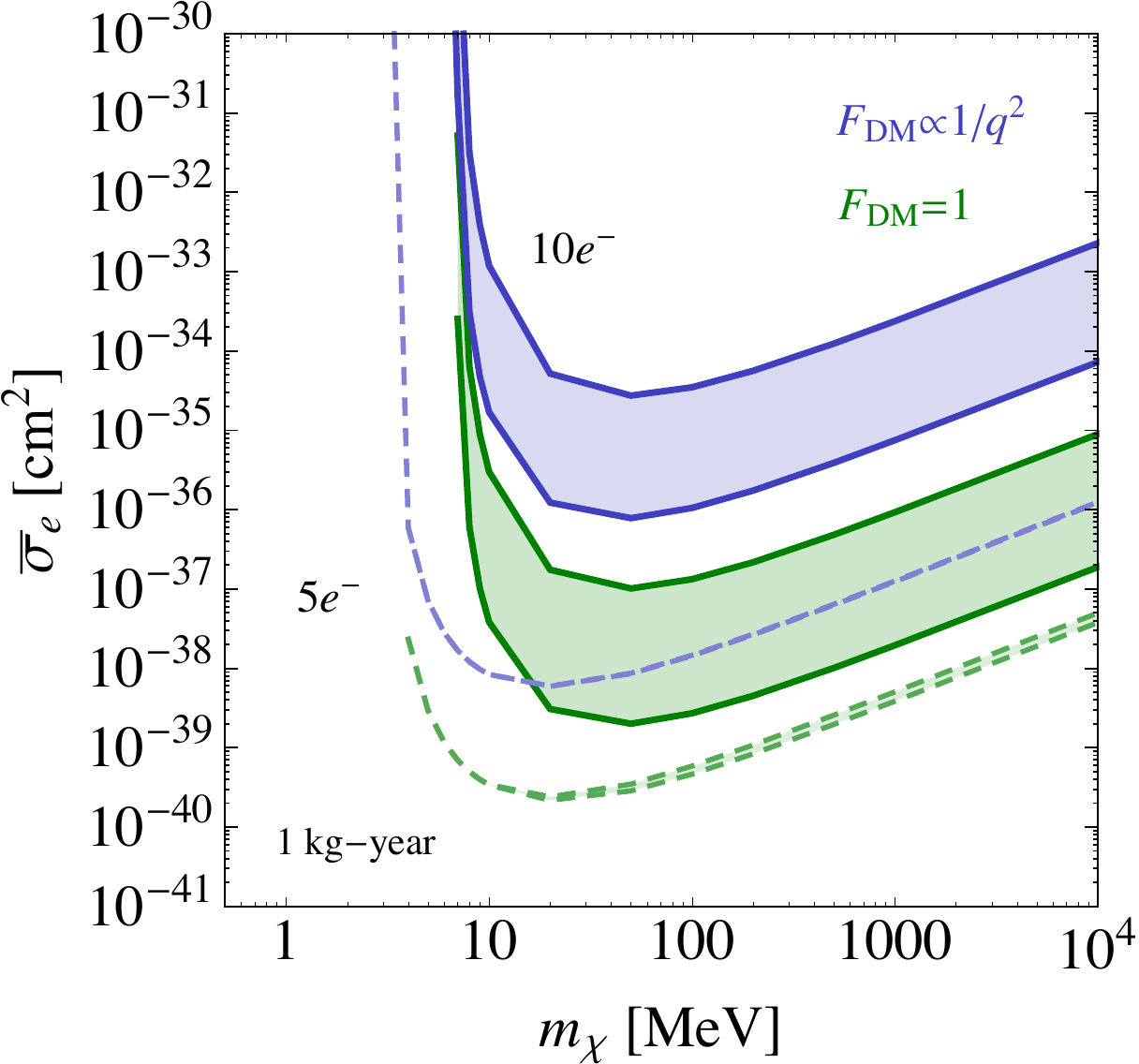}
\vspace{-10pt}
\caption{\footnotesize The importance of the 3d-shell electrons in germanium.  
{\bf Left:} Spectrum of events as function of the ionization signal $Q$, for $F_{\rm DM} = 1$ and $\overline\sigma_e = 10^{-37}$\,cm$^2$. 
The thick, upper lines show the rates including the 3d-shell electrons, while the thinner, lower lines include only the valence electrons 
(the thick and thin lines overlap for $Q\le 8$). 
The shading highlights the difference between the two.
{\bf Right:} The cross-section reach in germanium with a 1 kg-year background-free exposure, with an ionization threshold of $Q_{\rm th}=5$ for the dashed, lower lines and $Q_{\rm th}=10$ for the solid, upper lines.
The lower (upper) line of the shaded region show results with (without) the 3d-shell electrons (these overlap for the 5e threshold).
}
\label{fig:FDM_ER-core-valence}
\end{figure}

The importance of the 3d-shell electrons in germanium are illustrated in Fig.~\ref{fig:FDM_ER-core-valence}. 
We see that they dominate the rate at high recoil energies and thus for high thresholds.  
We discuss this in more detail in \S\ref{subsec:prospects}.

%%%%%%%%%%%%%%%%%%%%%%%%%%%%%%%%%%%%%%
%%%%%%%%%%%%%%%%%%%%%%%%%%%%%%%%%%%%%%
\section{A Monte Carlo model of secondary scattering}
\label{sec:MC-model}
%%%%%%%%%%%%%%%%%%%%%%%%%%%%%%%%%%%%%%

In the main results of this paper, we modeled the ionization response of a target crystal with the linear treatment described in \S\ref{sec:conversion-to-ionization-size}. 
For comparison, here we attempt to mock-up the secondary scattering with a Monte Carlo model, following~\cite{Bloom:1980}. 
The deposited energy $E_e$ is randomly split between an initial electron and hole. 
In each following step, each electron or hole with energy above a threshold $E_{\rm ion}$ then generates an extra electron-hole pair, with the energy being randomly split between the three particles. 
This is iterated until all particles have energy less than $E_{\rm ion}$. 

\begin{figure}[!t]
\begin{center}
\includegraphics[width=0.48\textwidth]{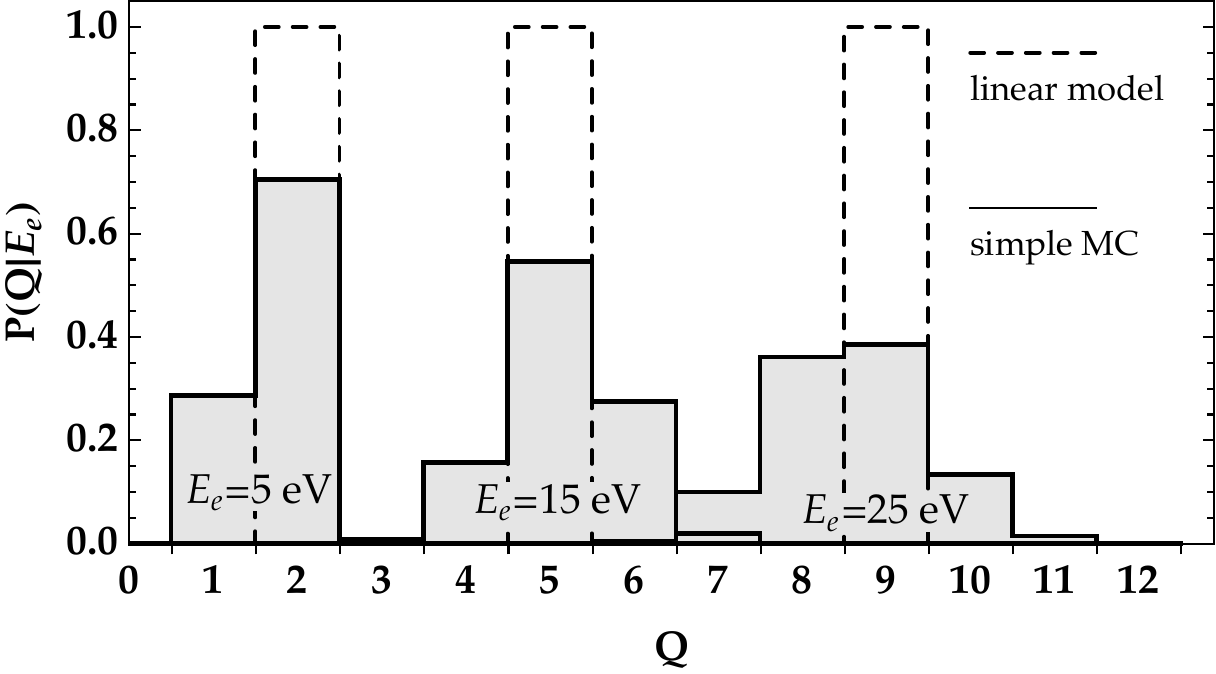}
\hfill
\includegraphics[width=0.48\textwidth]{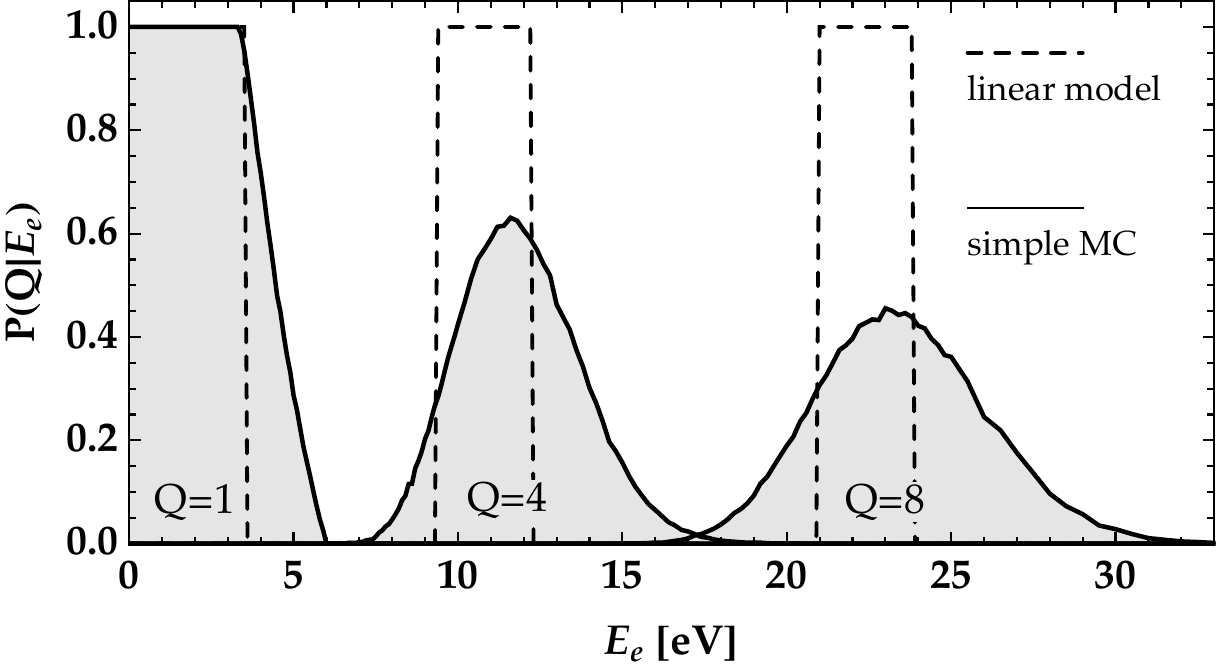}
\vspace{-5pt}
\caption{\footnotesize Probability distribution of the ionization signal $Q$ for a given energy deposition $E_e$ in germanium. The {\bf left} plot shows the distribution of $Q$ for the indicated fixed $E_e$, while the {\bf right} plot shows the probability to get a given $Q$ with varying $E_e$. Solid, filled lines show the cascade model discussed in \S\ref{sec:MC-model}, while dashed lines show the linear model described in \S\ref{sec:conversion-to-ionization-size} and used for our main results.}
\label{fig:MC-distribution}
\end{center}
\end{figure}

\begin{figure}[t]
\begin{center}
\includegraphics[width=0.48\textwidth]{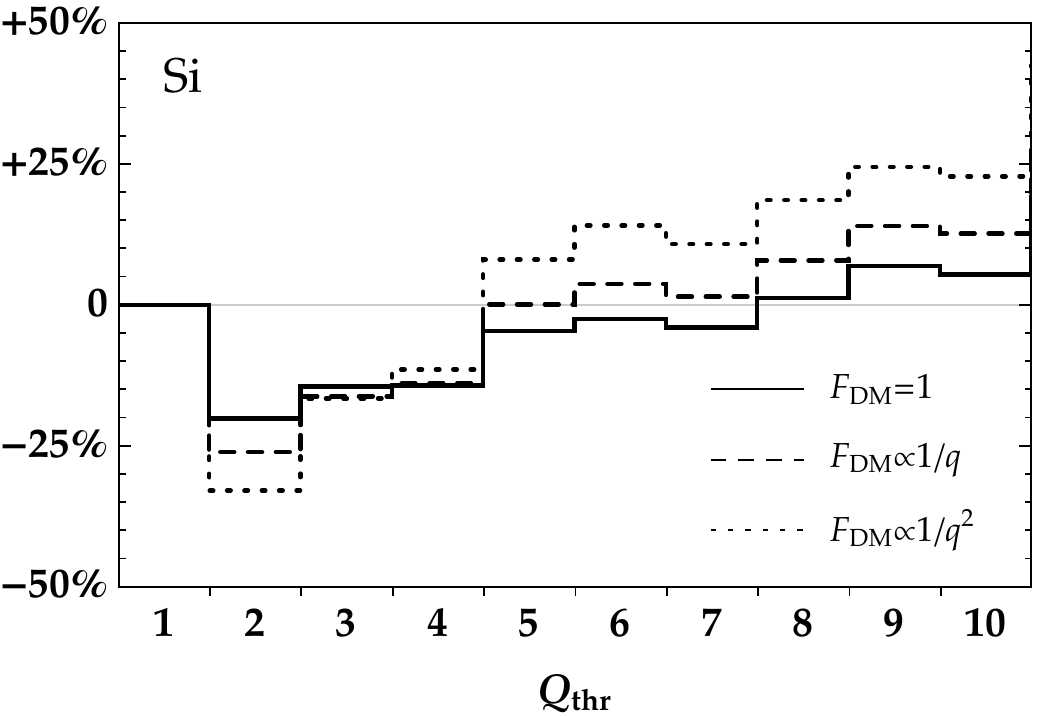}
\includegraphics[width=0.48\textwidth]{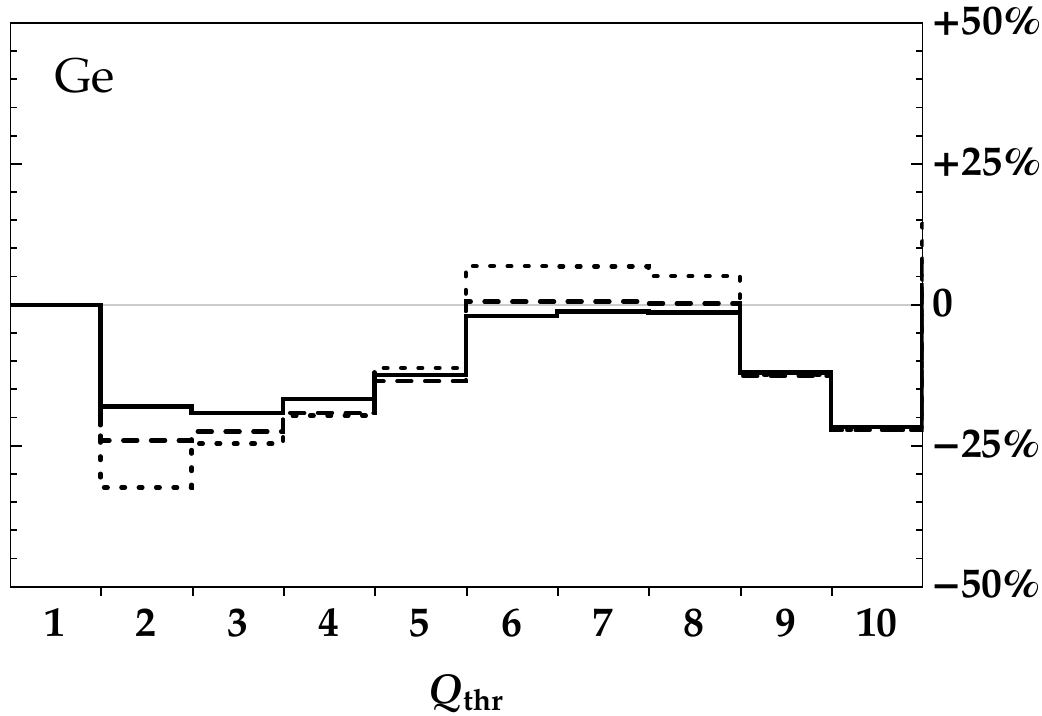}
\vspace{-5pt}
\caption{\footnotesize Fractional increase in the rate when modeling the secondary scattering with the cascade Monte Carlo instead of the linear model. (Explicitly, the $y$-axis is ($R_{\rm MC} - R_{\rm naive})/R_{\rm naive}$.) The rate here is the rate of event passing the ionization threshold $Q_{\rm thr}$, for a DM mass of 1\,GeV. 
See \S\ref{sec:MC-model} for more details.}
\label{fig:MC-rate-comparison}
\end{center}
\end{figure}

The random energy splittings follows a distribution that weights all phase space volume equally, with the density of states assumed to grow as $\sqrt{E}$ above and below the bandgap, as in a simple 2-band free electron/hole system. Explicitly, for the initial $1\to2$ splitting the probability distribution for energy $E_0$ to split into energies $E_1$ and $E_2$ has the form
\begin{equation} 
d P \propto \sqrt{E_1} \sqrt{E_2} \, \delta(E_0 - E_1 -E_2 - E_{\rm gap}) \, d E_1 \, d E_2 \, ,
\nonumber
\end{equation}
while for the subsequent $1\to3$ splittings it has the form
\begin{equation} 
d P \propto \sqrt{E_1} \sqrt{E_2} \sqrt{E_3} \, \delta(E_0 - E_1 -E_2 - E_3 - E_{\rm gap}) \, d E_1 \, d E_2 \, d E_3 \, ,
\nonumber
\end{equation}
where electon/hole energies are measured above/below the upper/lower edge of the band gap.
We ignore phonon losses during the cascade -- these are understood to be quantitatively fairly small, and should not affect the 
qualitative conclusions. 

The output of the Monte Carlo model is a probability distribution $P(Q|E_e)$ to get ionization $Q$ given a deposited energy $E_e$. 
Given the band-gap energy of $E_{\rm gap} = 0.67$\,eV (1.11\,eV) in germanium (silicon), we find that $E_{\rm ion} = 2.67$\,eV (3.1\,eV) reproduces the measured values of $\varepsilon$  for high energy recoils (see Eq.~\eqref{eq:epsilon}). The distributions for both elements have Fano factors of $F\approx 0.1$ for all energies above $\sim\!10$\,eV, consistent with measurements. 
We illustrate the probability distributions in Fig.~\ref{fig:MC-distribution}. 
In Fig.~\ref{fig:MC-rate-comparison}, we show the effect on the event rate of using this model rather than the naive linear model of \S\ref{sec:conversion-to-ionization-size}. 
For thresholds of 2 to 4 electron-hole pairs, downward fluctuations reduce the rate compared to the naive estimate. For higher thresholds, occasional upward fluctuations combined with the steeply falling recoil spectrum lead to an increase in the rate.
However, the two models are consistent within a few tens of percent.

%%%%%%%%%%%%%%%%%%%%%%%%%%%%%%%%%%%%
%%%%%%%%%%%%%%%%%%%%%%%%%%%%%%%%%%%%
%%%%%%%%%%%%%%%%%%%%%%%%%%%%%%%%%%%%
%%%%%%%%%%%%%%%%%%%%%%%%%%%%%%%%%%%%
\section{Review of Density Functional Theory and Pseudopotentials}
%%%%
\label{sec:DFT-review}

In this appendix, we review the formalism of density functional theory (DFT), explain in 
more detail the approximations used in the computation of the wavefunctions, and further explain the numerical methods.  

\subsection{Electronic structure within DFT}
Non-relativistic electrons interacting electrostatically with fixed nuclei are described by the electronic structure Schr\"odinger equation
\begin{equation}
\left[ - \frac{1}{2m_e} \sum_{\alpha} \nabla_\alpha^2 - \sum_{\alpha, I} \frac{Z_I e^2}{ \left| \vec{r}_\alpha - \vec{R}_I \right| } 
  +  \sum_{\alpha < \beta} \frac{e^2}{ \left| \vec{r}_\alpha - \vec{r}_\beta \right| } \right] 
  \Psi_i \left( \vec{r}_1 , \dots , \vec{r}_N \right) 
  = \varepsilon_i \Psi_i \left( \vec{r}_1 , \dots , \vec{r}_N \right),
\end{equation}
where $\alpha, \beta = 1, 2, \cdots, N$ label electrons, $I=1, 2, \cdots, M$ labels nuclei, and $Z_I$ is the atomic number of nuclei $I$. The first term in the Hamiltonian is the electron kinetic energy $T$, the second term is the Coulomb electron-nucleus attraction $V_{\text{ext}}$ and the third term is the electron-electron Coulomb repulsion $V_{ee}$. The constant nuclei-nuclei term has been omitted. Even though the question is well-posed, obtaining the many-electron wavefunctions $\Psi_i \left( \vec{r}_1 , \dots , \vec{r}_N \right) $ computationally is an extremely difficult task because of the exponential scaling of the problem with the number of electrons $N$. This method becomes then helpless for applications of interest, so in practice one needs to resort to approximate methods. 

DFT is a reformulation of the interacting quantum many-body problem in terms of functionals of the particle density $n(\mathbf{r})$. For the case of electrons, the Hohenberg-Kohn theorems \cite{Hohenberg:1964zz} imply that all properties of the interacting system are determined once the ground state electron density is known. Minimizing an energy functional $E \left[ n \right]$ will provide the ground state density $n_0(\mathbf{r})$ and the ground state energy $E_0$. Unfortunately this energy functional is not known in general. The Kohn-Sham method~\cite{Kohn:1965zzb} overcomes this obstacle by replacing the description strictly in terms of functionals for a wavefunction formulation: the system of interacting electrons with Hamiltonian $H = T + V_{ee} + V_{\text{ext}} $ is mapped into a system of independent electrons under the presence of an auxiliary potential $\tilde{H} = T + V_{aux} + V_{\text{ext}}$ which produces the same ground state density as $H$. This is of great advantage because, once this mapping is built, one has to solve the much simpler independent-particle system in order to obtain $E_0$ and $n_0(\mathbf{r})$. However, this comes at the expense of having to use an approximate auxiliary potential $V_{\text{aux}}$. Typically, $V_{\text{aux}}$ is split into the mean-field Hartree potential $V_{\text{Hartree}} (\vec{r}) = e \int d^3 \vec{r}' n(\vec{r}')/ \left| \vec{r} - \vec{r}' \right|$ and the so-called exchange-correlation potential $V_{xc}$, where the approximations are imposed. Once an approximation for $V_{xc}$ has been chosen, the non-interacting electron Schr\"odinger equation
\begin{equation}
	\label{eq:KS}
	\left[ - \frac{\nabla^2}{2m_e}  +V_{\text{ext}} (\vec{r})  
  + V_{\text{Hartree}} ( \vec{r} ) + V_{xc} ( \vec{r} ) \right] \psi_i (\vec{r} ) = \epsilon_i \psi_i (\vec{r} ),
\end{equation}
which are known as the Kohn-Sham equations, are solved to get the auxiliary Kohn-Sham wavefunctions $\psi_i (\vec{r} )$.  
From these, the density can be obtained as $n(\vec{r}) = \sum_i f_i \left| \psi_i (\vec{r} ) \right|^2$, where  $f_i$ are the occupation numbers ($f_i=2$ for spin-unpolarized systems) as well as the ground state energy by evaluating the energy density functional\footnote{The connection between an energy functional and its corresponding local potential is $E[n] = \int d^3 \vec{r} \,n(\vec{r}) \, V(\vec{r})$.} $E \left[ n \right] = T\left[ n \right] + E_{\text{ext}} \left[ n \right]+ E_{\text{Hartree}}\left[ n \right] + E_{xc} \left[ n \right] $,\footnote{The kinetic energy is calculated from the Kohn-Sham wavefunctions as $T=1/2m_e \sum_i \left| \nabla \psi_i \right|^2 $.} as well as a set of wavefunctions $\psi_i$ and eigenenergies $\varepsilon_i$. This problem is solved self-consistently until convergence is reached.

Expanding the wavefunctions in a finite plane-wave basis with elements labeled by the vectors $\vec G$ and $\vec G'$, the Kohn-Sham equations become the matrix equations
\begin{equation}
\label{eq:KSreciprocal}	
		\sum_{\vec{G}'} H_{\vec{G}, \vec{G}'} ( \vec{k} ) u_i(\vec{k} + \vec{G}) = E_i (\vec{k}) u_i(\vec{k} + \vec{G}).
\end{equation} 
where the Hamiltonian is 
\begin{equation} 
H_{\vec{G}, \vec{G}'} ( \vec{k} ) = \left\langle \vec{k} + \vec{G} \right| H \left| \vec{k} + \vec{G}' \right\rangle = \frac{1}{2m_e} \left| \vec{k} + \vec{G} \right|^2 \delta_{\vec{G}, \vec{G}'} + V( \vec{G} - \vec{G}')\,. 
\end{equation} 
It should be noted that, since the potential is local, its reciprocal space form does not depend on $\vec{k}$. Furthermore, the Kohn-Sham equations in reciprocal space Eq.~(\ref{eq:KSreciprocal}) decouple different $\vec{k}$'s, so the eigenvalue problem can be carried out independently at each $\vec{k}$. 

Despite all the successes of DFT, several notable shortcomings are known today. The most relevant one for us is that DFT is known to give an incorrect band gap. This is due to a discontinuity in the DFT exchange-correlation potential $\delta E_{xc}/\delta n(\vec r)$ when electrons are added above the gap~\cite{PhysRevLett.51.1888,PhysRevLett.51.1884}. There are methods based on many-body perturbation theory to improve the DFT band gap and band shapes, such as the GW method~\cite{HedinGW}. However, since the largest contribution to the scattering rate comes from the low energy excitations, we choose to follow an empirical ``scissor correction" approach~\cite{PhysRevLett.63.1719,PhysRevB.43.4187}. In this approach a rigid shift is imposed on the conduction bands with respect to the valence bands in order to set the band gap to the experimental values of $1.11$~eV for silicon and $0.67$~eV for germanium~\cite{ExptGaps}. 
It is worth noting that the semiconductor band gap features a temperature variation of around $10$ meV \cite{PhysRevB.31.2163}, but we are choosing the room temperature band gap values for our calculation.

\subsection{Energy Density Functionals}
In order to be able to use DFT, a choice for the exchange-correlation functional $ E_{xc}\left[ n \right]$ is required. 
%There exists a zoo of functionals in the literature, from highly empirical functionals to very fundamental ones based only on physical considerations.  
%We use the latter approach in this work. 
The Local Density Approximation (LDA) \cite{Kohn:1965zzb} has been remarkably successful because of its simplicity and transferability. In this method the exchange-correlation energy functional is based only on physical considerations and is approximated locally by the energy of a homogeneous electron gas (HEG) with the following density:
\begin{equation}
	E_{xc}^{\text{LDA}} \left[ n \right] = \int d^3 \vec r \, n ( \vec{r} ) \, \left[ \epsilon_x^{\text{HEG}} ( n ( \vec{r} ) ) + \epsilon_c^{\text{HEG}} ( n ( \vec{r} ) ) \right].
\end{equation}
The HEG exchange \cite{DiracExchange} and correlation \cite{CACorrelation} energy functionals are well established. There are some faults in the LDA which are known to be most dramatic where the electrons are highly localized and exchange repulsions are significant. In order to correct for that, the Generalized Gradient Approximations (GGA) introduce a dependence on the density gradient in the exchange-correlation energy density
\begin{equation}
	E_{xc}^{\text{GGA}} \left[ n \right] = \int d^3 \vec r \, n ( \vec{r} ) \, \epsilon_{xc}^{\text{GGA}} ( n ( \vec{r} ) , \left| \nabla n ( \vec{r} ) \right| ).
\end{equation}
In this work we choose the well-established PBE functional \cite{PhysRevLett.77.3865} which is known to produce a broad set of properties of materials to accuracies of order a few percent \cite{JChemPhys.110.5029}. Since LDA functionals tend to underestimate the energies of excited states compared to GGA functionals, we find a difference in cross-section sensitivity of around 10-20\%, with a larger difference at higher thresholds. 

\subsection{Pseudopotentials}
The valence electrons are responsible for the formation of interatomic bonds and their wavefunctions are in general delocalized, spanning over interatomic distances. The core electron wavefunctions, however, are very localized around the nucleus and they barely change from the isolated atom to the condensed matter phase. This fact allows to use the atomic core electron wavefunctions in the condensed matter phase by replacing the bare positive nuclear Coulomb potential and the negative Coulomb potential generated by the core wavefunctions with a \textit{pseudopotential} in the Kohn-Sham problem~Eq.~(\ref{eq:KS}). The advantage is two-fold: first, the number of electrons in the problem is reduced to the number of valence electrons and second, the only wavefunctions to be calculated are valence wavefunctions which, since they are rather smooth, do not require as fine a grid to represent them as a core electron wavefunction would, thus improving the computational efficiency. In this work we use Vanderbilt-type ultrasoft pseudopotentials \cite{PhysRevB.31.2163}. The pseudopotential for Si includes the 3s and 3p electrons in the valence, while in the case of germanium, we use a pseudopotential which includes the 3d, 4s and 4p electrons in the valence.

\bibliography{LDM-escattering-semiconductor.JHEP}
\bibliographystyle{JHEP}
\end{document}